\documentclass[twocolumn]{aastex631}

\usepackage{xspace}
\usepackage{amsmath}
\usepackage{mathrsfs}
\usepackage{pgffor}
\usepackage{graphicx}
\usepackage{ifthen}
\usepackage{booktabs}
\usepackage{afterpage}
\newenvironment{rotatepage}%
    {\clearpage\pagebreak[4]\global\pdfpageattr\expandafter{\the\pdfpageattr/Rotate 90}}%
    {\clearpage\pagebreak[4]\global\pdfpageattr\expandafter{\the\pdfpageattr/Rotate 0}}%

\newcommand{\Kepler}{\textit{Kepler}\xspace}
\newcommand{\TESS}{\textit{TESS}\xspace}
\newcommand{\Gaia}{\textit{Gaia}\xspace}

\newcommand{\Exofast}{\texttt{EXOFASTv2}\space}

\newcommand{\bjdtdb}{\ensuremath{\mathrm{BJD}_\mathrm{TDB}}\xspace}
\newcommand{\Rstar}{\ensuremath{R_{\star}}\xspace} 
\newcommand{\Mstar}{\ensuremath{M_{\star}}\xspace}
\newcommand{\Rjup}{\ensuremath{R_\mathrm{J}}\xspace} 
\newcommand{\Mjup}{\ensuremath{M_\mathrm{J}}\xspace}
\newcommand{\Rearth}{\ensuremath{R_\oplus}\xspace} 

\newcommand{\Rsun}{\ensuremath{R_\odot}\xspace} 
\newcommand{\Msun}{\ensuremath{M_\odot}\xspace}
\newcommand{\Rp}{\ensuremath{R_p}\xspace}
\newcommand{\Mp}{\ensuremath{M_p}\xspace}
\newcommand{\Teff}{\ensuremath{T_\mathrm{eff}}\xspace}
\newcommand{\logg}{\ensuremath{\log{g}}\xspace}
\newcommand{\feh}{\ensuremath{\mathrm{[Fe/H]}}\xspace}
\newcommand{\vsini}{\ensuremath{v\sin{i}}\xspace}
\newcommand{\jit}{\ensuremath{\sigma_{\mathrm{jit}}\xspace}}

\newcommand{\ms}{\ensuremath{\mathrm{m}\,\mathrm{s}^{-1}}\xspace}
\newcommand{\kms}{\ensuremath{\mathrm{km}\,\mathrm{s}^{-1}}\xspace}
\newcommand{\gcc}{\ensuremath{\mathrm{g}\,\mathrm{cm}^{-3}}\xspace}

\def\tois{2193,2207,2236,2421,2567,2570,3331,3540,3693,4137}

\begin{document}
\title{The TESS Grand Unified Hot Jupiter Survey.\ I.\ Ten TESS Planets}
\shorttitle{Ten Hot Jupiters from TESS}

\author[0000-0001-7961-3907]{Samuel W.\ Yee}
\email{swyee@princeton.edu}
\author[0000-0002-4265-047X]{Joshua N.\ Winn}
\author[0000-0001-8732-6166]{Joel D.\ Hartman}
\affiliation{Department of Astrophysical Sciences, Princeton University, 4 Ivy Lane, Princeton, NJ 08544, USA}
\author[0000-0001-8812-0565]{Joseph E. Rodriguez}
\affiliation{Department of Physics and Astronomy, Michigan State University, East Lansing, MI 48824, USA}
\author[0000-0002-4891-3517]{George Zhou}
\affiliation{University of Southern Queensland, Centre for Astrophysics, West Street, Toowoomba, QLD 4350, Australia}
\author[0000-0002-8964-8377]{Samuel~N.~Quinn}
\affiliation{Center for Astrophysics \textbar \ Harvard \& Smithsonian, 60 Garden St, Cambridge, MA 02138, USA}
\author[0000-0001-9911-7388]{David~W.~Latham}           % TESS Architect
\affiliation{Center for Astrophysics \textbar \ Harvard \& Smithsonian, 60 Garden St, Cambridge, MA 02138, USA}
\author[0000-0001-6637-5401]{Allyson~Bieryla}           % SG2
\affiliation{Center for Astrophysics \textbar \ Harvard \& Smithsonian, 60 Garden St, Cambridge, MA 02138, USA}
\author[0000-0001-6588-9574]{Karen~A.~Collins}          % SG1 Important People
\affiliation{Center for Astrophysics \textbar \ Harvard \& Smithsonian, 60 Garden St, Cambridge, MA 02138, USA}
% Alphabetized List below this
\author[0000-0003-3216-0626]{Brett C. Addison}			% MINERVA Builders
\affiliation{University of Southern Queensland, Centre for Astrophysics, West Street, Toowoomba, QLD 4350 Australia}
\author[0000-0002-9751-2664]{Isabel~Angelo}				% HIRES Observers
\affiliation{Department of Physics \& Astronomy, University of California Los Angeles, Los Angeles, CA 90095, USA}
\author[0000-0003-1464-9276]{Khalid~Barkaoui}			% SG1
\affiliation{Astrobiology Research Unit, Université de Liège, 19C Allée du 6 Août, 4000 Liège, Belgium}
\affiliation{Department of Earth, Atmospheric and Planetary Sciences, Massachusetts Institute of Technology, Cambridge, MA 02139, USA}
\affiliation{Instituto de Astrof\'isica de Canarias (IAC), E-38205 La Laguna, Tenerife, Spain}
\author[0000-0001-6981-8722]{Paul Benni}				% SG1
\affiliation{Acton Sky Portal (Private Observatory), Acton, MA, USA}
\author[0000-0001-6037-2971]{Andrew~W.~Boyle}			% SG3
\affiliation{Caltech/IPAC-NASA Exoplanet Science Institute, 770 S. Wilson Avenue, Pasadena, CA 91106, USA}
\author[0000-0002-9158-7315]{Rafael~Brahm}				% WINE Team
\affiliation{Facultad de Ingenier\'ia y Ciencias, Universidad Adolfo Iba\'a\~nez, Av. Diagonal las Torres 2640, Pen\~nalol\'en, Santiago, Chile}
\affiliation{Millennium Institute for Astrophysics, Chile}
\author[0000-0003-1305-3761]{R.~Paul~Butler}			% PFS Team
\affiliation{Earth and Planets Laboratory, Carnegie Institution for Science, 5241 Broad Branch Road, NW, Washington, DC 20015, USA}
\author[0000-0002-5741-3047]{David~ R.~Ciardi}			% SG3
\affiliation{Caltech/IPAC, NASA Exoplanet Science Institute, 770 S. Wilson Avenue, Pasadena, CA 91106, USA}
\author[0000-0003-2781-3207]{Kevin I.\ Collins}			% SG1 Key Project
\affiliation{George Mason University, 4400 University Drive, Fairfax, VA, 22030 USA}
\author[0000-0003-2239-0567]{Dennis M.\ Conti}			% SG1
\affiliation{American Association of Variable Star Observers, 49 Bay State Road, Cambridge, MA 02138, USA}
\author[0000-0002-5226-787X]{Jeffrey~D.~Crane}			% PFS Team
\affiliation{Earth and Planets Laboratory, Carnegie Institution for Science, 5241 Broad Branch Road, NW, Washington, DC 20015, USA}
\author[0000-0002-8958-0683]{Fei~Dai}					% HIRES Observers
\affiliation{Division of Geological and Planetary Science, California Institute of Technology, Pasadena, CA 91125, USA}
\author[0000-0001-8189-0233]{Courtney~D.~Dressing}		% SG3
\affiliation{Department of Astronomy,  University of California Berkeley, Berkeley, CA 94720, USA}
\author[0000-0003-3773-5142]{Jason~D.~Eastman}			% TESS Contributing Authors - TSO
\affiliation{Center for Astrophysics \textbar \ Harvard \& Smithsonian, 60 Garden St, Cambridge, MA 02138, USA}
\author[0000-0002-2482-0180]{Zahra~Essack}				% PFS Observers
\affiliation{Department of Earth, Atmospheric and Planetary Sciences, Massachusetts Institute of Technology, Cambridge, MA 02139, USA}
\affiliation{Department of Physics and Kavli Institute for Astrophysics and Space Research, Massachusetts Institute of Technology, Cambridge, MA 02139, USA}
\author[0000-0002-6482-2180]{Raquel For\'{e}s-Toribio}	% SG1
\affiliation{Departamento de Astronom\'{\i}a y Astrof\'{\i}sica, Universidad de Valencia, E-46100 Burjassot, Valencia, Spain}
\affiliation{Observatorio Astron\'omico, Universidad de Valencia, E-46980 Paterna, Valencia, Spain} 
\author[0000-0001-9800-6248]{Elise~Furlan}				% SG3
\affiliation{Caltech/IPAC, NASA Exoplanet Science Institute, 770 S. Wilson Avenue, Pasadena, CA 91106, USA}
\author[0000-0002-4503-9705]{Tianjun Gan}				% SG1
\affiliation{Department of Astronomy, Tsinghua University, Beijing 100084, China}
\author[0000-0002-8965-3969]{Steven~Giacalone}			% SG3
\affiliation{Department of Astronomy,  University of California Berkeley, Berkeley, CA 94720, USA}
\author[0000-0001-6171-7951]{Holden~Gill}
\affiliation{Department of Astronomy,  University of California Berkeley, Berkeley, CA 94720, USA}
\author[0000-0002-5443-3640]{Eric Girardin}									% SG1
\affiliation{Grand Pra Observatory, Switzerland}
\author[0000-0002-1493-300X]{Thomas~Henning}			% WINE Team
\affiliation{Max-Planck-Institut fur Astronomie, Konigstuhl 17, 69117 Heidelberg, Germany}
\author{Christopher~E.~Henze}							% TESS Architect
\affiliation{NASA Ames Research Center, Moffett Field, CA 94035, USA}
\author[0000-0002-5945-7975]{Melissa~J.~Hobson}			% WINE Team
\affiliation{Max-Planck-Institut fur Astronomie, Konigstuhl 17, 69117 Heidelberg, Germany}
\author[0000-0002-1160-7970]{Jonathan Horner}			% MINERVA Architects
\affiliation{University of Southern Queensland, Centre for Astrophysics, West Street, Toowoomba, QLD 4350 Australia}
\author[0000-0001-8638-0320]{Andrew~W.~Howard}			% HIRES Team		
\affiliation{Department of Astronomy, California Institute of Technology, Pasadena, CA 91125, USA}
\author[0000-0002-2532-2853]{Steve~B.~Howell}			% SG3
\affiliation{NASA Ames Research Center, Moffett Field, CA 94035, USA}
\author[0000-0003-0918-7484]{Chelsea~ X.~Huang}			% TESS Contributing Authors - TSO
\affiliation{University of Southern Queensland, Centre for Astrophysics, West Street, Toowoomba, QLD 4350, Australia}
\author[0000-0002-0531-1073]{Howard~Isaacson}			% HIRES Team
\affiliation{Department of Astronomy,  University of California Berkeley, Berkeley, CA 94720, USA}
\affiliation{University of Southern Queensland, Centre for Astrophysics, West Street, Toowoomba, QLD 4350, Australia}
\author[0000-0002-4715-9460]{Jon~M.~Jenkins}			% TESS Architect
\affiliation{NASA Ames Research Center, Moffett Field, CA 94035, USA}
\author[0000-0002-4625-7333]{Eric L.\ N.\ Jensen}		% SG1 Important People
\affiliation{Department of Physics \& Astronomy, Swarthmore College, Swarthmore PA 19081, USA}
\author[0000-0002-5389-3944]{Andr\'es Jord\'an}			% WINE Team
\affiliation{Facultad de Ingenier\'ia y Ciencias, Universidad Adolfo Iba\'a\~nez, Av. Diagonal las Torres 2640, Pen\~nalol\'en, Santiago, Chile}
\affiliation{Millennium Institute for Astrophysics, Chile}
\author[0000-0002-7084-0529]{Stephen~R.~Kane}			% MINERVA Architects
\affiliation{Department of Earth and Planetary Sciences, University of California, Riverside, CA 92521, USA}
\author[0000-0003-0497-2651]{John F.\ Kielkopf}			% SG1 Important People
\affiliation{Department of Physics and Astronomy, University of Louisville, Louisville, KY 40292, USA}
\author{Slawomir Lasota}								% SG1
\affiliation{Silesian University of Technology, Department of Electronics, Electrical Engineering and Microelectronics, Akademicka 16, 44-100 Gliwice, Poland}
\author[0000-0001-8172-0453]{Alan~M.~Levine}			% TESS Contributing Authors - POC
\affiliation{Department of Physics and Kavli Institute for Astrophysics and Space Research, Massachusetts Institute of Technology, Cambridge, MA 02139, USA}
\author[0000-0001-7047-8681]{Jack~Lubin}				% HIRES Observers
\affiliation{Department of Physics \& Astronomy, University of California, Irvine, Irvine, CA 92697, USA}
\author[0000-0003-3654-1602]{Andrew~W.~Mann}			% SG3
\affiliation{Department of Physics and Astronomy, The University of North Carolina at Chapel Hill, Chapel Hill, NC 27599-3255, USA}
\author[0000-0001-8879-7138]{Bob Massey}				% SG1
\affiliation{Villa '39 Observatory, Landers, CA 92285, USA}
\author[0000-0001-9504-1486]{Kim K. McLeod}				% SG1
\affiliation{Department of Astronomy, Wellesley College, Wellesley, MA 02481, USA}
\author[0000-0002-7830-6822]{Matthew W. Mengel}			% MINERVA Builders
\affiliation{University of Southern Queensland, Centre for Astrophysics, West Street, Toowoomba, QLD 4350 Australia}
\author[0000-0001-9833-2959]{Jose A. Mu\~noz}			% SG1
\affiliation{Departamento de Astronom\'{\i}a y Astrof\'{\i}sica, Universidad de Valencia, E-46100 Burjassot, Valencia, Spain}
\affiliation{Observatorio Astron\'omico, Universidad de Valencia, E-46980 Paterna, Valencia, Spain}
\author{Felipe~Murgas}									% SG1 Key Project
\affiliation{Instituto de Astrof\'isica de Canarias (IAC), E-38205 La Laguna, Tenerife, Spain}
\affiliation{Departamento de Astrof\'isica, Universidad de La Laguna (ULL), E-38206 La Laguna, Tenerife, Spain}
\author[0000-0003-0987-1593]{Enric~Palle}									% SG1 Key Project
\affiliation{Instituto de Astrof\'isica de Canarias (IAC), E-38205 La Laguna, Tenerife, Spain}
\affiliation{Departamento de Astrof\'isica, Universidad de La Laguna (ULL), E-38206 La Laguna, Tenerife, Spain}
\author[0000-0002-8864-1667]{Peter Plavchan}			% MINERVA Architects
\affiliation{George Mason University, 4400 University Drive MS 3F3, Fairfax, VA 22030, USA}
\author[0000-0003-3184-5228]{Adam Popowicz}				% SG1
\affiliation{Silesian University of Technology, Department of Electronics, Electrical Engineering and Microelectronics, Akademicka 16, 44-100 Gliwice, Poland}
\author[0000-0002-3940-2360]{Don J. Radford}			% SG1
\affiliation{Brierfield Observatory, New South Wales, Australia}
\author[0000-0003-2058-6662]{George~R.~Ricker}			% TESS Architect
\affiliation{Department of Physics and Kavli Institute for Astrophysics and Space Research, Massachusetts Institute of Technology, Cambridge, MA 02139, USA}
\author[0000-0002-4829-7101]{Pamela~Rowden}				% TESS Contributing Authors - POC
\affiliation{Royal Astronomical Society, Burlington House, Piccadilly, London W1J 0BQ, UK}
\author[0000-0003-1713-3208]{Boris~S.~Safonov}			% SG3
\affiliation{Sternberg Astronomical Institute, Lomonosov Moscow State University, 119992, Universitetskij prospekt 13, Moscow, Russia}
\author[0000-0002-2454-768X]{Arjun~B.~Savel}			% SG3
\affiliation{Department of Astronomy, University of Maryland, College Park, College Park, MD, USA}
\author[0000-0001-8227-1020]{Richard P. Schwarz}		% SG1
\affiliation{Patashnick Voorheesville Observatory, Voorheesville, NY 12186, USA}
\author[0000-0002-6892-6948]{S.~Seager}					% TESS Architect
\affiliation{Department of Physics and Kavli Institute for Astrophysics and Space Research, Massachusetts Institute of Technology, Cambridge, MA 02139, USA}
\affiliation{Department of Earth, Atmospheric and Planetary Sciences, Massachusetts Institute of Technology, Cambridge, MA 02139, USA}
\affiliation{Department of Aeronautics and Astronautics, MIT, 77 Massachusetts Avenue, Cambridge, MA 02139, USA}
\author{Ramotholo~Sefako}								% SG1 Key Project
\affiliation{South African Astronomical Observatory, P.O. Box 9, Observatory, Cape Town 7935, South Africa}
\author[0000-0002-1836-3120]{Avi~Shporer}				% SG1 Key Project
\affiliation{Department of Physics and Kavli Institute for Astrophysics and Space Research, Massachusetts Institute of Technology, Cambridge, MA 02139, USA}
\author{Gregor Srdoc}									% SG1
\affiliation{Kotizarovci Observatory, Sarsoni 90, 51216 Viskovo, Croatia}
\author{Ivan~S.~Strakhov}								% SG3
\affiliation{Sternberg Astronomical Institute, Lomonosov Moscow State University, 119992, Universitetskij prospekt 13, Moscow, Russia}
\author{Johanna~K.~Teske}								% PFS Team
\affiliation{Earth and Planets Laboratory, Carnegie Institution for Science, 5241 Broad Branch Road, NW, Washington, DC 20015, USA}
\author{C.G. Tinney}									% MINERVA Architects
\affiliation{Exoplanetary Science at UNSW, School of Physics, UNSW Sydney, NSW 2052, Australia}
\author[0000-0003-0298-4667]{Dakotah~Tyler}				% HIRES Observers
\affiliation{Department of Physics \& Astronomy, University of California Los Angeles, Los Angeles, CA 90095, USA}
\author[0000-0001-9957-9304]{Robert A. Wittenmyer} 		% MINERVA Architects
\affiliation{University of Southern Queensland, Centre for Astrophysics, West Street, Toowoomba, QLD 4350 Australia}
\author{Hui Zhang}										% MINERVA Architects
\affiliation{Shanghai Astronomical Observatory, Chinese Academy of Sciences, Shanghai 200030, China}
\author[0000-0002-0619-7639]{Carl~Ziegler}				% SG3
\affiliation{Department of Physics, Engineering and Astronomy, Stephen F. Austin State University, 1936 North Street, Nacogdoches, TX 75962, USA}

\correspondingauthor{Samuel W.\ Yee}

\begin{abstract}
Hot Jupiters -- short-period giant planets -- were the first extrasolar planets to be discovered, but many questions about their origin remain.
NASA's Transiting Exoplanet Survey Satellite (\TESS), an all-sky search for transiting planets, presents an opportunity to address these questions by constructing a uniform sample of hot Jupiters for demographic study through new detections and unifying the work of previous ground-based transit surveys. As the first results of an effort to build this large sample of planets, we report here the discovery of ten new hot Jupiters (TOI-2193A\,b, TOI-2207\,b, TOI-2236\,b, TOI-2421\,b, TOI-2567\,b, TOI-2570\,b, TOI-3331\,b, TOI-3540A\,b, TOI-3693\,b, TOI-4137\,b).
All of the planets were identified as planet candidates based on periodic flux dips observed by \TESS, and were subsequently confirmed using ground-based time-series photometry, high angular resolution imaging, and high-resolution spectroscopy coordinated with the \TESS Follow-up Observing Program. 
The ten newly discovered planets orbit relatively bright F and G stars ($G<12.5$, $\Teff$ between 4800 and 6200\,K).
The planets' orbital periods range from 2 to 10 days, and their masses range from 0.2 to 2.2 Jupiter masses. TOI-2421\,b is notable for being a Saturn-mass planet and TOI-2567\,b for being a ``sub-Saturn', with masses of $0.322\pm0.073$ and $0.195\pm0.030$ Jupiter masses, respectively.
We also measured a detectably eccentric orbit ($e=0.17\pm0.05$) for TOI-2207\,b, a planet on an 8-day orbit, while placing an upper limit of $e<0.052$ for TOI-3693\,b, which has a 9-day orbital period.
The ten planets described here represent an important step toward using \TESS to create a large and statistically useful sample of hot Jupiters.
\end{abstract}

\section{Introduction} \label{sec:intro}

The origin of hot Jupiters is one of the longest-standing unresolved problems in exoplanet science.
Prior to the discovery of the first hot Jupiter 51~Pegasi~b \citep{Mayor1995}, our understanding of planet formation was entirely based on our knowledge of our solar system.
It was thought that giant planets could only form beyond a few astronomical units from their host stars, where the surface density of solids within the protoplanetary disk would be high enough to allow
for the formation of a solid body massive enough to undergo runaway gas accretion.
The existence of hot Jupiters implies that either this expectation was incorrect and giant planets can form close in to their host stars, or that the initially wide orbits of giant planets can sometimes shrink by a factor of 100 (see the reviews by \citealt{Dawson2018,Fortney2021}, and references therein).
In the latter scenario, the orbital shrinkage might be due to gravitational interactions with the gaseous protoplanetary disk, and would therefore need to occur within the first few million years after the formation of the star \citep{Lin1986}, or the orbital alterations might be caused by eccentricity excitation followed by tidal orbital circularization, which need not occur early in the system's history \citep{Rasio1996,Fabrycky2007}.
Each of these possible formation pathways would shape the population of hot Jupiters in different ways, and hence studying the demographics of hot Jupiters and the distributions of their orbital properties may help us understand their relative importance.

NASA's ongoing \TESS mission \citep{TESS_Ricker15} was designed to detect smaller planets --- super-Earths and sub-Neptunes --- but \TESS is also capable of revolutionizing our knowledge of hot Jupiter demographics.
As a nearly all-sky space-based photometric survey that dwells on a given star field for 27 days at a time, the \TESS survey should be able to identify
nearly all of the hot Jupiters that transit stars which are bright
and nearby enough for detailed follow-up observations and characterization \citep{Zhou2019a}.
Simulations have predicted that the \TESS planet catalog will eventually
contain $\approx$400 hot Jupiters around FGK stars brighter than $G = 12.5$ \citep{Yee2021b}.
Such a sample would be an order of magnitude larger than the sample of 40 hot Jupiters found during the original \Kepler mission, which is the largest statistically useful sample of such planets currently available.
The process of constructing a very large sample of hot Jupiters around bright stars is greatly facilitated and accelerated by the fact that hundreds of these planets have already been discovered 
by the ground-based transit surveys such as TrES \citep{TrES_Alonso2004}, XO \citep{XO_McCullough2005}, WASP \citep{WASP_Pollacco2006}, HATNet and HATSouth \citep{HAT_Bakos2004,HATS_Bakos2013}, KELT \citep{KELT_Pepper2007,KELT_Pepper2012}, and NGTS \citep{NGTS_Wheatley2018}, accounting for roughly 40\% of the expected 400 planets in a $G < 12.5$ magnitude-limited sample.
The hard work of these previous surveyors over many years led to the discovery
of many important and well-studied systems, but the selection functions
of these surveys are complex, different from each other, and not well documented.
\TESS will provide a homogeneous dataset that encompasses essentially all of
the previously known planets and has a selection function that should be easier
to model. 
Hence, \TESS offers the possibility of unifying all of the previous surveys and leveraging over two decades of observational effort.

Still, even with \TESS, expanding and completing a statistical sample suitable for demographic study will require a significant follow-up effort to rule out astrophysical false positives (generally eclipsing binaries that masquerade as transiting planets), measure planet masses, and model the selection function.
Ground-based imaging and time-series photometric follow-up with higher angular resolution than \TESS can be used to check for nearby eclipsing binaries and confirm that the transit signal belongs to the identified star and not a foreground or background star.
Additional transit observations help to improve our knowledge of the planet parameters and transit ephemerides, while observations in multiple bandpasses can also be used to check for chromatic effects that are indicative of eclipsing binaries.
High-resolution Doppler spectroscopy provides final confirmation of the planet's existence
and measures its mass, as well as providing a high signal-to-noise
spectrum that is useful for characterizing the host star.

Given all the different types of observations and resources that are needed,
confirming hundreds of new hot Jupiters from \TESS within a reasonable
amount of time is only feasible with a large community effort.

The lead authors of this paper have begun to organize such an effort,
the Grand Unified Hot Jupiter Survey, by playing a linking role between the new planet discoveries from various \TESS follow-up groups, and the planets (and false positives) that have already been investigated in previous planet searches.
This work is being conducted as part of the \TESS Follow-Up Observing Program (TFOP; \citealt{TFOP_Collins2018,ExoFoPTESS})\footnote{\url{https://heasarc.gsfc.nasa.gov/docs/tess/followup.html}}\textsuperscript{,}\footnote{\url{https://exofop.ipac.caltech.edu/tess/}}, which provides a platform for coordinating observations and sharing data, and is open to any interested astronomer.

The 10 planets described in this paper are the first newly discovered planets
from this survey, and are based on data contributed by many TFOP members.
The planets are known by their \TESS Object of Interest (TOI) numbers:
TOI-2193A\,b, TOI-2207\,b, TOI-2236\,b, TOI-2421\,b, TOI-2567\,b, TOI-2570\,b, TOI-3331\,b, TOI-3540A\,b, TOI-3693\,b, and TOI-4137\,b.
Section \ref{ssec:tess} presents time-series photometry from \TESS,
and Sections \ref{ssec:sg1} through \ref{ssec:spec} describe ground-based follow-up photometry, imaging, and spectroscopic follow-up observations.
Section \ref{sec:stellar_char} presents a characterization of each host star, and Section \ref{sec:planet_char} describes the application of \Exofast \citep{ExoFASTv2_Eastman19} to jointly model all of the data and determine the parameters of each system.
Section \ref{sec:discussion} examines the properties of the new planets within the context of the entire known sample of hot Jupiters,
and Section \ref{sec:conclusion} draws some conclusions.

\section{Observations and Data} \label{sec:obs}

\subsection{TESS Photometry} \label{ssec:tess}

All of the new planets described here were first detected in \TESS photometry.
\TESS observes a $24^\circ \times 96^\circ$ region of the sky for 27 days at a time, before rotating its field of view to a new sector.
During its initial two-year Prime Mission (Sectors 1 -- 26, running from July 2018 to July 2020), \TESS obtained high-precision photometry of $\gtrsim$\,200,000 preselected stars with a two-minute cadence, with the remainder of the field-of-view being observed in the full-frame images (FFIs) at 30-minute cadence.
With the Extended Mission beginning in 2020 (Sectors 27 onward), two-minute cadence data continues to be obtained for $\approx$\,20,000 targets per sector, while the FFI cadence has been reduced to 10 minutes.

The 2-minute \TESS photometry was extracted and reduced by the \TESS Science Processing Operations Center (SPOC) pipeline, as described by \cite{TESS_SPOC_Jenkins2016}.
Only one of the 10 planet-hosting stars presented in this paper had been preselected for 2-minute cadence observations (TOI-2207, Sector 27). However, several others were added to the short-cadence target list in the Extended Mission following their identification as planet candidates in the FFIs from the Prime Mission.
The longer-cadence FFI data were calibrated with the \texttt{tica} software \citep{TESS_QLP_Fausnaugh2020}, and the light-curves were extracted with the MIT Quick-Look Pipeline (QLP; \citealt{TESS_QLP_Huang2020a,TESS_QLP_Huang2020b,TESS_QLP_Kunimoto2021}).

Both the SPOC pipeline and QLP search the extracted light-curves for transit-like signals (``Threshold Crossing Events'' or TCEs) which are then vetted by the \TESS Science Office (TSO).
Objects with signals surviving the vetting process are designated \TESS Objects of Interest (TOIs; \citealt{TESS_TOIs_Guerrero2021}) and public notifications are distributed.
All of the planets in this paper were alerted as TOIs, including some from the recent search of QLP light-curves for stars fainter than $T > 10.5$ by \citet{TESS_Faint_Kunimoto2021b}.
Table \ref{tab:tess_summary} summarizes the ten targets and the \TESS sectors they were observed in.

In addition, four of the targets had previously been identified as ``community TOIs'' (cTOIs) by separate investigators.
TOI-2421\,b was first flagged as a planet candidate by \citet{Montalto2020}, who used the DIAmante pipeline to extract difference imaging light-curves from the first year of TESS FFIs.
They then identified transit events in the light-curves with the Box-Least Squares algorithm \citep{BLS_Kovacs2002}, and vetted the candidate events with a Random Forest classifier.
TOI-2567\,b, TOI-2570\,b, and TOI-4137\,b were identified as planet candidates by \citet{Olmschenk2021}, who classified light-curves from the SPOC-calibrated FFIs extracted by the \texttt{eleanor} pipeline \citep{Feinstein2019} with a convolutional neural network.
These light-curves were then subjected to the Quasi-Automated Transit Search (QATS; \citealt{QATS_Kruse2019}) pipeline and vetted with the Discovery and Vetting of Exoplanets pipeline (DAVE; \citealt{DAVE_Kostov2019}).
All four cTOIs were later promoted to TOIs following vetting by the TESS Science Office \citep{TESS_cTOIs_Mireles2021}.

\begin{figure}
\epsscale{1.2}
\plotone{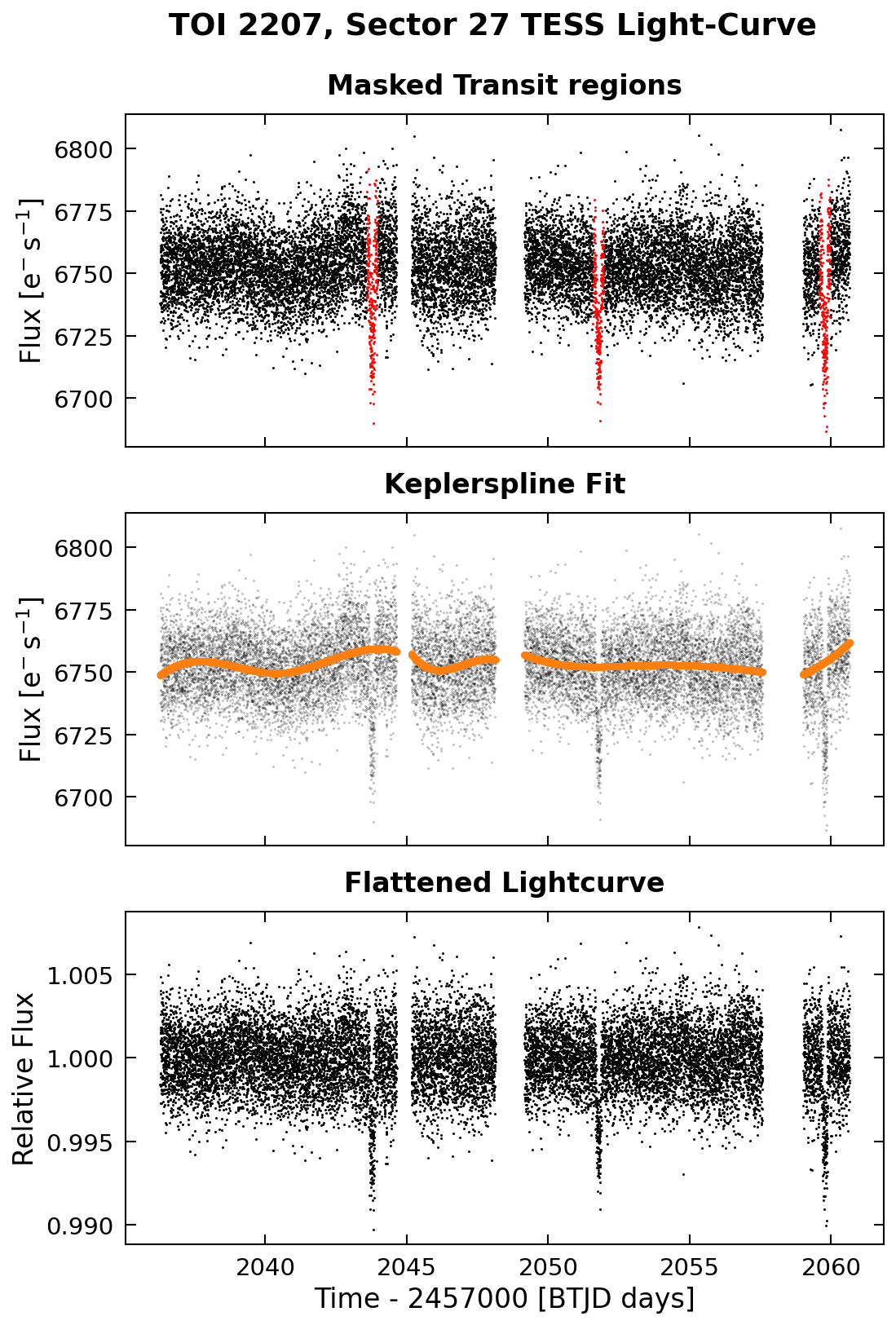}
\caption{TESS Sector 27 light-curve for TOI-2207, extracted by the SPOC pipeline.
Top panel: SPOC PDC light-curve, with red points highlighting transit events, which are masked out when performing the spline fit.
Middle panel: \texttt{Keplerspline} fit to the masked light-curve.
Bottom panel: Flattened light-curve.
\label{fig:toi2207_tess_lightcurve}}
\end{figure}

We identified these targets as candidate hot Jupiters ($R_p > 8$~\Rearth, $P < 10$~days) from the TOI catalog, with selection based on catalog photometry and astrometry indicating that they orbit FGK stars.
We then began performing follow-up observations to determine whether the transit-like signals are truly from transiting planets or are instead
from eclipsing binaries or other ``false positives.''
When analyzing each planetary system (Section \ref{sec:planet_char}), we used all of the available photometry from TESS.
We used the \texttt{lightkurve} package \citep{Lightkurve18} to download the TESS light-curves from the Mikulski Archive for Space Telescopes (MAST).
When available, we used light-curves produced from the SPOC pipeline -- this applies to all the short-cadence data, as well as some of the long-cadence FFI data, for which the SPOC pipeline has recently begun processing 160{,}000 targets per sector \citep{TESS_SPOC_Caldwell2020}.
We used the Presearch Data Conditioning (PDC; \citealt{TESS_PDC_Stumpe2012,TESS_PDC_Smith2012,TESS_PDC_Stumpe2014}) light-curves, which have been corrected for instrumental effects.
We additionally ``flattened'' the SPOC PDC light-curves with the \texttt{Keplerspline}\footnote{\url{https://github.com/avanderburg/keplersplinev2}} routine \citep{Keplerspline_Vanderburg2014,Keplerspline_Shallue2018}, which fits a spline to the light-curve (with transit events masked) to correct for stellar or instrumentally induced variability (Figure \ref{fig:toi2207_tess_lightcurve}).
When SPOC light-curves were unavailable for the long-cadence data, we used the light-curves extracted by the QLP, which were also flattened with \texttt{Keplerspline}.
We provide the flattened and normalized \TESS photometry in Table \ref{tab:tess_data}.

\begin{deluxetable}{cccr}
\tablecolumns{4}
\tablecaption{Summary of TESS Observations \label{tab:tess_summary}}
\tablehead{
	\colhead{Target} & \colhead{Sector} & \colhead{Source} & \colhead{Cadence (s)}
}
\startdata
% target, sector, source, exp_time 
TOI-2193 & 13 & SPOC & 1800 \\
 $\cdots$ & 27 & SPOC & 120 \\
TOI-2207 & 27 & SPOC & 120 \\
TOI-2236 & 12,13 & SPOC & 1800 \\
 $\cdots$ & 27,39 & SPOC & 600 \\
TOI-2421 & 3 & SPOC & 1800 \\
 $\cdots$ & 30 & SPOC & 600 \\
TOI-2567 & 14--26 & SPOC & 1800 \\
 $\cdots$ & 40,41 & SPOC & 120 \\
TOI-2570 & 19 & SPOC & 1800 \\
 $\cdots$ & 44,45 & SPOC & 120 \\
TOI-3331 & 13 & QLP & 1800 \\
TOI-3540 & 15 & QLP & 1800 \\
TOI-3693 & 17,18 & SPOC & 1800 \\
TOI-4137 & 19,26 & SPOC & 1800 \\

\enddata
\end{deluxetable}

\begin{deluxetable}{lcccc}
\tablecolumns{5}
\tablecaption{Flattened \& Normalized TESS Photometry \label{tab:tess_data}}
\tablehead{
	\colhead{TOI} & \colhead{Sector} & \colhead{\bjdtdb} & \colhead{Flux\tablenotemark{a}} & \colhead{Flux Err.}
}
\startdata
% target, sector, bjdtdb, flux, flux_err 
2193 & 13 & 2458657.514245 & 0.999797 & 0.000731\\
2193 & 13 & 2458657.535079 & 0.999887 & 0.000734\\
2193 & 13 & 2458657.555913 & 1.000377 & 0.000736\\
2193 & 13 & 2458657.597580 & 0.996574 & 0.000740\\
2193 & 13 & 2458657.618414 & 0.992282 & 0.000740\\
2193 & 13 & 2458657.680915 & 1.000831 & 0.000741\\

\enddata
\tablenotemark{a}{Flux has been detrended and normalized \replaced{to a baseline of 1.0.}{such that the mean out-of-transit flux has a baseline of 1.0.}}
\tablecomments{This table is available in its entirety in machine-readable form.}
\end{deluxetable}

\subsection{Ground-Based Photometry} \label{ssec:sg1}

Additional ground-based photometry of each of our targets was obtained as part of TFOP's Seeing-limited Photometry Sub-Group 1 (SG1).
These multi-band photometric observations of individual transits helped to rule out false-positive scenarios such as nearby eclipsing binaries.
We also used the ground-based light-curves to refine the transit parameters and ephemerides.
Observations were obtained from the Brierfield Observatory; KeplerCam on the Fred Lawrence Whipple Observatory (FLWO) 1.2m telescope; the Hazelwood Observatory; the Acton Sky Portal; the Villa '39 observatory; the Observatori Astronòmic de la Universitat de València (OAUV) TURIA2 0.3m telescope; the Grand-Pra Observatory; the Silesian University of Technology Observatories (SUTO) OTIVAR 0.3m telescope; the MEarth-South telescope array \citep{MEarth_Nutzman2008,MEarth_Berta2012} at the Cerro Tololo Inter-American Observatory (CTIO);
as well as the 0.4m and 1.0m telescopes of the Las Cumbres Observatory Global Telescope (LCOGT; \citealt{LCOGT_Brown2013}) global network, using sites at the Observatorio del Teide, CTIO, the Siding Spring Observatory (SSO), and the South African Astronomical Observatory (SAAO).
Data reduction and aperture photometry for all follow-up observations, except for the MEarth-South observations, was done with the \texttt{AstroImageJ} software \citep{AstroImageJ_Collins17}.
The MEarth-South data were reduced according to the procedures described in \citet{MEarth_Berta2012} and \citet{MEarth_Irwin2007}.
We summarize the observations and facilities used for each target in Table \ref{tab:sg1_lcs}, and present the complete set of photometric observations in Table \ref{tab:sg1_lc_data}.

\begin{deluxetable*}{cccccrrc}
\tablecolumns{8}
\tablecaption{Summary of Ground-Based Photometric Follow-Up Observations \label{tab:sg1_lcs}}
\tablehead{
	\colhead{Target} & \colhead{Facility/Instrument} & \colhead{Aperture} & \colhead{Filter} &
	\colhead{Date} & \colhead{Cadence} & \colhead{Precision$^\mathrm{a}$} & \colhead{Detrending Vectors} \\
	& & \colhead{(m)} & & \colhead{(UT)} & \colhead{(s)} & \colhead{(mmag)} &
}
\startdata
% target, observatory, size, filter, ut, cadence, precision, detrend_cols
TOI-2193 b & Brierfield & 0.36 & $I$ &2020 Sep 23 & 195 & 4.0 & Airmass, Y (T1) \\
$\cdots$ & LCO SAAO/Sinistro & 1.0 & $z^\prime$ &2020 Sep 27 & 52 & 2.5 & Width, Total Counts \\
$\cdots$ & LCO SAAO/Sinistro & 1.0 & $i^\prime$ &2020 Oct 14 & 52 & 2.0 & Sky/Pixel \\
$\cdots$ & LCO CTIO/Sinistro & 1.0 & $B$ &2021 Jul 07 & 100 & 2.3 & Airmass \\
$\cdots$ & LCO SAAO/Sinistro & 1.0 & $B$ &2021 Jul 07 & 100 & 1.2 & Airmass \\
$\cdots$ & LCO CTIO/Sinistro & 1.0 & $z^\prime$ &2021 Jul 26 & 110 & 1.3 & Width \\
TOI-2207 b & CTIO/MEarth-South & 0.4 & $I$ &2021 Jun 14 & 50 & 3.7 & FWHM, SKY, Meridian Flip \\
$\cdots$ & LCO CTIO/Sinistro & 1.0 & $z^\prime$ &2021 Jun 21 & 90 & 1.2 & FWHM, Total Counts \\
TOI-2236 b & Brierfield & 0.36 & $I$ &2020 Oct 05 & 195 & 3.0 & Airmass \\
$\cdots$ & LCO CTIO/Sinistro & 1.0 & $B$ &2021 Jun 27 & 166 & 2.6 & None \\
$\cdots$ & LCO CTIO/Sinistro & 1.0 & $z^\prime$ &2021 Jun 27 & 166 & 3.1 & Total Counts, FWHM \\
TOI-2421 b & LCO SAAO/Sinistro & 1.0 & $i^\prime$ &2020 Dec 16 & 53 & 2.0 & Airmass \\
$\cdots$ & LCO CTIO/Sinistro & 1.0 & $i^\prime$ &2020 Dec 17 & 53 & 1.8 & Airmass \\
$\cdots$ & LCO CTIO/Sinistro & 1.0 & $g^\prime$ &2021 Sep 21 & 116 & 1.4 & Airmass, Sky/Pixel \\
$\cdots$ & LCO CTIO/Sinistro & 1.0 & $i^\prime$ &2021 Sep 21 & 116 & 1.8 & Airmass, Sky/Pixel \\
TOI-2567 b & FLWO/KeplerCam & 1.2 & $i^\prime$ &2021 May 20 & 32 & 3.0 & Airmass \\
TOI-2570 b & Grand-Pra Observatory/RCO & 0.4 & $i^\prime$ &2021 Dec 11 & 133 & 2.1 & Y (T1) \\
$\cdots$ & LCO Teide/SBIG-6303 & 0.4 & $g^\prime$ &2021 Dec 18 & 198 & 3.9 & BJD$_\mathrm{TDB}$ \\
$\cdots$ & SUTO/OTIVAR & 0.3 & $B$ &2022 Jan 10 & 366 & 4.8 & Airmass, Sky/Pixel \\
TOI-3331 b & LCO CTIO/SBIG-6303 & 0.4 & $i^\prime$ &2021 Jun 06 & 86 & 5.4 & Airmass \\
$\cdots$ & LCO CTIO/Sinistro & 1.0 & $g^\prime$ &2021 Jun 12 & 104 & 1.5 & Airmass, Sky/Pixel \\
$\cdots$ & LCO CTIO/Sinistro & 1.0 & $i^\prime$ &2021 Jun 12 & 104 & 1.5 & Width, BJD$_\mathrm{TDB}$ \\
$\cdots$ & LCO CTIO/SBIG-6303 & 0.4 & $i^\prime$ &2021 Jun 14 & 81 & 5.2 & Airmass, FWHM \\
$\cdots$ & LCO CTIO/SBIG-6303 & 0.4 & $i^\prime$ &2021 Jun 18 & 81 & 4.6 & FWHM \\
$\cdots$ & Brierfield & 0.36 & $I$ &2021 Jul 18 & 193 & 2.5 & Airmass \\
$\cdots$ & Hazelwood & 0.318 & $g^\prime$ &2021 Aug 01 & 250 & 2.0 & Airmass \\
TOI-3540 b & Acton Sky Portal & 0.28 & $R$ &2021 Aug 07 & 26 & 5.3 & Airmass \\
$\cdots$ & Villa '39 & 0.355 & $I$ &2021 Aug 13 & 179 & 3.0 & Sky/Pixel \\
TOI-3693 b & OAUV/TURIA2 & 0.3 & $R$ &2021 Jul 19 & 106 & 6.5 & None \\
TOI-4137 b & Grand-Pra Observatory/RCO & 0.4 & $i^\prime$ &2021 Sep 30 & 73 & 2.7 & Airmass, Width \\

\enddata
\tablenotetext{a}{Precision is computed as the rms of the residuals after subtracting the transit and detrending model.}
\end{deluxetable*}

\begin{deluxetable*}{lccccccccc}
\tablecolumns{10}
\tablewidth{\columnwidth}
\tablecaption{TFOP SG1 Photometry \label{tab:sg1_lc_data}}
\tablehead{
	\colhead{TOI} & \colhead{Facility} & \colhead{Filter} & \colhead{Date (UT)} & 
	\colhead{\bjdtdb} & \colhead{Flux\tablenotemark{a}} & \colhead{Flux Err.} &
	\colhead{Detrend Var. 1\tablenotemark{b}} & \colhead{Detrend Var. 2} & \colhead{Detrend Var. 3}
}
\startdata
% target, observatory, filter, utdate, bjdtdb, flux, flux_err, detrend_0, detrend_1, detrend_2 
2193 & Brierfield & I & 2020-09-23 & 2459115.874574 & 0.987369 & 0.002610 & -0.268009 & -0.087072 & -- \\
2193 & Brierfield & I & 2020-09-23 & 2459115.876842 & 0.992802 & 0.002590 & -0.273435 & -0.093719 & -- \\
2193 & Brierfield & I & 2020-09-23 & 2459115.879110 & 0.996050 & 0.002600 & -0.278568 & -0.125733 & -- \\
2193 & Brierfield & I & 2020-09-23 & 2459115.881379 & 0.992478 & 0.002590 & -0.283524 & -0.164315 & -- \\
2193 & Brierfield & I & 2020-09-23 & 2459115.883647 & 0.993419 & 0.002580 & -0.288294 & -0.183758 & -- \\
2193 & Brierfield & I & 2020-09-23 & 2459115.885915 & 0.992296 & 0.002580 & -0.292864 & -0.213525 & -- \\

\enddata
\tablenotetext{a}{Flux has been normalized \replaced{to a baseline of 1.0}{such that the mean out-of-transit flux has a baseline of 1.0}, but is not yet detrended.}
\tablenotetext{b}{The detrend variables are as listed in Table \ref{tab:sg1_lcs}.}
\tablecomments{This table is available in its entirety in machine-readable form.}
\end{deluxetable*}

\subsection{High-Resolution Imaging} \label{ssec:imaging}

\begin{deluxetable*}{ccccccr}
\tablecolumns{7}
\tablecaption{Summary of High-Resolution Imaging Observations \label{tab:imaging_obs}}
\tablehead{
	\colhead{Target} & \colhead{Telescope} & \colhead{Instrument} & \colhead{Filter} &
	\colhead{Date} & \colhead{Image Type} & \colhead{Contrast} 
}
\startdata
% target, telescope, instrument, filter, ut, image_type, contrast
TOI-2193 & SOAR (4.1 m) & HRCam & $I_c$ & 2020 Oct 31 & Speckle & $\Delta$mag = 4.1 at $1\farcs0$  \\
 TOI-2207 & SOAR (4.1 m) & HRCam & $I_c$ & 2020 Oct 31 & Speckle & $\Delta$mag = 6.3 at $1\farcs0$  \\
 TOI-2236 & Gemini-S (8 m) & Zorro & 832 nm & 2021 Jul 20 & Speckle & $\Delta$mag = 5.3 at $0\farcs5$  \\
 TOI-2421 & SOAR (4.1 m) & HRCam & $I_c$ & 2020 Dec 03 & Speckle & $\Delta$mag = 6.5 at $1\farcs0$  \\
 TOI-2567 & Gemini-N (8 m) & 'Alopeke & 562 nm & 2021 Jun 25 & Speckle & $\Delta$mag = 4.22 at $0\farcs5$  \\
 $\cdots$ & Gemini-N (8 m) & 'Alopeke & 832 nm & 2021 Jun 25 & Speckle & $\Delta$mag = 6.18 at $0\farcs5$  \\
 TOI-2570 & Palomar (5 m) & PHARO & Br$\gamma$ & 2021 Nov 11 & AO & $\Delta$mag = 5.325 at $0\farcs5$  \\
 $\cdots$ & Shane (3 m) & ShARCS & $J$ & 2021 Mar 28 & AO & $\Delta$mag = 3.41 at $1\farcs0$  \\
 $\cdots$ & Shane (3 m) & ShARCS & $K_s$ & 2021 Mar 28 & AO & $\Delta$mag = 4.52 at $1\farcs0$  \\
 TOI-3331 & SOAR (4.1 m) & HRCam & $I_c$ & 2021 Jul 14 & Speckle & $\Delta$mag = 5.8 at $1\farcs0$  \\
 TOI-3540 & SOAR (4.1 m) & HRCam & $I_c$ & 2021 Oct 01 & Speckle & $\Delta$mag = 6.7 at $1\farcs0$  \\
 $\cdots$ & Palomar (5 m) & PHARO & Br$\gamma$ & 2021 Aug 24 & AO & $\Delta$mag = 6.621 at $0\farcs5$  \\
 $\cdots$ & Palomar (5 m) & PHARO & $H$cont & 2021 Aug 24 & AO & $\Delta$mag = 7.485 at $0\farcs5$  \\
 TOI-3693 & SAI-2.5m (2.5 m) & Speckle Polarimeter & $I_c$ & 2021 Jul 20 & Speckle & $\Delta$mag = 6.1 at $1\farcs0$  \\
 TOI-4137 & SAI-2.5m (2.5 m) & Speckle Polarimeter & $I_c$ & 2021 Oct 29 & Speckle & $\Delta$mag = 5.6 at $1\farcs0$  \\
 
\enddata
\end{deluxetable*}

\begin{figure}
\epsscale{1.2}
\plotone{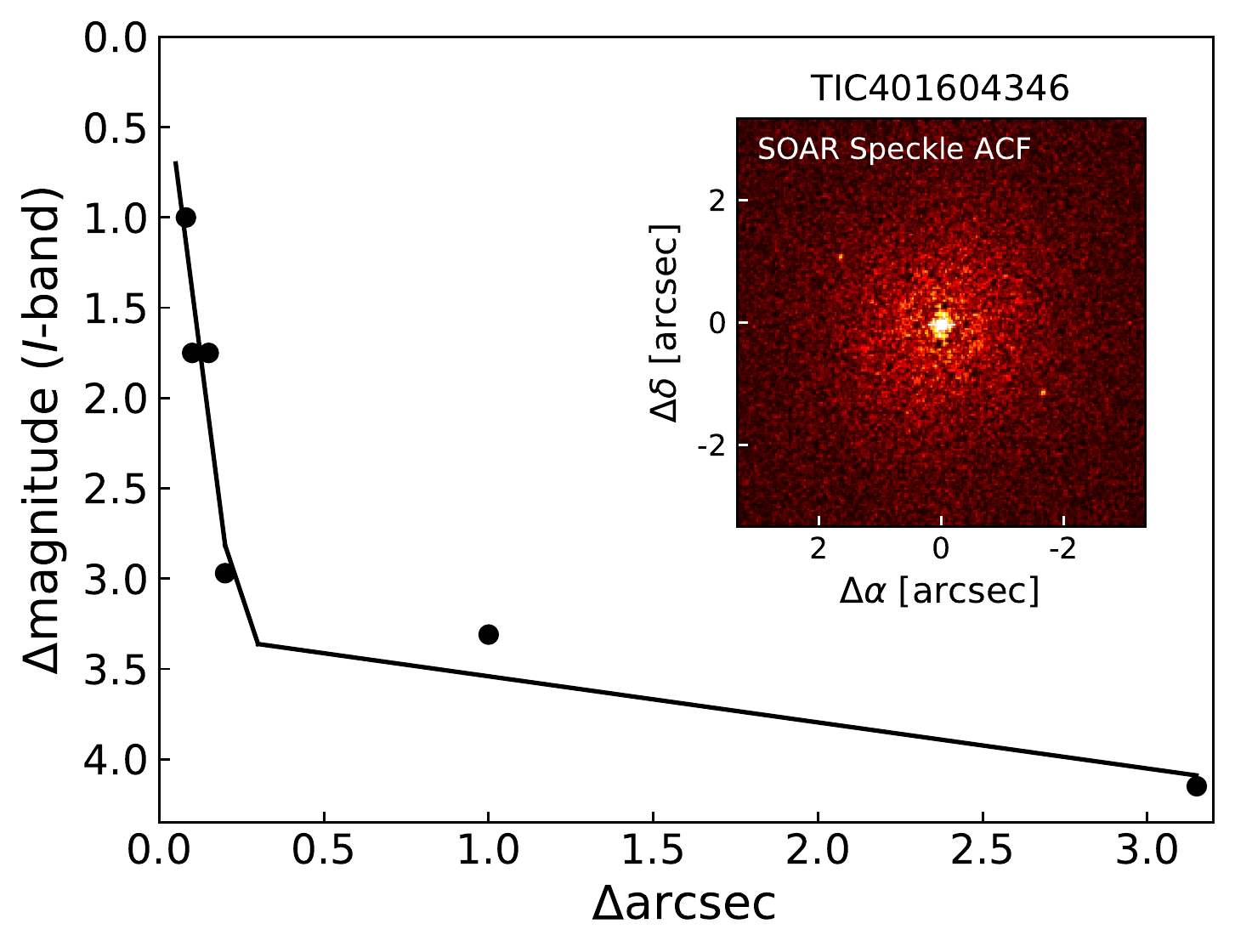}
\caption{Speckle sensitivity curve (solid line) and auto-correlation function (ACF, inset image) from SOAR HRCam observations of TOI-2193.
A nearby stellar companion is detected at $1\farcs89$, seen as the two bright points to the northwest and southeast of the primary in the ACF image.
\label{fig:toi2193_speckle}}
\end{figure}

\begin{figure}
\epsscale{1.2}
\plotone{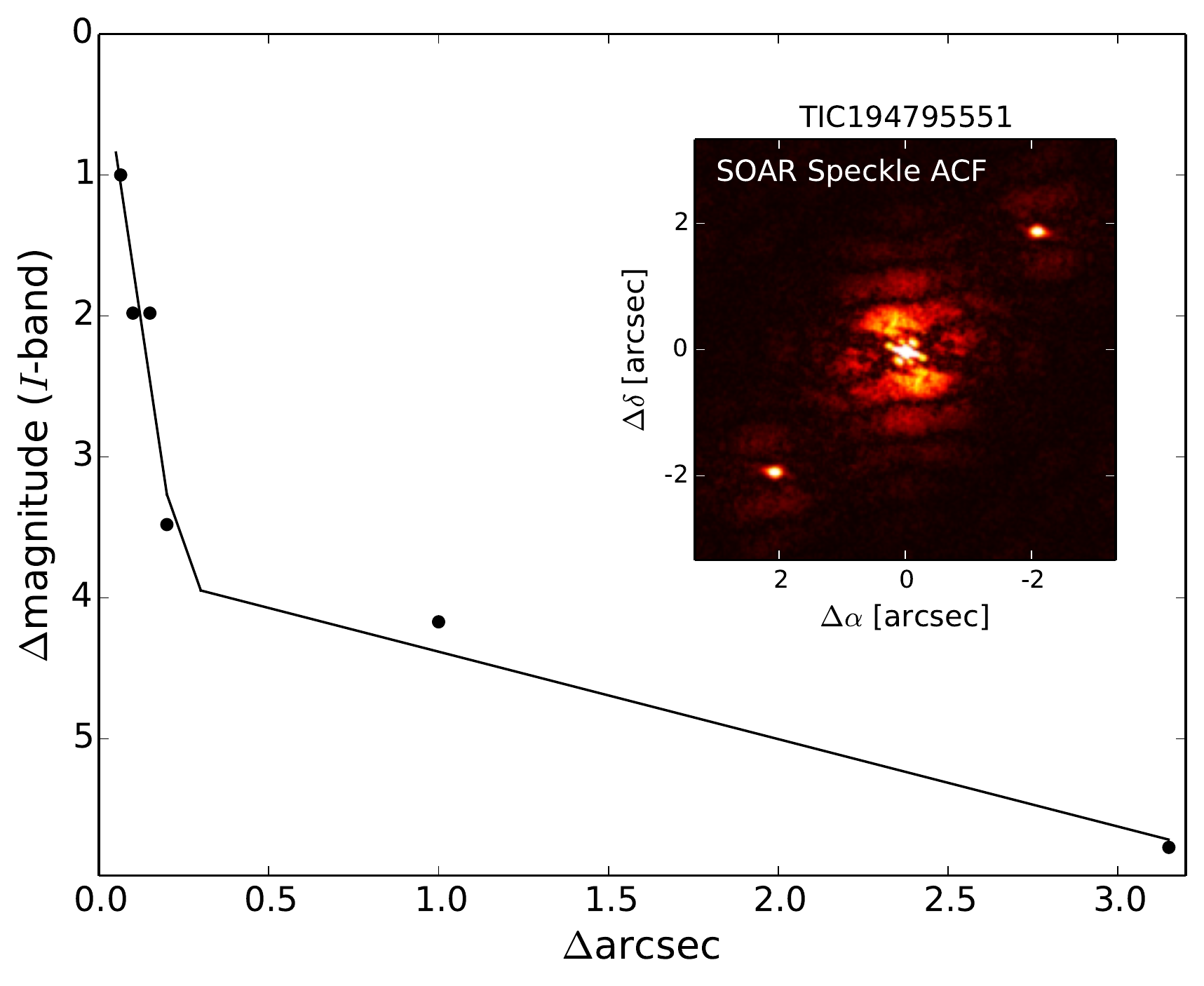}
\caption{Same as Figure \ref{fig:toi2193_speckle}, but for TOI-3331.
A nearby stellar companion is detected at $2\farcs663$, visible in the inset image to the NE and SW of the primary star.
\label{fig:toi3331_speckle}}
\end{figure}

\begin{figure}
\epsscale{1.0}
\plotone{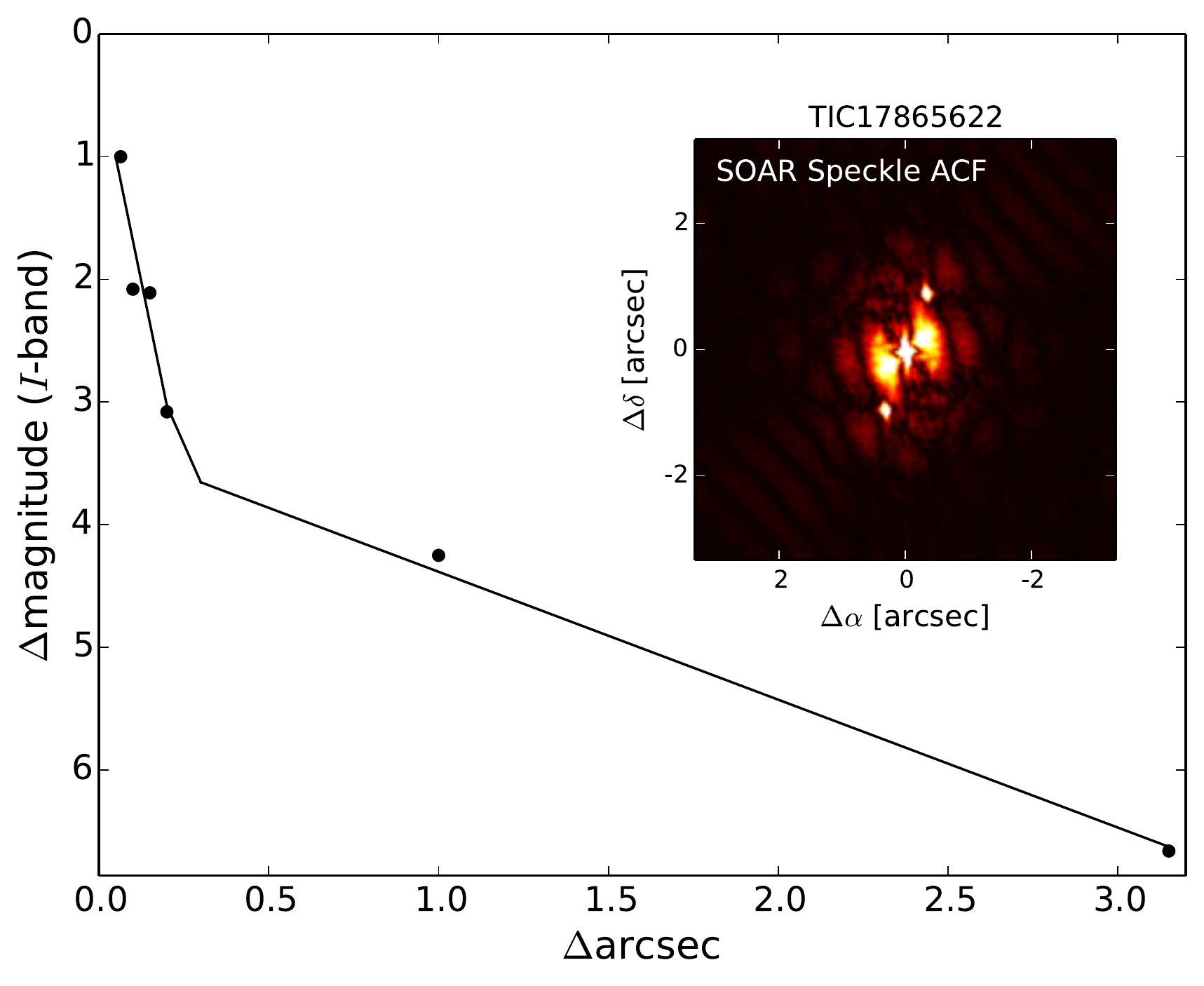}
\epsscale{1.2}
\plotone{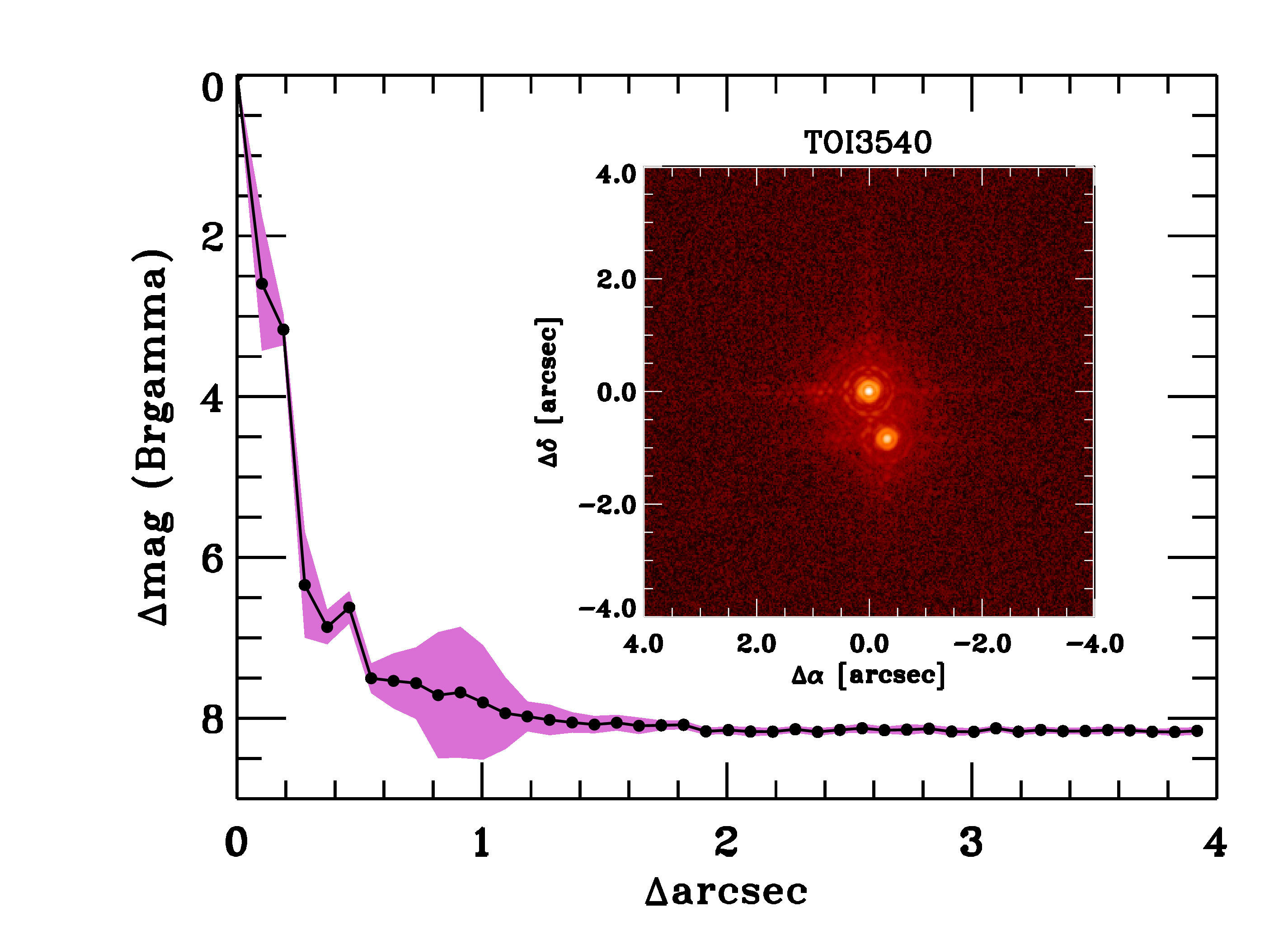}
\plotone{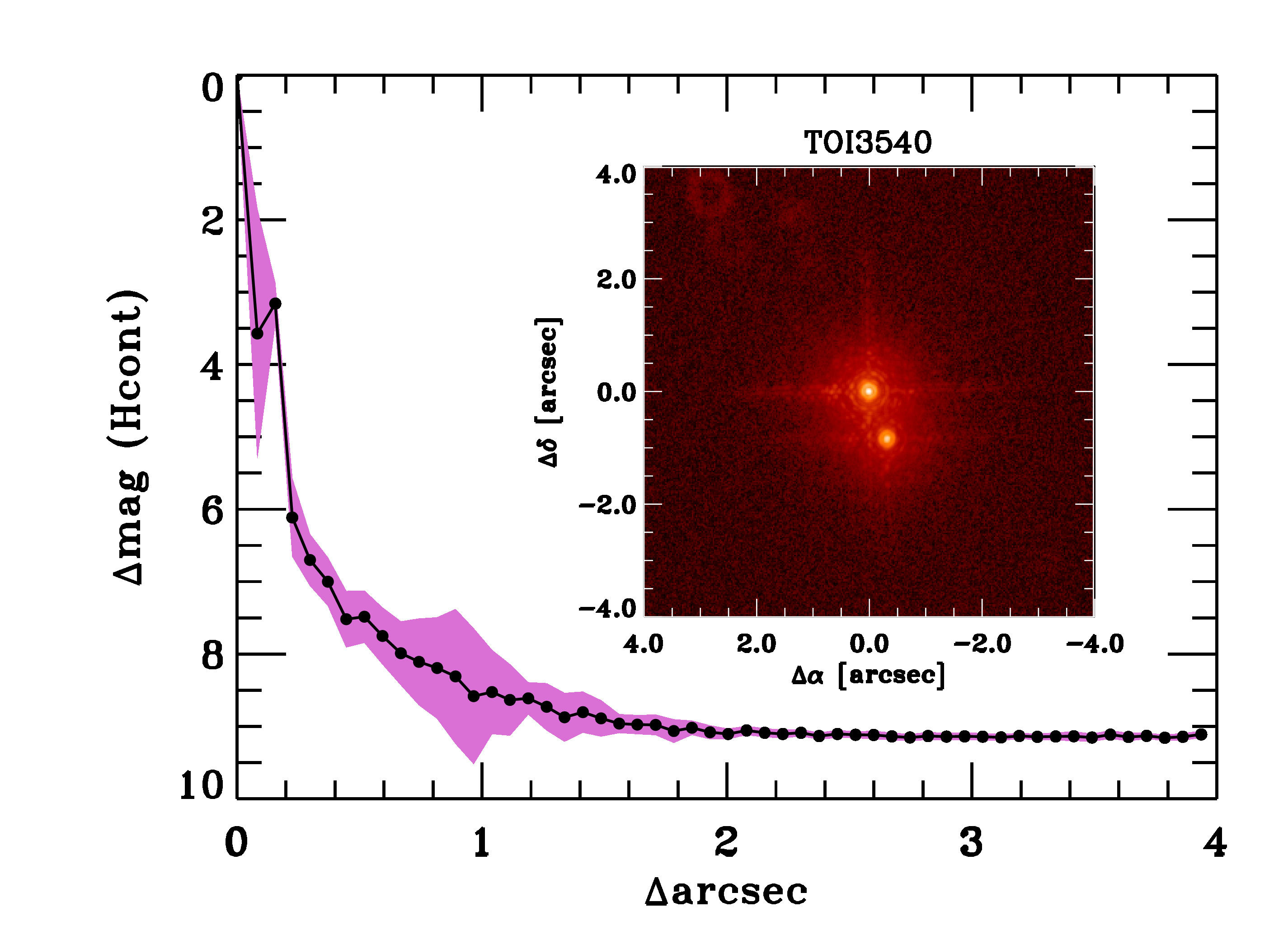}
\caption{From top to bottom: SOAR HRCam speckle imaging, Palomar PHARO Br$\gamma$ AO imaging, Palomar PHARO $H$-cont AO imaging of TOI-3540.
The close companion at $0\farcs92$ is detected by all three observations.
The line plots in each panel show the sensitivity curves for the corresponding observation\added{, while the shaded regions indicate the uncertainty on the contrast curve}.
\label{fig:toi3540_ao}}
\end{figure}

\begin{figure}
\centering
\includegraphics[width=0.35\linewidth]{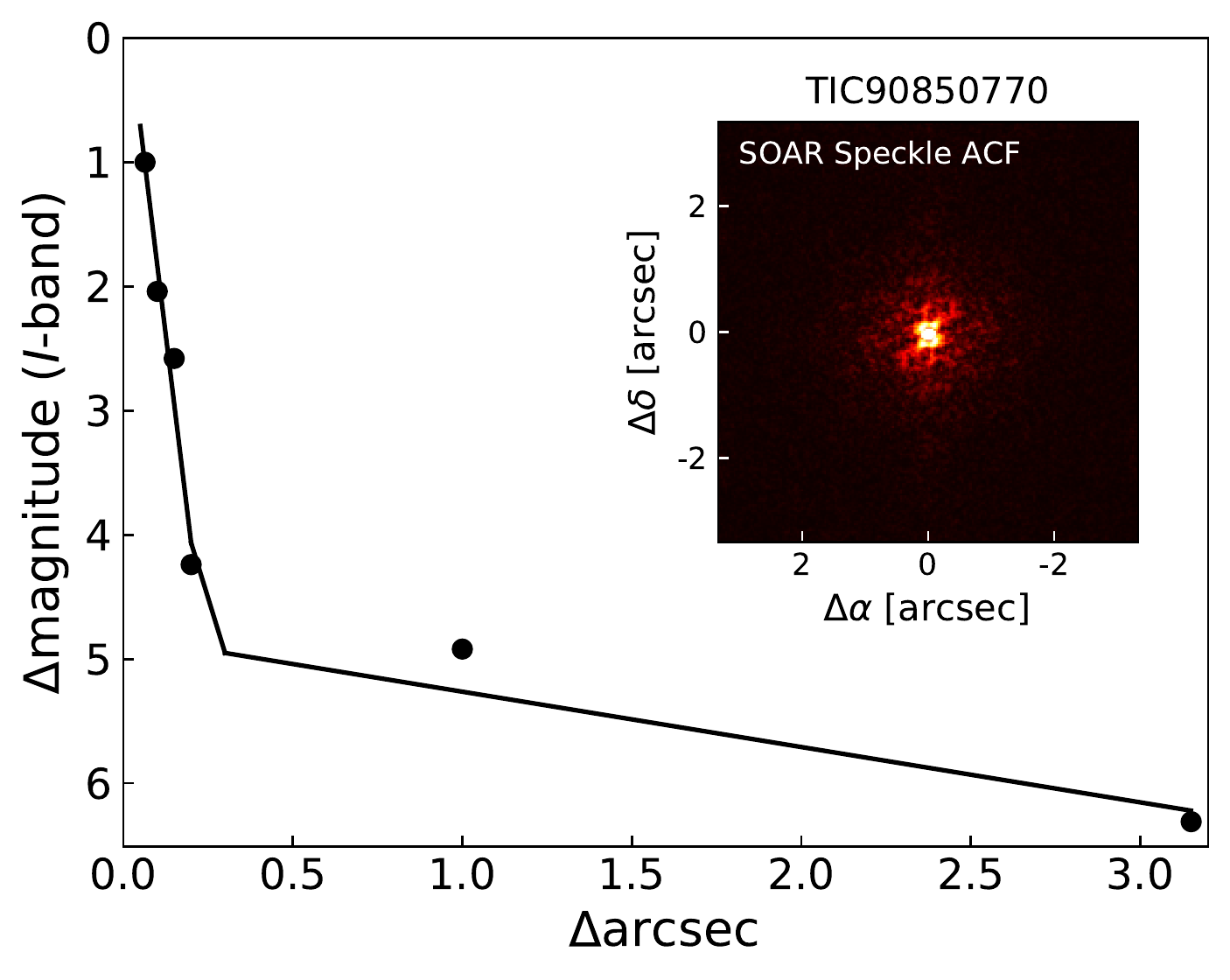}
\includegraphics[width=0.35\linewidth]{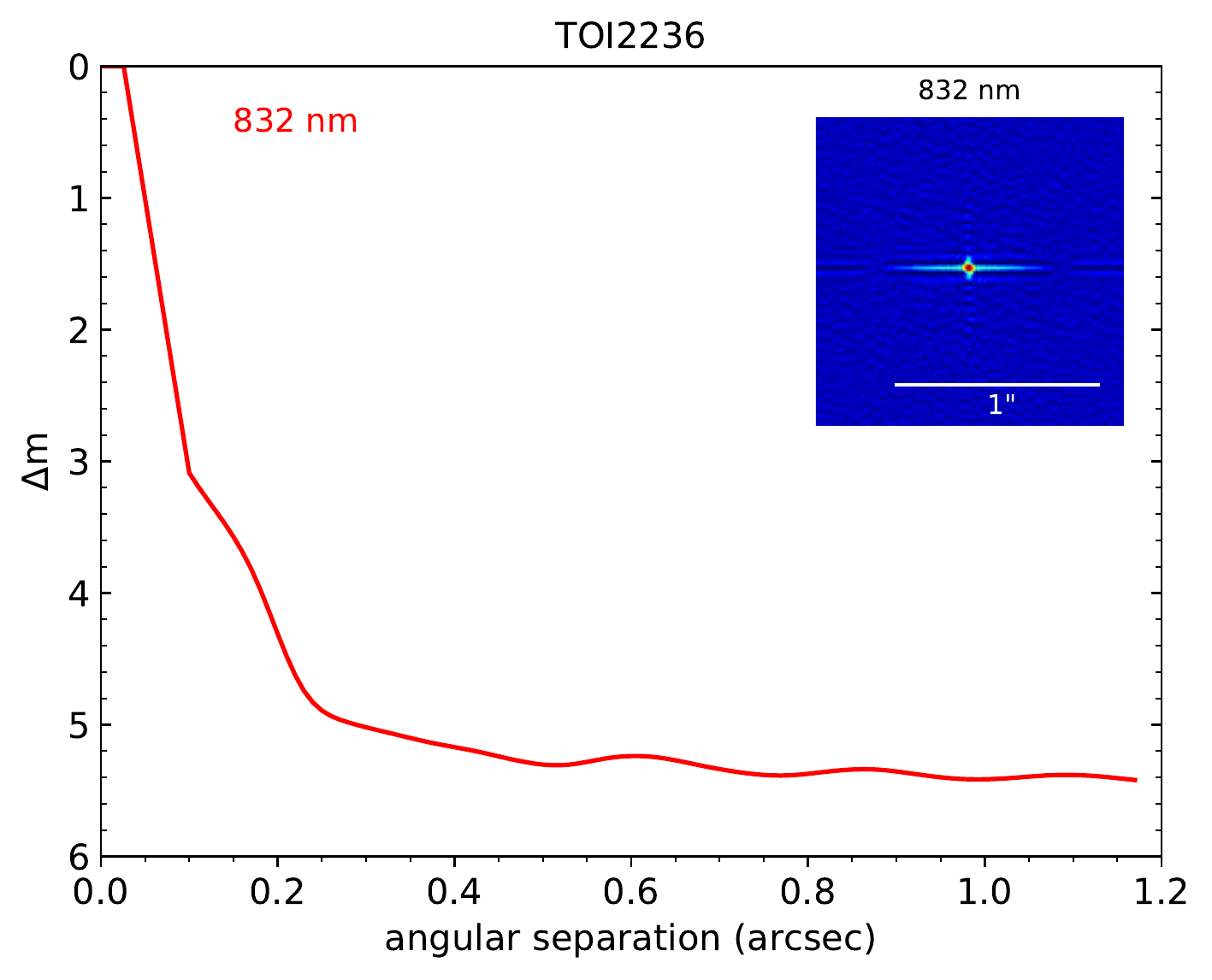} \\
\includegraphics[width=0.35\linewidth]{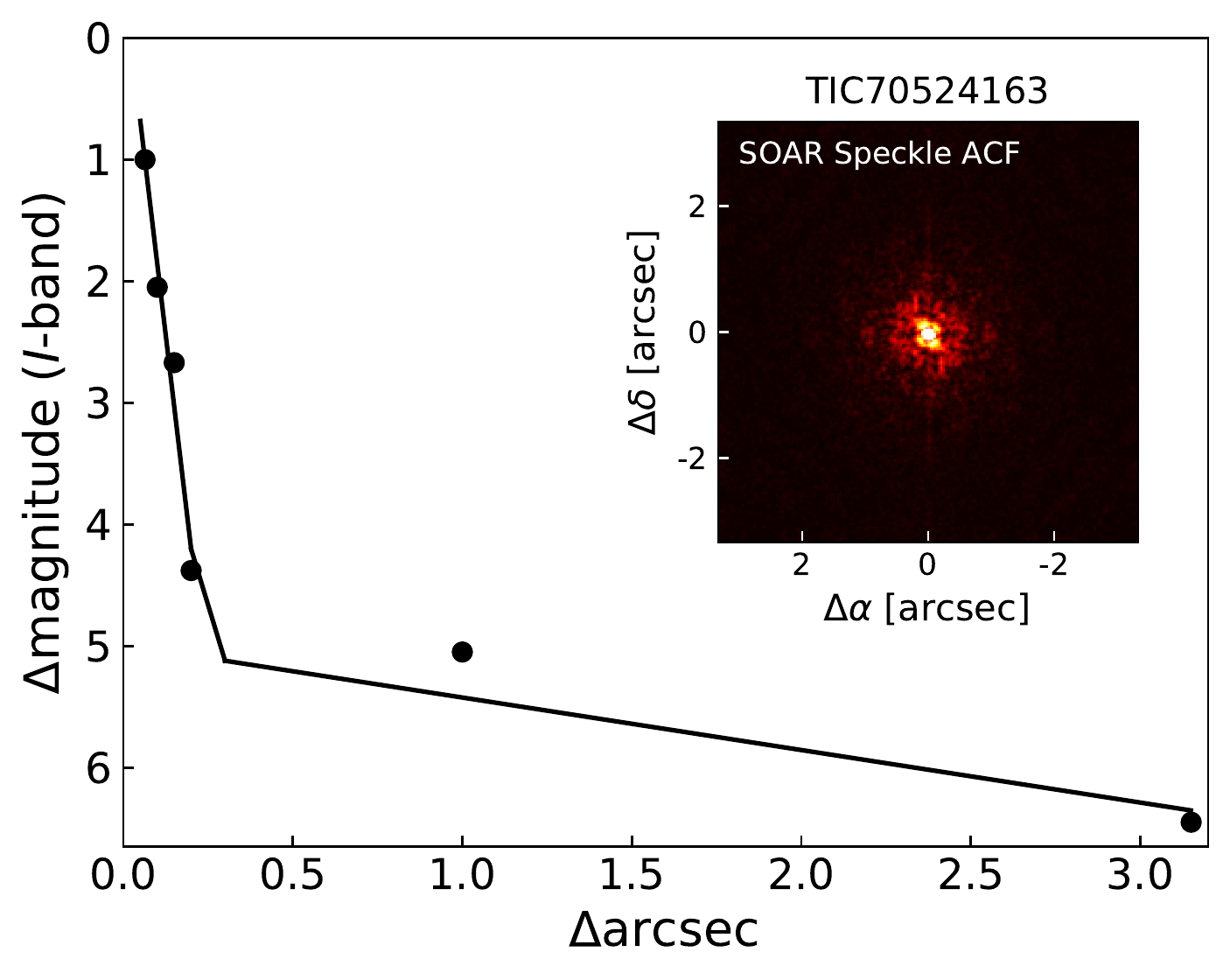}
\includegraphics[width=0.35\linewidth]{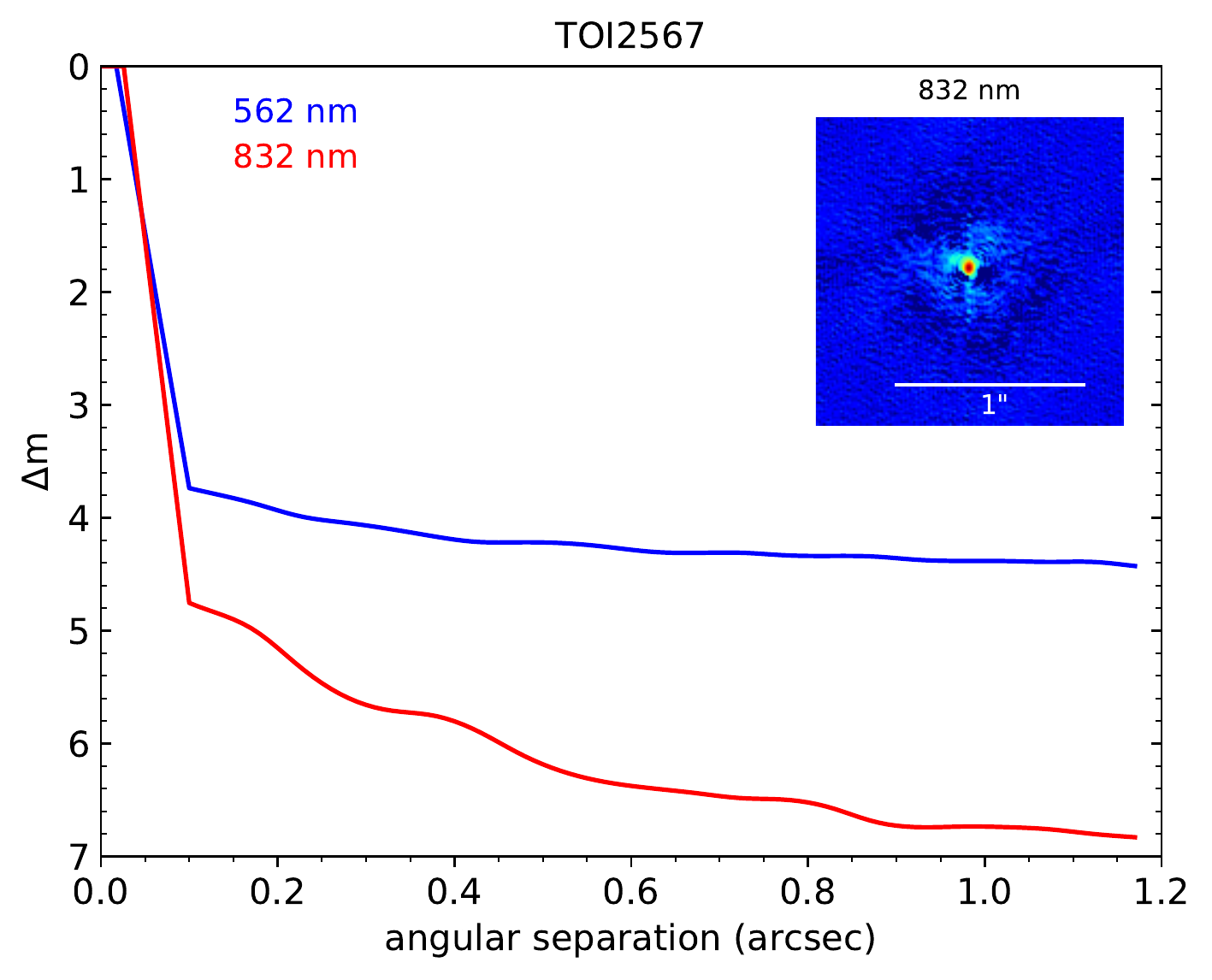} \\
\includegraphics[width=0.32\linewidth]{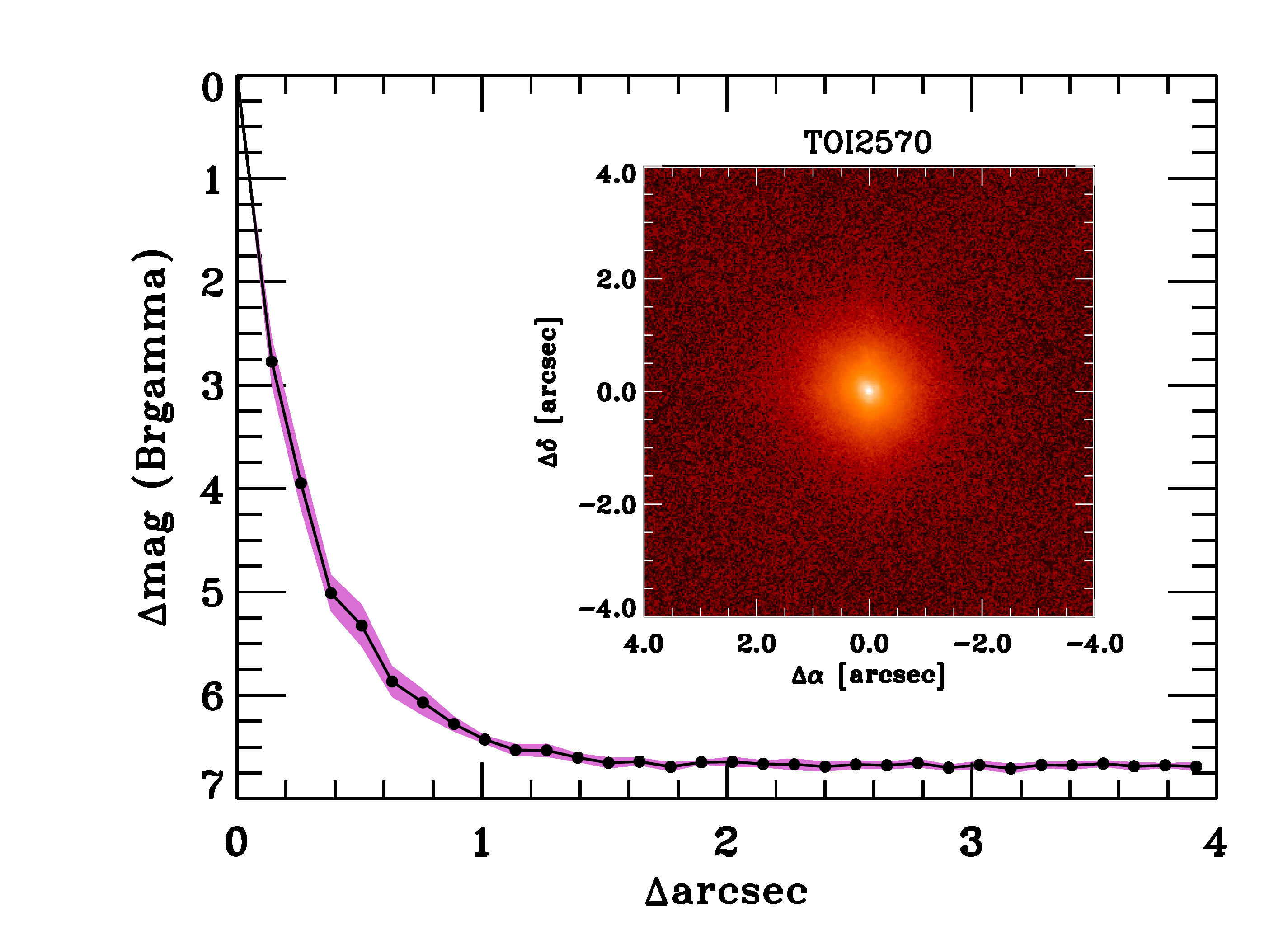}
\includegraphics[width=0.25\linewidth]{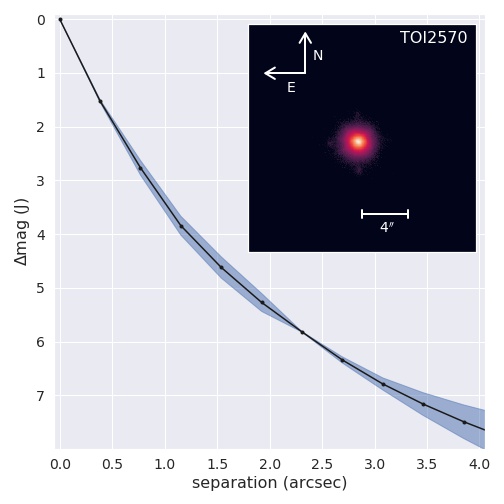}
\includegraphics[width=0.25\linewidth]{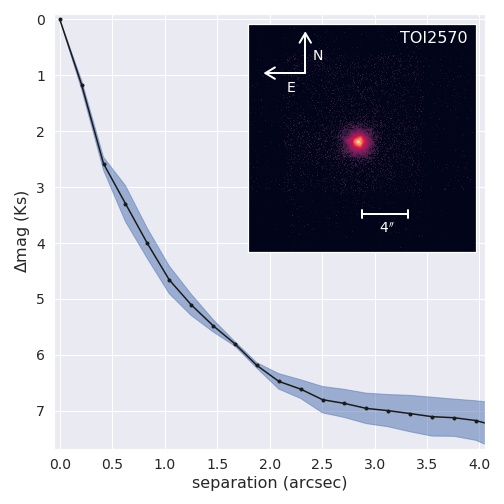}
\includegraphics[width=0.35\linewidth]{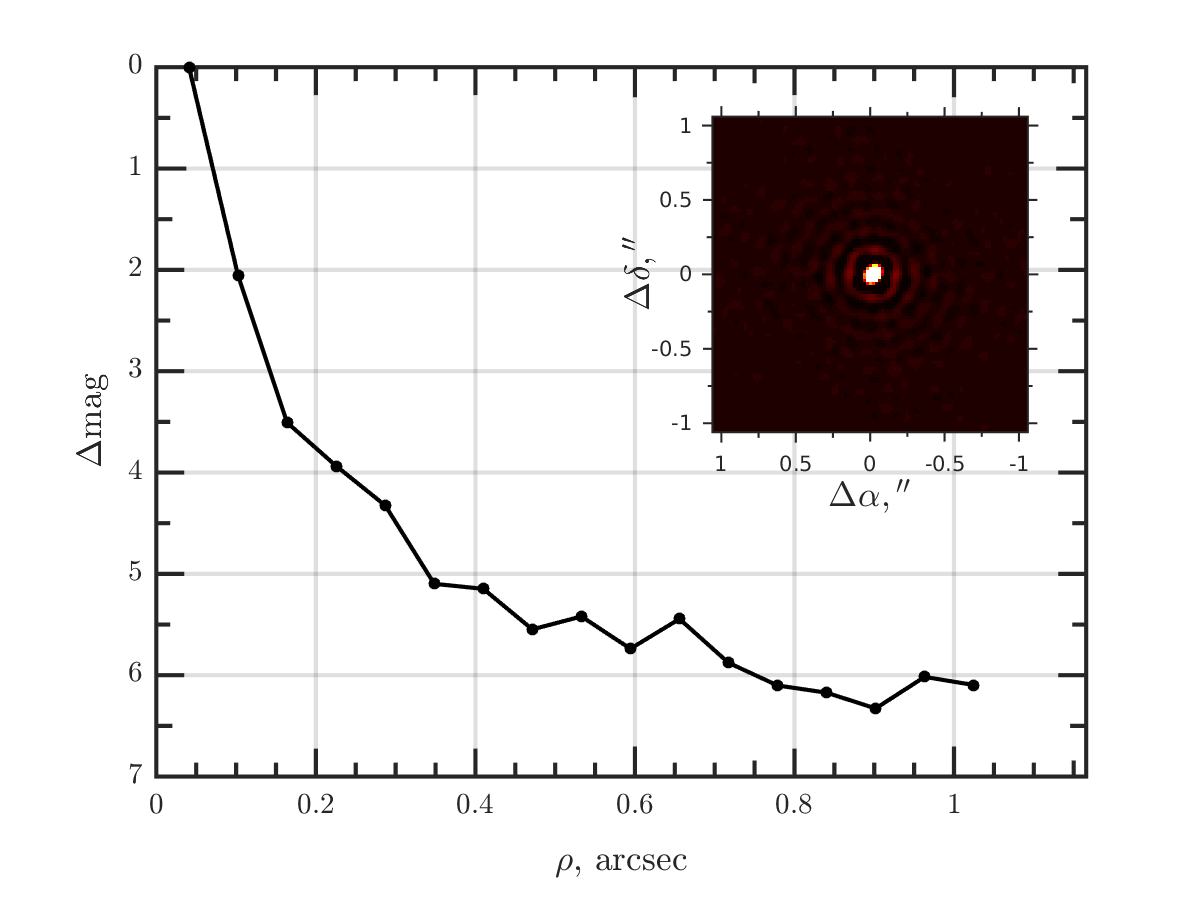} 
\includegraphics[width=0.35\linewidth]{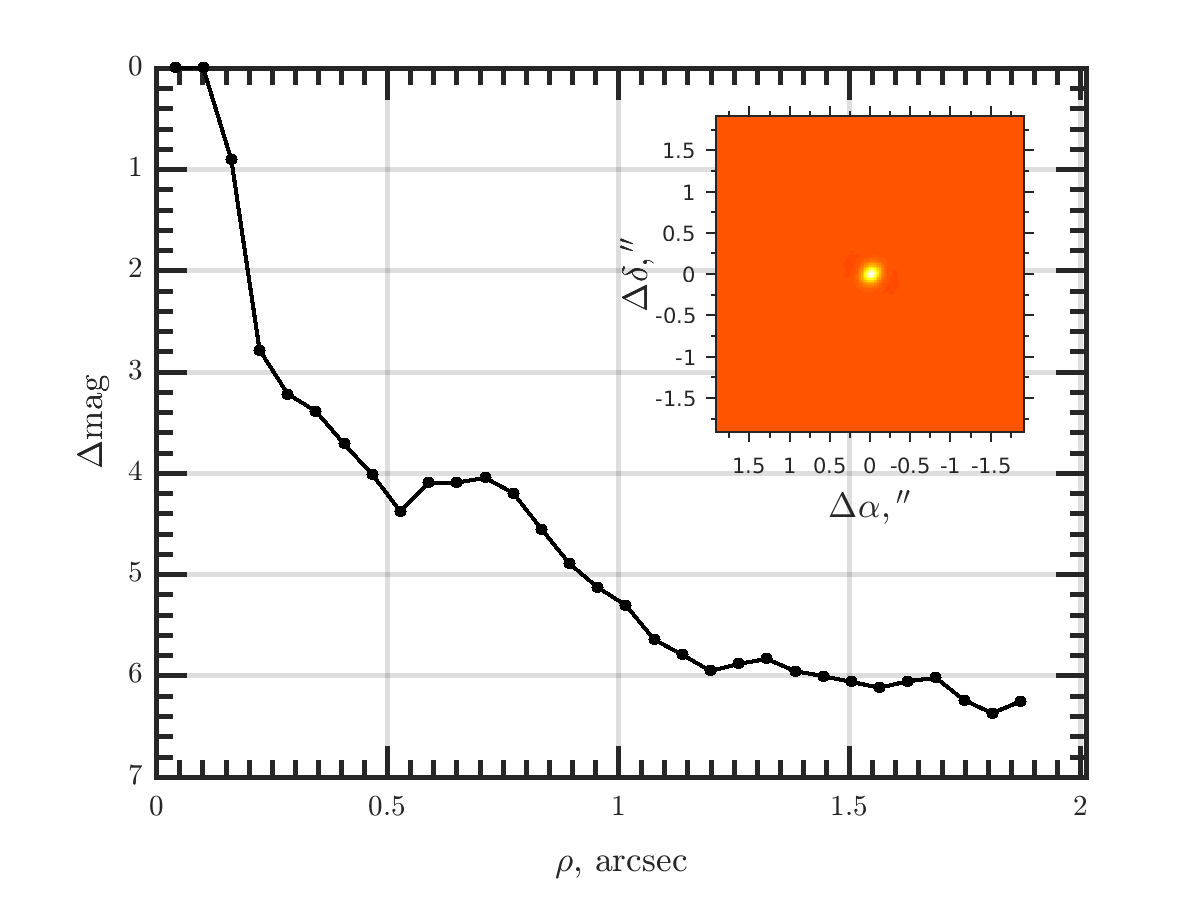}
\caption{High-Resolution imaging of hot Jupiter hosts described in this paper.
From top to bottom, left to right:
\textbf{Row 1:}
SOAR HRCam Observation of TOI-2207;
Gemini-South Zorro Observation of TOI-2236;
\textbf{Row 2:}
SOAR HRCam Observation of TOI-2374;
SOAR HRCam Observation of TOI-2421;
\textbf{Row 3:}
Gemini-North 'Alopeke Observation of TOI-2567;
ShaneAO/SHARCS Observation($J$-band) of TOI-2570;
ShaneAO/SHARCS Observation($K_s$-band) of TOI-2570;
\textbf{Row 4:}
SAI-2.5m Speckle Polarimeter Observation of TOI-3693;
SAI-2.5m Speckle Polarimeter Observation of TOI-4137.
No companions were detected down to the instrumental detection limits for any of these targets.
\label{fig:high_res_imaging}}
\end{figure}

We also obtained high angular-resolution imaging of all of our target systems, in order to detect and characterize stellar companions that might have been blended with the primary target in the \TESS images.
This allows us to assess the spectroscopic follow-up potential of a target star, eliminate false-positive scenarios, and evaluate the impact of any contamination on the fitted planetary properties.
Table \ref{tab:imaging_obs} summarizes the observations made for each system, which were coordinated by the TFOP High-Resolution Imaging Sub-Group 3 (SG3).
Nearby companions were detected in three cases\added{: TOI-2193, TOI-3331, and TOI-3540, with imaging} shown in Figures \ref{fig:toi2193_speckle}, \ref{fig:toi3331_speckle} and \ref{fig:toi3540_ao} \added{respectively}.
No companions were found for the other targets down to detection limits, and plots illustrating these observations can be found at the end of this paper, in Figure \ref{fig:high_res_imaging}.

TOI-2193, TOI-2207, TOI-2421, TOI-3331, and TOI-3540 were observed in the $I$ band with the High-Resolution Camera (HRCam; \citealt{SOAR_Tokovinin2008}), a speckle imaging instrument on the Southern Astrophysical Research (SOAR) 4.1m telescope.
The general observing strategy for \TESS targets and data reduction procedure are described in \cite{SOAR_TESS_Ziegler2019,SOAR_TESS_Ziegler2021} and \cite{SOAR_Tokovinin2018} respectively, while an example SOAR observation for TOI-2193 is shown in Figure \ref{fig:toi2193_speckle}.
Nearby stellar companions were detected in the SOAR observations for TOI-2193 (Sep. = $1\farcs89$, $\Delta I = 3.8$~mag); TOI-3331 (Sep. = $2\farcs66$, $\Delta I = 2.6$~mag); and TOI-3540 (Sep. = $0\farcs92$, $\Delta I = 1.8$~mag).

TOI-2570 was observed using the ShARCS camera on the Shane 3-m telescope at Lick Observatory \citep{ShaneAO_Kupke2012,ShaneAO_Gavel2014,ShaneAO_McGurk2014}.
Two sequences of observations were taken with the Shane adaptive optics system in natural guide star mode, one with a $Ks$ filter ($\lambda_0 = 2.150$~$\mu$m, $\Delta \lambda = 0.320$~$\mu$m) and one with a $J$ filter ($\lambda_0 = 1.238$~$\mu$m, $\Delta \lambda = 0.271$~$\mu$m).
The data were reduced using the publicly available \texttt{SImMER} pipeline \citep{SIMmER_Savel2020}.\footnote{\url{https://github.com/arjunsavel/SImMER}}
We found no evidence for stellar companions within our detection limits.

TOI-2570 and TOI-3540 were also observed in the near-infrared with the Palomar High Angular Resolution Observer (PHARO; \citealt{PHARO_Hayward2001}) on the 200-in Hale telescope at Palomar Observatory.
Both targets were observed in the Br$\gamma$ filters, while TOI-3540 was also observed in the $H$-cont filter.
These observations improved the bound for the lack of stellar companions to TOI-2570, while also identifying the $0\farcs92$ stellar companion to TOI-3540, as seen in the SOAR speckle images.

We observed TOI-2236 and TOI-2567 with the Zorro and 'Alopeke imaging instruments on the Gemini-South and Gemini-North telescopes respectively \citep{Gemini_NESSI_Alopeke_Scott2018,Gemini_Zorro_Alopeke_Scott2021}.
No companions were detected in either set of observations, down to the instrumental detection limits.

Finally, the targets TOI-3693 and TOI-4137 were observed with the speckle polarimeter on the 2.5m telescope at the Caucasian Observatory of Sternberg Astronomical Institute (SAI) of Lomonosov Moscow State University \citep{SAI_Safonov2017}.
The observations were made in the $I$-band, and no stellar companions were detected for either target.

\subsection{High-Resolution Spectroscopy} \label{ssec:spec}

We obtained high-resolution spectroscopy for each of the ten target systems to measure precise relative radial-velocities (RVs) for each system, allowing us to confirm the planetary nature of the transiting companion and measure its mass.
We sought to obtain 6--8 observation epochs per target, primarily at orbital quadratures.
We describe the instruments used and data analysis procedures in the rest of this section, and summarize the observations for each target in Table \ref{tab:rvs}.

In addition to extracting RV measurements from each spectroscopic observation, we measured the bisector inverse slope (BIS) of the spectral line profiles, using the procedures described by \cite{Hartman2019}, for the observations from PFS, CHIRON, FEROS, HIRES and NEID.
For spectra that were observed through an iodine cell, we only used the iodine-free orders to perform this measurement.
The complete RV and BIS data for each target are presented in Table \ref{tab:rv_data}, and plotted in Figures \ref{fig:toi2193_multiplot} and \ref{fig:toi2207_multiplot}-\ref{fig:toi4137_multiplot}.

\begin{deluxetable*}{ccrrcc}
\tablecolumns{6}
\tablewidth{\columnwidth}
\tablecaption{Summary of Radial-Velocity Measurements \label{tab:rvs}}
\tablehead{
	\colhead{Target} & \colhead{Instrument} & \colhead{N$_\mathrm{obs}$} & \colhead{Median $\sigma_\mathrm{RV}$ } & \colhead{First Observation Date} & \colhead{Last Observation Date} \\
	&  &  & \colhead{(m/s)$^{a}$} & \colhead{(UT)} & \colhead{(UT)}
}
\startdata
% target, observatory, num_rvs, median_err, ut_start, ut_end
TOI-2193 & Magellan-Clay/PFS & 8 & 5.8 & 2021 Aug 21 & 2021 Sep 18 \\
TOI-2207 & CTIO-1.5m/CHIRON & 16 & 42.5 & 2021 Aug 14 & 2021 Oct 05 \\
$\cdots$ & MPG/FEROS & 6 & 10.6 & 2021 Jul 22 & 2021 Oct 24 \\
$\cdots$ & Magellan-Clay/PFS & 6 & 5.5 & 2021 Aug 26 & 2021 Oct 19 \\
TOI-2236 & CTIO-1.5m/CHIRON & 8 & 70.5 & 2021 Aug 27 & 2021 Sep 29 \\
TOI-2421 & CTIO-1.5m/CHIRON & 14 & 29.5 & 2021 Aug 05 & 2021 Sep 29 \\
$\cdots$ & Minerva-Australis-3 & 8 & 25.0 & 2021 Jul 07 & 2021 Aug 16 \\
$\cdots$ & Minerva-Australis-6 & 3 & 52.0 & 2021 Jun 26 & 2021 Aug 06 \\
TOI-2567 & Keck-I/HIRES & 6 & 5.4 & 2021 Oct 11 & 2021 Nov 23 \\
$\cdots$ & FLWO/TRES & 2 & 41.6 & 2021 Apr 04 & 2021 Apr 19 \\
TOI-2570 & WIYN/NEID & 7 & 6.3 & 2021 Nov 19 & 2022 Jan 07 \\
$\cdots$ & FLWO/TRES & 2 & 41.5 & 2015 Nov 30 & 2016 Dec 18 \\
TOI-3331 & Magellan-Clay/PFS & 6 & 4.5 & 2021 Aug 26 & 2021 Oct 18 \\
TOI-3540 & Keck-I/HIRES & 6 & 6.0 & 2021 Oct 10 & 2021 Nov 26 \\
$\cdots$ & FLWO/TRES & 2 & 32.8 & 2021 Aug 01 & 2021 Aug 03 \\
TOI-3693 & Keck-I/HIRES & 9 & 6.3 & 2021 Oct 24 & 2021 Dec 26 \\
$\cdots$ & FLWO/TRES & 3 & 29.4 & 2021 Sep 05 & 2021 Dec 21 \\
TOI-4137 & WIYN/NEID & 6 & 15.8 & 2021 Oct 30 & 2021 Nov 27 \\
$\cdots$ & FLWO/TRES & 2 & 41.3 & 2021 Sep 18 & 2021 Oct 09 \\

\enddata
\tablenotetext{a}{Median instrumental RV uncertainty for each target and instrument.}
\end{deluxetable*}	

\begin{deluxetable*}{clrrrrc}
\tablecolumns{7}
\tablewidth{\columnwidth}
\tablecaption{Radial-Velocity Measurements \label{tab:rv_data}}
\tablehead{
	\colhead{Target} & \colhead{Mid-Time} & \colhead{RV} & \colhead{$\sigma(\mathrm{RV})$}
	& \colhead{BIS} & \colhead{$\sigma(\mathrm{BIS})$} & \colhead{Instrument} \\
	& \colhead{\bjdtdb} & \colhead{\ms} & \colhead{\ms} & \colhead{\ms} & \colhead{\ms} &
}
\startdata
% target, bjdtdb, rv, rverr, bis, bis_err, inst 
TOI-2193 & 2459447.667662 & -155.5 & 5.6 & 122 & 178 & PFS \\
TOI-2193 & 2459449.648993 & -51.0 & 9.2 & -520 & 348 & PFS \\
TOI-2193 & 2459452.611481 & 39.8 & 5.9 & -122 & 222 & PFS \\
TOI-2193 & 2459471.612512 & -25.0 & 5.5 & -483 & 294 & PFS \\
TOI-2193 & 2459472.610162 & 17.7 & 5.6 & -27 & 81 & PFS \\

\enddata
\tablecomments{The complete table of RV measurements is available in machine-readable form. \added{The RVs presented here are relative RVs, where an independent offset for each target and instrument $\gamma_\mathrm{rel}$, has been subtracted}.}
\end{deluxetable*}

\subsubsection{PFS Spectroscopy}

We observed TOI-2193, TOI-2207 and TOI-3331 with the Planet Finder Spectrograph (PFS) on the 6.5-meter Magellan II Clay Telescope at the Las Campanas Observatory in Chile \citep{PFS_Crane2006,PFS_Crane2008,PFS_Crane2010}.
PFS is a high-resolution echelle spectrograph that uses an iodine absorption cell to produce precise radial-velocities.
We observed each target in the 3x3 binning mode with the iodine cell in the optical path, choosing short exposure times of ten minutes or less, which allowed us to attain a typical RV precision of about 5~\ms, well above the instrument's demonstrated long-term precision of $\lesssim 1$~\ms.
The spectra were reduced and velocities extracted with the custom pipeline described by \cite{PFS_Butler1996}.
A high S/N, iodine-free, template spectrum was also obtained for each target, as required for the RV extraction procedure.

\subsubsection{CHIRON Spectroscopy}

TOI-2207, TOI-2236, and TOI-2421 were observed with the CTIO High Resolution Spectrometer (CHIRON; \citealt{CHIRON_Tokovinin2013,CHIRON_Paredes2021}) on the CTIO 1.5m telescope on Cerro Tololo in Chile.
CHIRON is an optical, fiber-fed echelle spectrometer with a spectral resolution of $R \approx$80,000 when used with an image slicer.
We observed with typical exposure times of 1200 to 1800s, bracketed by calibration observations of a ThAr lamp.
The spectra were flat-fielded, bias subtracted, and wavelength calibrated using the standard CHIRON pipeline.
Radial-velocities were extracted via least-squares deconvolution of the spectra against synthetic templates \citep{CHIRON_Donati1997,CHIRON_Zhou2020}, achieving a typical RV precision of 30-70~\ms.
We also used the Stellar Parameter Classification (SPC; \citealt{SPC_Buchhave2012}) code to derive atmospheric properties from the spectra, with the results shown in Table \ref{tab:spec_props}.

\subsubsection{FEROS Spectroscopy}
TOI-2207 was also observed on six epochs with the FEROS spectrograph \citep{FEROS_Kaufer1999} mounted on the MPG 2.2m telescope at the ESO La Silla Observatory, in Chile, in the context of the Warm gIaNts with tEss collaboration \citep[WINE,][]{WINE_Brahm2019,WINE_Trifonov2021}.
FEROS is a stabilized fiber-fed high resolution spectrograph configured with a comparison fiber to trace instrumental radial-velocity drift during the scientific exposures.
The six observations of TOI-2207 were performed between 2021-07-22 and 2021-10-24 with a typical exposure time of 1200 seconds, achieving an S/N per resolution element ranging from 70 to 100.
All FEROS data were processed with the CERES pipeline \citep{CERES_Brahm2017}, which executed all steps involved in obtaining high precision radial-velocities with the cross-correlation technique, starting from the raw images.
The typical radial-velocity error of these observations was $\approx 10$ m/s.
CERES also estimates the stellar atmospheric parameters from the spectra, which are tabulated in Table \ref{tab:spec_props}. 

\subsubsection{MINERVA-Australis Spectroscopy}

We also obtained 11 observations of TOI-2421 between 2021-06-26 and 2021-08-16 using the \textsc{Minerva}-Australis telescope array \citep{MinervaAustralis_Wittenmyer2018,MinervaAustralis_Addison2019}, located at Mt. Kent Observatory, Australia.
\textsc{Minerva}-Australis is an array of four identical 0.7 m telescopes linked via fiber feeds to a single KiwiSpec echelle spectrograph at a spectral resolving power of $R\sim$80,000 over the wavelength region of 5000-6300\AA.
The array is wholly dedicated to radial-velocity follow-up of TESS planet candidates \citep[e.g.,][]{Nielsen2019,Addison2021,MinervaAustralis_Wittenmyer2018}.
Simultaneous wavelength calibration is provided via two calibration fibers illuminated by a quartz lamp through an iodine cell.
The spectra were extracted for each telescope individually and the radial-velocities were extracted via the same techniques as those described above for the CHIRON observations.

\subsubsection{TRES Spectroscopy} \label{ssec:tres_spec}

We observed all five targets in the Northern hemisphere (TOI-2567, TOI-2570, TOI-3540, TOI-3693, and TOI-4137) with the Tillinghast Reflector Echelle Spectrograph (TRES; \citealt{TRES_Furesz2008}).
TRES has a spectral resolution of $R \sim 44,000$ and is located on the FLWO 1.5m Tillinghast Reflector telescope on Mount Hopkins, Arizona.
The observations for each target were scheduled near the two opposite quadratures, ensuring maximum sensitivity to the planet's orbital motion.
Two TRES spectra were taken of each target except for TOI-3693, for which we collected three observations.
The data were reduced and radial-velocities extracted using the pipeline described in \citet{Buchhave2010} and \citet{Quinn2012}.

The spectra were also analyzed with SPC to derive the stellar atmospheric parameters \Teff, \logg, \feh, and \vsini.
The results from each observation were weighted according to the cross-correlation function and averaged together, with the final stellar properties presented in Table \ref{tab:spec_props}.

\subsubsection{HIRES Spectroscopy}

We observed TOI-2567, TOI-3540 and TOI-3693 with the High Resolution Echelle Spectrometer (HIRES; \citealt{HIRES_Vogt94}) on the Keck-I 10m telescope on Maunakea, Hawaii.
We obtained 6--9 observations for each target with an iodine cell that provides a precise wavelength calibration and allows for radial-velocity extraction.
These observations were made through the queue system operated by the California Planet Search (CPS), and reduced with the standard CPS procedures \citep{CPS_Howard2010,CPS_Howard2016}.

To reduce the observation time necessary to obtain a high-S/N, iodine-free template spectrum, we used the matched template technique developed by \citet{CPS_Dalba2020a}.
Briefly, we first obtained an iodine-free but low S/N ($\sim 40$/pixel) reconnaissance spectrum, which was then matched against a library of archival HIRES template spectra described in \citet{SpecMatchEmp_Yee2017}.
A Deconvolved Stellar Spectral Template (DSST) was then derived for the target using the high S/N spectrum of the best-matching library star, which could then be used in the radial-velocity extraction procedure.
For the slowly rotating F and G dwarfs observed here, \citet{CPS_Dalba2020a} found that the technique introduced a median error of 4.7\,\ms to the RV measurements, compared with obtaining a high-S/N template of the target star.
We included this introduced error by adding it in quadrature to the internal RV precision of the observations.
For the massive planets targeted by our work, this additional error should not significantly affect the characterization of the planetary systems, especially given that we chose observation exposure times of $\lesssim 10$~minutes to yield an RV precision of a similar magnitude, $5$--$10$~\ms.

\subsubsection{NEID Spectroscopy}

TOI-2570 and TOI-4137 were observed with the NEID spectrograph on the WIYN 3.5m telescope at Kitt Peak National Observatory (KPNO).
NEID is a newly commissioned stabilized, fiber-fed optical spectrograph with a resolving power of $R\approx$110,000 spanning the wavelength range from 3800 to 9300\,\AA. \citep{NEID_Schwab2016,NEID_Halverson2016}.
The data were reduced and RVs extracted using v1.1.2 of the standard NEID Data Reduction Pipeline (NEID-DRP)\footnote{\url{https://neid.ipac.caltech.edu/docs/NEID-DRP}}, which derives velocities through cross-correlation with a weighted numerical stellar mask based on spectral type \citep{Baranne1996,Pepe2002}.
We used relatively short exposure times for our observations, obtaining RV precisions of about 5~\ms for TOI-2570 and 15~\ms for the relatively fainter TOI-4137.

\section{Stellar Characterization} \label{sec:stellar_char}

We collected literature photometry and astrometry of each of the planet host stars from the \TESS Input Catalog (TIC; \citealp{TIC_Stassun2018,TIC_Stassun2019}), \Gaia EDR3 \citep{GaiaEDR3_Brown2021,GaiaEDR3_Riello2021,GaiaEDR3_Lindegren2021}, \textit{2MASS} \citep{TMASS_Cutri2003}, \textit{WISE} \citep{WISE_Cutri2012}, and Tycho-2 \citep{Tycho2_Hog2000} catalogs.
These properties are displayed in Tables \ref{tab:stellar_props} and \ref{tab:stellar_props_2}, and were used in our global modelling of each system (\S\ref{sec:planet_char}).

\subsection{Spectroscopic Parameters} \label{ssec:spec_char}
For each system, we further characterized the host star properties using the high resolution stellar spectra obtained as part of our spectroscopic follow-up program.
In the case of systems observed with the stabilized spectrographs CHIRON and NEID, we used a spectrum from a single epoch with the highest signal-to-noise ratio (S/N).
For systems observed with PFS, we used the high S/N iodine-free template spectrum.
For those observed with HIRES, we used the low S/N iodine-free reconnaissance spectrum.

We used the publicly available code \texttt{SpecMatch-Emp} \citep{SpecMatchEmp_Yee2017}\footnote{\url{https://github.com/samuelyeewl/specmatch-emp}} to obtain a homogeneous set of stellar properties for our targets.
This code works by comparing a target spectrum to a library of observed high-resolution ($R \approx $60,000), high S/N ($\mathrm{S/N} \approx 150$/pixel) spectra from Keck/HIRES.
The library stars have well-determined empirical stellar properties from a variety of sources, including asteroseismology, interferometry, and spectrophotometry.
The code finds the five library spectra that best match the target, accounting for rotational broadening, and interpolates between those stars' properties to derive (\Teff, \Rstar, \feh) for the target, with uncertainties of $\sigma(\Teff) = 100$~K, $\sigma(\Delta \Rstar/\Rstar) = 15\%$, $\sigma(\feh) = 0.09$~dex, which are robust even when the S/N is as low as 20/pixel.

While \texttt{SpecMatch-Emp} was developed for use with Keck/HIRES spectra, it has been successfully used with spectra from other instruments \citep[e.g.,][]{Teske2018}.
To account for the narrow line spread functions of the other spectrographs used compared with that of HIRES, we modified the code to allow the target star's spectrum to be broadened relative to the library spectra.
We found that this produced sharper $\chi^2$ minima, and a better match to the target spectrum.
Previous testing of this approach using a cross-validation technique with the library Keck/HIRES spectra showed no degradation in the accuracy of the derived parameters.
We compared the \texttt{SpecMatch-Emp}-derived parameters to those derived from the TRES reconnaissance spectra when available (\S\ref{ssec:tres_spec}), and found them to be within 1-$\sigma$ agreement.

We measured \vsini for each of the targets using the \texttt{SpecMatch-Synth}\footnote{\url{https://github.com/petigura/specmatch-syn}} \citep{SpecMatchSynth_Petigura2015} code.
This code works similarly to \texttt{SpecMatch-Emp}, but matches the target spectrum to a synthetic spectral library from \cite{Coelho2005} instead.
A set of eight best-matching spectra are selected from the synthetic library, which spans a range of \Teff, \logg, and \feh.
These are then combined using trilinear interpolation and convolved with a rotational-macroturbulent profile and Gaussian instrumental profile to create a better match to the target spectrum.
During this process, the macroturbulent broadening is assumed to follow the relationship from \citet{Valenti2005}:
\begin{equation}
    v_\mathrm{mac} = \left(3.98 - \frac{\Teff - 5770\,\mathrm{K}}{650\,\mathrm{K}}\right)\,\kms.
\end{equation}
The code optimizes over the interpolation weights and \vsini to derive the target stellar atmospheric properties.
We report only the \vsini from this code, but we found that the other parameters were typically within 1-$\sigma$ agreement with those derived by \texttt{SpecMatch-Emp}, which is more robust for low S/N spectra.
\replaced{The combined results from \texttt{SpecMatch-Emp} and \texttt{SpecMatch-Synth} are presented in Table \ref{tab:spec_props}.}
{We report \Teff, \logg, and \feh from \texttt{SpecMatch-Emp}, and $\vsini$ and $v_\mathrm{mac}$ from \texttt{SpecMatch-Synth} in Table \ref{tab:spec_props}.}

For those targets with observations from TRES, we also used SPC \citep{SPC_Buchhave2012} to derive stellar atmospheric properties.
SPC cross-correlates an observed spectrum against a grid of synthetic spectra from \cite{Kurucz1993}, allowing \Teff, \logg, \feh, and \vsini to be determined.
The SPC stellar parameters are also tabulated in Table \ref{tab:spec_props}, and we found that in all cases, the derived properties did not differ from those from \texttt{SpecMatch} by more than 1.5-$\sigma$.
For targets with CHIRON spectra, we derived stellar properties by matching the spectra against a library of $\sim$10,000 observed spectra previously classified by SPC.
This procedure is described in more detail in \cite{Rodriguez2021}, with results shown in Table \ref{tab:spec_props}.
Finally, for TOI-2207, we used the CERES code \citep{CERES_Brahm2017} to estimate stellar properties from the FEROS spectra.
In all cases, the stellar properties derived by these different codes do not differ significantly, giving us greater confidence in these results.
For consistency, we used the \texttt{SpecMatch} results for \Teff, \Rstar and \feh for all targets as prior constraints in our \Exofast fits, as described in Section \ref{sec:planet_char}.

% \afterpage{\clearpage}
\makeatletter\onecolumngrid@push\makeatother
\begin{rotatepage}
\movetabledown=1.5in
\begin{rotatetable}
\begin{deluxetable}{lcccccc}
\tablecaption{Catalog Photometry and Astrometry of Planet Host Stars \label{tab:stellar_props}}
\tabletypesize{\small}
\tablecolumns{7}
\tablehead{
\colhead{\textbf{Target}} & \colhead{TOI-2193} & \colhead{TOI-2207} & \colhead{TOI-2236} & \colhead{TOI-2421} & \colhead{TOI-2567} & \colhead{Source} 
}
\startdata
\multicolumn{7}{l}{\textbf{Identifiers}}\\
TIC & 401604346 & 90850770 & 394722182 & 70524163 & 258920431 & \\
GAIA EDR3 & 6373308503181838592 & 6675883485986480256 & 4613145315172329984 & 4968289907406396160 & 2254929887069708160 & \\
2MASS & 20544592-7248166 & 20302318-4453150 & 01203986-8658476 & 02123692-3523272 & 19135180+6620524 & \\
Tycho-2 & 9329-00167-1 & 7962-00808-1 & 9498-00283-1 & 7009-01148-1 & -- & \\
WISE & J205445.92-724816.8 & J203023.20-445315.2 & J012039.88-865847.8 & J021236.95-352327.2 & J191351.77+662052.5 & \\
\multicolumn{7}{l}{\textbf{Astrometric Measurements}}\\
R.A. (J2000) & 20:54:45.895 & 20:30:23.207 & 01:20:39.841 & 02:12:36.966 & 19:13:51.759 & 1\\
Decl. (J2000) & -72:48:16.71 & -44:53:15.44 & -86:58:47.88 & -35:23:27.31 & +66:20:52.58 & 1\\
$\mu_{{\alpha}}\cos{\delta}$ (mas yr$^{{-1}}$) & -2.465 $\pm$ 0.009 & 15.782 $\pm$ 0.021 & -3.653 $\pm$ 0.017 & 29.004 $\pm$ 0.012 & -16.973 $\pm$ 0.011 & 1\\
$\mu_{{\delta}}$ (mas yr$^{{-1}}$) & -0.911 $\pm$ 0.011 & -25.806 $\pm$ 0.018 & -7.860 $\pm$ 0.015 & -3.689 $\pm$ 0.017 & 9.021 $\pm$ 0.011 & 1\\
Parallax (mas) & 2.896 $\pm$ 0.009 & 2.626 $\pm$ 0.020 & 2.837 $\pm$ 0.012 & 3.055 $\pm$ 0.020 & 1.929 $\pm$ 0.009 & 1\\
$b$ ($^{{\circ}}$) & -34.782 & -35.849 & -30.123 & -70.835 & 22.507 & 1\\
$l$ ($^{{\circ}}$) & 320.935 & 355.480 & 302.488 & 243.467 & 97.326 & 1\\
\multicolumn{7}{l}{\textbf{Photometric Measurements}}\\
$T$ (mag) & 11.400 $\pm$ 0.006 & 10.965 $\pm$ 0.006 & 10.867 $\pm$ 0.006 & 10.692 $\pm$ 0.006 & 11.749 $\pm$ 0.008 & 2\\
$G$ (mag) & 11.806 $\pm$ 0.003 & 11.352 $\pm$ 0.003 & 11.289 $\pm$ 0.003 & 11.157 $\pm$ 0.003 & 12.210 $\pm$ 0.003 & 1\\
$G_\mathrm{{BP}}$ (mag) & 12.109 $\pm$ 0.003 & 11.637 $\pm$ 0.003 & 11.601 $\pm$ 0.003 & 11.510 $\pm$ 0.003 & 12.573 $\pm$ 0.003 & 1\\
$G_\mathrm{{RP}}$ (mag) & 11.331 $\pm$ 0.004 & 10.906 $\pm$ 0.004 & 10.806 $\pm$ 0.004 & 10.637 $\pm$ 0.004 & 11.683 $\pm$ 0.004 & 1\\
$B_T$ (mag) & 12.857 $\pm$ 0.202 & 12.519 $\pm$ 0.207 & 12.267 $\pm$ 0.135 & 12.149 $\pm$ 0.089 & -- & 3\\
$V_T$ (mag) & 12.057 $\pm$ 0.143 & 11.422 $\pm$ 0.102 & 11.664 $\pm$ 0.105 & 11.488 $\pm$ 0.068 & -- & 3\\
$J$ (mag) & 10.767 $\pm$ 0.024 & 10.438 $\pm$ 0.024 & 10.261 $\pm$ 0.023 & 10.090 $\pm$ 0.024 & 11.118 $\pm$ 0.021 & 4\\
$H$ (mag) & 10.462 $\pm$ 0.023 & 10.201 $\pm$ 0.023 & 10.006 $\pm$ 0.023 & 9.752 $\pm$ 0.024 & 10.764 $\pm$ 0.019 & 4\\
$K_s$ (mag) & 10.383 $\pm$ 0.021 & 10.133 $\pm$ 0.023 & 9.960 $\pm$ 0.021 & 9.667 $\pm$ 0.021 & 10.717 $\pm$ 0.017 & 4\\
$W1$ (mag) & 10.343 $\pm$ 0.023 & 10.091 $\pm$ 0.022 & 9.900 $\pm$ 0.023 & 9.624 $\pm$ 0.024 & 10.659 $\pm$ 0.023 & 5\\
$W2$ (mag) & 10.354 $\pm$ 0.021 & 10.149 $\pm$ 0.020 & 9.925 $\pm$ 0.020 & 9.678 $\pm$ 0.020 & 10.714 $\pm$ 0.020 & 5\\
$W3$ (mag) & 10.361 $\pm$ 0.068 & 10.191 $\pm$ 0.069 & 9.884 $\pm$ 0.037 & 9.623 $\pm$ 0.040 & 10.629 $\pm$ 0.054 & 5\\
\enddata

\tablerefs{(1) - \Gaia EDR3 \citep{GaiaEDR3_Brown2021};
(2) - \TESS Input Catalog \citep{TIC_Stassun2019};
(3) - Tycho-2 \citep{Tycho2_Hog2000};
(4) - 2MASS \citep{TMASS_Cutri2003};
(5) - WISE \citep{WISE_Cutri2012}.}
% (6) - This work (\texttt{SpecMatch-Emp} derived);
% (7) - This work (\texttt{SpecMatch-Synth} derived), $v_\mathrm{mac}$ is assumed based on effective temperature and the relation from \citet{Valenti2005}.}
\tablecomments{The catalog photometry presented here has not been corrected for contamination by nearby stellar companions (\S\ref{ssec:companions}).\\
The data in this table is available in machine-readable form.}
\end{deluxetable}
\end{rotatetable}
% \makeatletter\onecolumngrid@pop\makeatother
% \pdfpageattr\expandafter{\the\pdfpageattr/Rotate 90}

% \makeatletter\onecolumngrid@push\makeatother
\movetabledown=1.5in
\begin{rotatetable}
\begin{deluxetable}{lcccccc}
\tablecaption{Catalog Photometry and Astrometry of Planet Host Stars \label{tab:stellar_props_2}}
\tabletypesize{\small}
\tablecolumns{7}
\tablehead{
\colhead{\textbf{Target}} & \colhead{TOI-2570} & \colhead{TOI-3331} & \colhead{TOI-3540} & \colhead{TOI-3693} & \colhead{TOI-4137} & \colhead{Source} 
}
\startdata
\multicolumn{7}{l}{\textbf{Identifiers}}\\
TIC & 239816546 & 194795551 & 17865622 & 240823272 & 417646390 & \\
GAIA EDR3 & 3445148131761842944 & 4042548116644168832 & 1896138833241139584 & 404433018447476096 & 497750842338402560 & \\
2MASS & 05484513+3205028 & 18051781-3406254 & 21553871+2810460 & 01023706+5118143 & 05102703+7023279 & \\
Tycho-2 & -- & 7399-00615-1 & -- & 3275-01223-1 & 4346-00736-1 & \\
WISE & J054845.12+320502.6 & J180517.84-340624.5 & J215538.71+281045.9 & J010237.05+511814.1 & J051027.06+702327.9 & \\
\multicolumn{7}{l}{\textbf{Astrometric Measurements}}\\
R.A. (J2000) & 05:48:45.129 & 18:05:17.800 & 21:55:38.729 & 01:02:37.054 & 05:10:27.093 & 1\\
Decl. (J2000) & +32:05:02.55 & -34:06:25.82 & +28:10:46.20 & +51:18:14.00 & +70:23:27.88 & 1\\
$\mu_{{\alpha}}\cos{\delta}$ (mas yr$^{{-1}}$) & -6.531 $\pm$ 0.014 & 0.445 $\pm$ 0.032 & 6.306 $\pm$ 0.044 & -6.962 $\pm$ 0.016 & 19.686 $\pm$ 0.013 & 1\\
$\mu_{{\delta}}$ (mas yr$^{{-1}}$) & -15.848 $\pm$ 0.009 & -17.028 $\pm$ 0.022 & -1.296 $\pm$ 0.043 & -21.307 $\pm$ 0.014 & -3.005 $\pm$ 0.015 & 1\\
Parallax (mas) & 2.765 $\pm$ 0.014 & 4.446 $\pm$ 0.026 & 3.541 $\pm$ 0.040 & 5.656 $\pm$ 0.018 & 2.891 $\pm$ 0.017 & 1\\
$b$ ($^{{\circ}}$) & 2.207 & -6.219 & -20.456 & -11.529 & 17.587 & 1\\
$l$ ($^{{\circ}}$) & 177.656 & 357.625 & 82.170 & 124.715 & 141.500 & 1\\
\multicolumn{7}{l}{\textbf{Photometric Measurements}}\\
$T$ (mag) & 11.977 $\pm$ 0.006 & 11.321 $\pm$ 0.006 & 10.954 $\pm$ 0.007 & 11.419 $\pm$ 0.006 & 10.856 $\pm$ 0.009 & 2\\
$G$ (mag) & 12.450 $\pm$ 0.003 & 11.824 $\pm$ 0.003 & 11.428 $\pm$ 0.003 & 11.955 $\pm$ 0.003 & 11.246 $\pm$ 0.003 & 1\\
$G_\mathrm{{BP}}$ (mag) & 12.821 $\pm$ 0.003 & 12.212 $\pm$ 0.003 & 11.674 $\pm$ 0.003 & 12.386 $\pm$ 0.003 & 11.534 $\pm$ 0.003 & 1\\
$G_\mathrm{{RP}}$ (mag) & 11.915 $\pm$ 0.004 & 11.255 $\pm$ 0.004 & 10.835 $\pm$ 0.005 & 11.357 $\pm$ 0.004 & 10.797 $\pm$ 0.004 & 1\\
$B_T$ (mag) & -- & 12.989 $\pm$ 0.372 & -- & 12.783 $\pm$ 0.174 & 12.243 $\pm$ 0.093 & 3\\
$V_T$ (mag) & -- & 11.753 $\pm$ 0.204 & -- & 12.025 $\pm$ 0.120 & 11.395 $\pm$ 0.058 & 3\\
$J$ (mag) & 11.303 $\pm$ 0.022 & 10.348 $\pm$ 0.026 & 10.237 $\pm$ 0.021 & 10.657 $\pm$ 0.022 & 10.314 $\pm$ 0.023 & 4\\
$H$ (mag) & 10.960 $\pm$ 0.021 & 10.068 $\pm$ 0.029 & 9.899 $\pm$ 0.022 & 10.246 $\pm$ 0.023 & 10.048 $\pm$ 0.031 & 4\\
$K_s$ (mag) & 10.905 $\pm$ 0.020 & 9.914 $\pm$ 0.026 & 9.784 $\pm$ 0.017 & 10.151 $\pm$ 0.020 & 10.004 $\pm$ 0.022 & 4\\
$W1$ (mag) & 10.860 $\pm$ 0.023 & 9.565 $\pm$ 0.022 & 9.742 $\pm$ 0.023 & 10.132 $\pm$ 0.023 & 9.957 $\pm$ 0.023 & 5\\
$W2$ (mag) & 10.918 $\pm$ 0.020 & 9.653 $\pm$ 0.020 & 9.792 $\pm$ 0.019 & 10.178 $\pm$ 0.020 & 9.999 $\pm$ 0.021 & 5\\
$W3$ (mag) & 11.178 $\pm$ 0.167 & 9.895 $\pm$ 0.075 & 9.678 $\pm$ 0.062 & 10.141 $\pm$ 0.049 & 9.930 $\pm$ 0.093 & 5\\
\enddata

\end{deluxetable}
\end{rotatetable}
% \end{rotatepage}
% \makeatletter\onecolumngrid@pop\makeatother
% \clearpage

% \cleardoublepage

% \makeatletter\onecolumngrid@push\makeatother
% \begin{rotatepage}
\movetabledown=1in
\addtolength{\tabcolsep}{4pt}
\begin{splitdeluxetable}{lcccccccccBlccccccccc@{\hspace{6em}}}
\rotate
\tabletypesize{\small}
\tablecaption{Spectroscopic Stellar Properties \label{tab:spec_props}}
\tablecolumns{20}
\tablehead{
\colhead{\textbf{Target}} & \multicolumn{1}{c}{TOI-2193} & \multicolumn{2}{c}{TOI-2207} & \multicolumn{2}{c}{TOI-2236} & \multicolumn{2}{c}{TOI-2421} & \multicolumn{2}{c}{TOI-2567} & \colhead{\textbf{Target}} & \multicolumn{2}{c}{TOI-2570} & \multicolumn{1}{c}{TOI-3331} & \multicolumn{2}{c}{TOI-3540} & \multicolumn{2}{c}{TOI-3693} & \multicolumn{2}{c}{TOI-4137} \\
\cmidrule(lr){2-2}\cmidrule(lr){3-4}\cmidrule(lr){5-6}\cmidrule(lr){7-8}\cmidrule(lr){9-10}\cmidrule(lr){12-13}\cmidrule(lr){14-14}\cmidrule(lr){15-16}\cmidrule(lr){17-18}\cmidrule(lr){19-20}
 \colhead{Code} & SpecMatch & SpecMatch & CERES & SpecMatch & SPC & SpecMatch & SPC & SpecMatch & SPC & \colhead{Code} & SpecMatch & SPC & SpecMatch & SpecMatch & SPC & SpecMatch & SPC & SpecMatch & SPC \\
\colhead{Instrument} & PFS & PFS & FEROS & CHIRON & CHIRON & CHIRON & CHIRON & HIRES & TRES & \colhead{Instrument} & NEID & TRES & PFS & HIRES & TRES & HIRES & TRES & NEID & TRES 
}
\startdata
$T_\mathrm{eff}$ (K) & 5974 $\pm$ 110 & 6075 $\pm$ 110 & 6050 $\pm$ 150 & 6228 $\pm$ 110 & 6164 $\pm$ 50 & 5645 $\pm$ 110 & 5577 $\pm$ 50 & 5609 $\pm$ 110 & 5650 $\pm$ 50 & $T_\mathrm{eff}$ (K) & 5756 $\pm$ 110 & 5765 $\pm$ 50 & 5521 $\pm$ 110 & 5865 $\pm$ 110 & 5969 $\pm$ 50 & 5246 $\pm$ 110 & 5274 $\pm$ 50 & 6125 $\pm$ 110 & 6134 $\pm$ 50\\
$R_\star$ ($R_\odot$) & 1.21 $\pm$ 0.18 & 1.49 $\pm$ 0.22 & -- & 1.66 $\pm$ 0.25 & -- & 1.92 $\pm$ 0.29 & -- & 1.83 $\pm$ 0.27 & -- & $R_\star$ ($R_\odot$) & 1.35 $\pm$ 0.20 & -- & 1.06 $\pm$ 0.16 & 1.34 $\pm$ 0.20 & -- & 0.85 $\pm$ 0.13 & -- & 1.65 $\pm$ 0.25 & --\\
$\log{g}$ (cgs) & -- & -- & 4.20 $\pm$ 0.20 & -- & 4.17 $\pm$ 0.10 & -- & 3.82 $\pm$ 0.10 & -- & 4.15 $\pm$ 0.10 & $\log{g}$ (cgs) & -- & 4.43 $\pm$ 0.10 & -- & -- & 4.40 $\pm$ 0.10 & -- & 4.61 $\pm$ 0.10 & -- & 4.13 $\pm$ 0.10\\
$[\mathrm{Fe/H}]$ (dex) & -0.04 $\pm$ 0.09 & 0.15 $\pm$ 0.09 & 0.10 $\pm$ 0.10 & 0.08 $\pm$ 0.09 & 0.00 $\pm$ 0.10 & 0.18 $\pm$ 0.09 & 0.01 $\pm$ 0.10 & 0.24 $\pm$ 0.09 & 0.41 $\pm$ 0.08 & $[\mathrm{Fe/H}]$ (dex) & 0.18 $\pm$ 0.09 & 0.22 $\pm$ 0.08 & 0.12 $\pm$ 0.09 & 0.18 $\pm$ 0.09 & 0.30 $\pm$ 0.08 & 0.12 $\pm$ 0.09 & -0.09 $\pm$ 0.08 & 0.08 $\pm$ 0.09 & 0.14 $\pm$ 0.08\\
$v\sin{i}$ (km/s)\tablenotemark{a} & 4.6 $\pm$ 1.0 & 6.9 $\pm$ 1.0 & 7.5 $\pm$ 2.0 & 9.8 $\pm$ 1.0 & 10.2 $\pm$ 0.5 & 4.2 $\pm$ 1.0 & 5.2 $\pm$ 0.5 & 2.5 $\pm$ 1.0 & 4.6 $\pm$ 0.5 & $v\sin{i}$ (km/s) & 2.3 $\pm$ 1.0 & 4.0 $\pm$ 0.5 & 3.5 $\pm$ 1.0 & 3.9 $\pm$ 1.0 & 6.3 $\pm$ 0.5 & 4.4 $\pm$ 1.0 & 5.1 $\pm$ 0.5 & 8.8 $\pm$ 1.0 & 9.2 $\pm$ 0.5\\
$v_\mathrm{mac}$ (km/s)\tablenotemark{b} & 3.6 $\pm$ 0.2 & 3.8 $\pm$ 0.2 & -- & 3.2 $\pm$ 0.2 & -- & 4.1 $\pm$ 0.2 & -- & 4.1 $\pm$ 0.2 & -- & $v_\mathrm{mac}$ (km/s) & 4.2 $\pm$ 0.2 & -- & 4.1 $\pm$ 0.2 & 3.7 $\pm$ 0.2 & -- & 4.7 $\pm$ 0.2 & -- & 5.0 $\pm$ 0.2 & --\\
\\\hline
\enddata

\tablenotetext{a}{The \vsini in the \texttt{SpecMatch} column was computed from \texttt{SpecMath-Synth}.}
\tablenotetext{b}{$v_\mathrm{mac}$ is assumed based on effective temperature and the relation from \citet{Valenti2005}.}
\end{splitdeluxetable}
\addtolength{\tabcolsep}{-4pt}
% \pdfpageattr\expandafter{\the\pdfpageattr/Rotate 0}
\end{rotatepage}
\makeatletter\onecolumngrid@pop\makeatother

% \clearpage
% \cleardoublepage
% \pdfpageattr\expandafter{\the\pdfpageattr/Rotate 0}

\subsection{SED Fitting for Stars with Nearby Companions} \label{ssec:companions}

\begin{deluxetable}{lrr}
\tablecaption{Observed Properties of Stellar Companions \label{tab:stellar_companions}}
\tablehead{
	& \colhead{Primary} & \colhead{Secondary}
}
\startdata
\hline
\textbf{TOI-2193} & & \\
\hline
% row, primary, secondary 
Gaia EDR3 ID & 6373308503181838592 & 6373308503181838080 \\
TIC ID & 401604346 & 1988059412 \\
Ang. Sep. ($"$) & -- & 1.885 \\
PA ($^\circ$) & -- & 124 \\
$\Delta I $ (mag) & -- & 3.8 \\
Parallax (mas) & $2.938 \pm 0.021$ & $2.926 \pm 0.087$ \\
$\mu_\alpha \cos{\delta}$ (mas/yr) & $-2.376 \pm 0.029$ & $-2.378 \pm 0.120$ \\
$\mu_\delta$ (mas/yr) & $-0.809 \pm 0.037$ & $-1.050 \pm 0.165$ \\
RV (km/s) & $-17.8 \pm 1.9$ & -- \\
$G$ (mag) & $11.813 \pm 0.000$ & $16.047 \pm 0.004$ \\
$G_{\mathrm{BP}}$ (mag) & $12.127 \pm 0.001$ & -- \\
$G_{\mathrm{RP}}$ (mag) & $11.340 \pm 0.001$ & -- \\

\\\hline
\textbf{TOI-3331} & & \\
\hline
% row, primary, secondary 
Gaia EDR3 ID & 4042548116644168832 & 4042548120990244096 \\
TIC ID & 194795551 & 1565174683 \\
Ang. Sep. ($"$) & -- & 2.663 \\
PA ($^\circ$) & -- & 48 \\
$\Delta I $ (mag) & -- & 2.6 \\
Parallax (mas) & $4.577 \pm 0.057$ & $5.388 \pm 0.171$ \\
$\mu_\alpha \cos{\delta}$ (mas/yr) & $0.421 \pm 0.118$ & $10.766 \pm 0.604$ \\
$\mu_\delta$ (mas/yr) & $-16.943 \pm 0.090$ & $-22.713 \pm 0.475$ \\
RV (km/s) & $-46.5 \pm 0.6$ & -- \\
$G$ (mag) & $11.831 \pm 0.001$ & $14.264 \pm 0.002$ \\
$G_{\mathrm{BP}}$ (mag) & $12.249 \pm 0.003$ & $14.999 \pm 0.004$ \\
$G_{\mathrm{RP}}$ (mag) & $11.268 \pm 0.002$ & $13.301 \pm 0.014$ \\

\\\hline
\textbf{TOI-3540} & & \\
\hline
% row, primary, secondary 
Gaia EDR3 ID & 1896138833241139584 & -- \\
TIC ID & 17865622 & -- \\
Ang. Sep. ($"$) & -- & 0.917 \\
PA ($^\circ$) & -- & 200 \\
$\Delta I$ (mag) & -- & 1.8 \\
$\Delta \mathrm{Br}\gamma$ (mag) & -- & 1.022 \\
$\Delta H\mathrm{cont}$ (mag) & -- & 1.144 \\
Parallax (mas) & $3.633 \pm 0.259$ & -- \\
$\mu_\alpha \cos{\delta}$ (mas/yr) & $5.722 \pm 0.496$ & -- \\
$\mu_\delta$ (mas/yr) & $-4.567 \pm 0.592$ & -- \\
RV (km/s) & $-4.1 \pm 0.7$ & -- \\
$G$ (mag) & $11.404 \pm 0.003$ & -- \\
$G_{\mathrm{BP}}$ (mag) & $11.698 \pm 0.002$ & -- \\
$G_{\mathrm{RP}}$ (mag) & $10.839 \pm 0.003$ & -- \\

\enddata
\end{deluxetable}

\begin{figure}
\epsscale{1.08}
\plotone{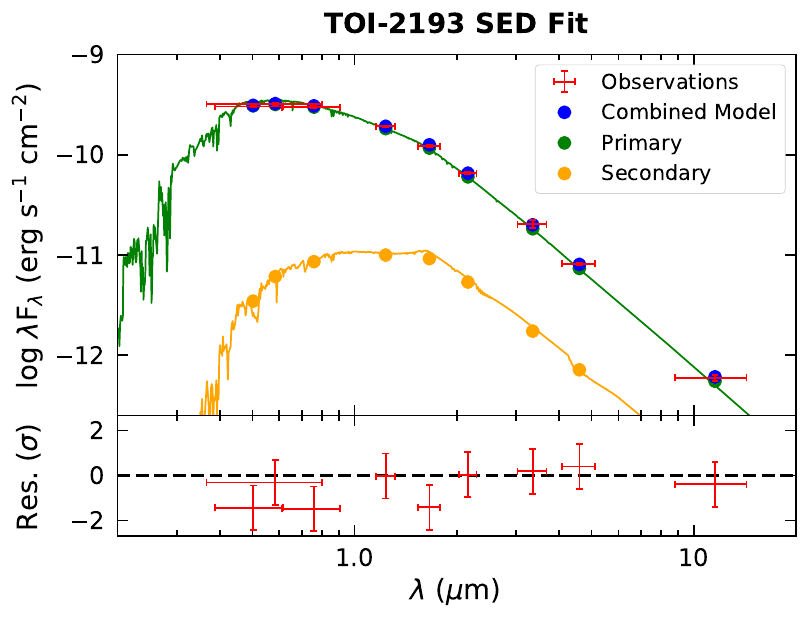}
\plotone{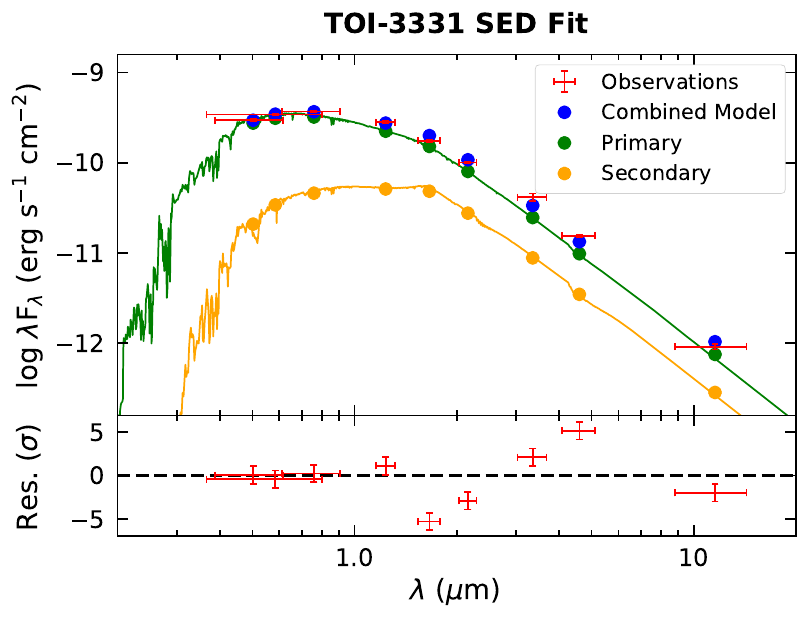}
\plotone{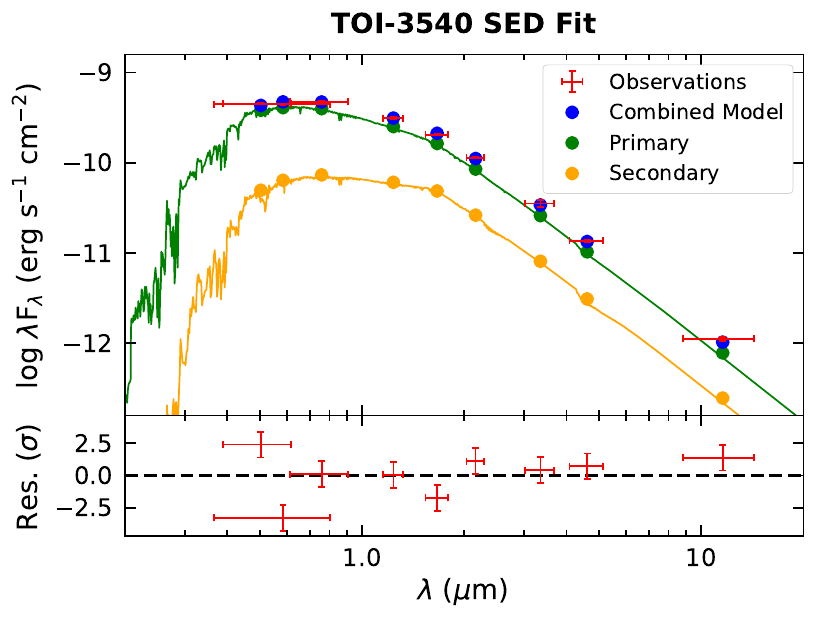}
\caption{Multi-component SED fits for the systems TOI-2193 (top), TOI-3331 (middle), and TOI-3540 (bottom), for which we detected nearby stellar companions.
The green points show the best-fit model for the primary, while the orange points show the model for the secondary, as derived from isochrone fitting.
The blue points show the combined model fluxes from both components, while the red points with error bars show the observed catalog fluxes.
Extinction-corrected atmospheric models from \cite{Kurucz1993} are plotted for the two stellar components for illustration, though these are not used directly in the fit, which is performed with the MIST bolometric correction tables.
The lower panel of each plot shows the residuals of the combined observed fluxes to the combined model fit, in units of the uncertainty on each measurement.
\label{fig:multi_sed_plots}}
\end{figure}

\begin{deluxetable}{lccc}
\tablecaption{Secondary Properties from SED Fit \label{tab:sec_properties}}
\tablehead{
	& \colhead{TOI-2193} & \colhead{TOI-3331} & \colhead{TOI-3540}
}
\startdata
\\[-\normalbaselineskip]\multicolumn{4}{l}{\textbf{Stellar Properties}}\\
\Teff (K) & $3913 \pm 19$ & $4172^{+260}_{-94}$ & $4819^{+67}_{-65}$\\
\feh (dex) & $-0.069^{+0.061}_{-0.064}$ & $-0.16^{+0.13}_{-0.21}$ & $0.151^{+0.073}_{-0.078}$\\
Age (Gyr) & $7.2 \pm 1.4$ & $7.3^{+4.2}_{-4.5}$ & $6.3^{+1.6}_{-1.3}$\\
\Mstar (\Msun) & $0.54 \pm 0.01$ & $0.599^{+0.025}_{-0.022}$ & $0.800^{+0.020}_{-0.021}$\\
\Rstar (\Rsun) & $0.5126^{+0.0080}_{-0.0079}$ & $0.580^{+0.022}_{-0.020}$ & $0.761 \pm 0.013$\\
$\log{{g}}$ (cgs) & $4.7480 \pm 0.0054$ & $4.689^{+0.017}_{-0.018}$ & $4.5772^{+0.0078}_{-0.0087}$\\
\multicolumn{4}{l}{\textbf{Synthetic Photometry}}\\
$G$ (mag) & $16.09 \pm 0.02$\tablenotemark{a} & $14.22 \pm 0.01$\tablenotemark{a} & $13.54^{+0.10}_{-0.09}$\\
$G_\mathrm{{BP}}$ (mag) & $16.96 \pm 0.03$\tablenotemark{a} & $15.01 \pm 0.02$\tablenotemark{a} & $14.1 \pm 0.1$\\
$G_\mathrm{{RP}}$ (mag) & $15.19 \pm 0.01$\tablenotemark{a} & $13.36 \pm 0.02$\tablenotemark{a} & $12.86^{+0.09}_{-0.08}$\\
$T$ (mag) & $15.15 \pm 0.01$\tablenotemark{a} & $13.33 \pm 0.02$\tablenotemark{a} & $12.84^{+0.09}_{-0.08}$\\
$B_T$ (mag) & $18.38 \pm 0.03$ & $16.40 \pm 0.04$ & $15.1 \pm 0.1$\\
$V_T$ (mag) & $16.87 \pm 0.03$ & $14.90 \pm 0.02$ & $13.9 \pm 0.1$\\
$J$ (mag) & $13.98 \pm 0.03$ & $12.20^{+0.05}_{-0.06}$ & $12.02^{+0.07}_{-0.06}$\\
$H$ (mag) & $13.27 \pm 0.03$ & $11.47^{+0.06}_{-0.07}$ & $11.46 \pm 0.05$\\
$K$ (mag) & $13.10 \pm 0.03$ & $11.32^{+0.06}_{-0.08}$ & $11.38 \pm 0.05$\\
$W1$ (mag) & $13.01 \pm 0.03$ & $11.25^{+0.07}_{-0.09}$ & $11.34 \pm 0.05$\\
$W2$ (mag) & $12.99 \pm 0.03$ & $11.28^{+0.06}_{-0.09}$ & $11.40 \pm 0.05$\\
$W3$ (mag) & $12.88 \pm 0.03$ & $11.16^{+0.06}_{-0.09}$ & $11.32 \pm 0.05$\\

\enddata
\tablenotetext{a}{We did not correct the catalog photometry for the secondary fluxes in these bands, as the primary and secondary were resolved in the \Gaia catalog.}
\end{deluxetable}

Three of the target systems (TOI-2193, TOI-3331, TOI-3540) had nearby companions detected in high-resolution imaging (\S\ref{ssec:imaging}, Figures \ref{fig:toi2193_speckle}--\ref{fig:toi3540_ao}).
The magnitude difference and angular separation of the primary and secondary derived from this imaging are presented in Table \ref{tab:stellar_companions}.
In the case of TOI-2193 and TOI-3331, these companions were also detected by \Gaia as they are relatively bright and at separations $\gtrsim1\farcs0$, which \Gaia can reliably resolve.

To correct the catalog photometry and photometric timeseries data for contamination by these companions, we used the \texttt{isochrones} package \citep{Isochrones_Morton2015} to perform a multi-component Spectral Energy Distribution (SED) fit.
For each system, we fitted the blended catalog photometry, together with the $\Delta$mag between the primary and secondary stars obtained from high-resolution imaging, to synthetic photometry derived from the MIST isochrones \citep{MIST0_Dotter2016,MISTI_Choi2016}.
We placed an error floor of 0.02~mag for the \Gaia and 2MASS photometry, and 0.03~mag for the WISE photometry, to account for possible systematic errors in the isochrones in reproducing the broad-band photometry measurements.
The fit was additionally constrained by the parallax measurements from \Gaia, the spectroscopic parameters derived for the primary in Section \ref{ssec:spec_char}, and an upper limit on the line-of-sight extinction from \citet{Schlegel1998} and \citet{Schlafly2011}.

We provide the best-fit stellar properties and MIST isochrone synthetic photometry for the secondary, along with corresponding uncertainties from a Markov Chain Monte Carlo analysis, in Table \ref{tab:sec_properties}.
We then subtracted the synthetic photometry from the blended catalog photometry (as listed in Tables \ref{tab:stellar_props} and \ref{tab:stellar_props_2}), and use these corrected fluxes for our global modelling (\S\ref{sec:planet_char}).
\replaced{We also computed dilution factors for the bands in which time-series photometry was obtained, and used those in the global \texttt{ExoFAST} fits.}
{We also computed flux dilution factors (defined in \S\ref{sec:planet_char}) for each band in which time-series photometry was obtained, to correct for the contribution from the nearby stars to the light-curve.
These dilution factors were used in the global \texttt{ExoFAST} fits.}

In addition, as a further check that the spectroscopic parameters derived in Section \ref{ssec:spec} were not biased by possible contamination of the observed spectrum by the companion, we performed a series of tests on the \texttt{SpecMatch-Emp} code.
We injected a diluted spectrum matching the companion's spectral type into the target spectrum and performed the stellar characterization procedure.
We found that even if the companion was responsible for contaminating the target spectrum by up to 10\%, the derived stellar properties did not vary by more than the uncertainties.
Furthermore, contamination down to $\sim$1\% would be detected in the residuals as a poor $\chi^2$ match, which was not the case for these three targets.

We discuss each of these three targets in more detail in the rest of this section.

\subsubsection{TOI-2193 Companion}
The TOI-2193 system contains two stars (TOI-2193A and TOI-2193B) separated by $1\farcs885$ and a magnitude difference of $\Delta I=3.8$~mag.
The \Gaia EDR3 catalog \citep{GaiaEDR3_Brown2021,GaiaEDR3_Riello2021} contains astrometry for both components, as well as 3-band photometry ($G, G_\mathrm{BP}, G_\mathrm{RP}$) for the primary and $G$-band photometry for the secondary.
The parallaxes and proper motions for the two components are identical within the uncertainties, suggesting that they are a bound system with a projected separation of $\approx 640$~AU (Table \ref{tab:stellar_companions}).
We therefore performed the \texttt{isochrone} fit assuming that the two stars have the same age and had the same initial metallicity.

The best-fit SED model found the secondary to have a mass of $\Mstar = 0.54\pm0.01\Msun$, and the estimated fluxes are shown in the top panel of Figure \ref{fig:multi_sed_plots}.
We used this best-fit model to correct the 2MASS and WISE catalog photometry, but not the \Gaia and TIC photometry, since the two components were resolved in \Gaia as well as the \TESS Input Catalog (TIC; \citealt{TIC_Stassun2018,TIC_Stassun2019}).
For the same reason, we do not impose an additional dilution factor for the \TESS light-curve, since it was already accounted for in the crowding corrections performed by the \TESS SPOC.

\subsubsection{TOI-3331 Companions}

In the case of the TOI-3331 system, SOAR imaging detected a nearby star with an angular separation of $2\farcs663$ and $\Delta I = 2.6$~mag.
This system was also resolved by \Gaia. The \Gaia EDR3 catalog gives a parallax of $\Pi_2 = 5.39 \pm 0.17$ for the secondary, compared with $\Pi_1 = 4.577 \pm 0.057$ for the primary.
The proper motions for the two stars differ significantly (Table \ref{tab:stellar_companions}).
Thus, the two stars are most likely a chance alignment along the line of sight.
\Gaia photometry also resolved a third star at an angular separation of $4\farcs89$, but we ignored this object in our fit due to its faintness ($G = 20.1$).
In our isochrone fit, we model the two stars with independent ages and metallicities, with individual parallaxes as constrained by \Gaia, and show the results in the middle panel of Figure \ref{fig:multi_sed_plots}.
As with the previous case, we did not correct the \Gaia and \TESS photometry, as the secondary was resolved in those catalogs.

\subsubsection{TOI-3540 Companion}

For TOI-3540, SOAR speckle imaging as well as PHARO AO imaging both detected a companion at $0\farcs917$, with $\Delta I = 1.8$~mag.
This companion was not resolved by \Gaia, so we do not have parallax or proper motion measurements for this object.
However, studies \citep[e.g.,][]{Horch2014,Matson2018} have shown that most nearby companions within $1"$ are likely to be bound, leading us to assume that this is the case for TOI-3540.
This would give the pair of stars a projected separation of $\approx 250$~AU.
We then performed the isochrone fit under this assumption, finding that the catalog photometry is well-described by a two-component system in which the secondary has a mass of $\Mstar = 0.79\pm0.02\Msun$.
We corrected the fluxes in all photometric bands, and computed the appropriate dilution factors for the \TESS and ground-based time-series photometry.

\section{Planetary System Characterization} \label{sec:planet_char}
\pdfpageattr\expandafter{\the\pdfpageattr/Rotate 0}

We characterized each planetary system with the exoplanet fitting code \Exofast \citep{ExoFAST_Eastman2013,ExoFASTv2_Eastman19}.
This software models the star and planet in a self-consistent manner, fitting transit and radial-velocity observations as well as the broad-band photometry, with constraints on the stellar properties from the MIST stellar evolutionary models \citep{MIST0_Dotter2016,MISTI_Choi2016}.
\Exofast uses a differential evolution Markov Chain Monte Carlo (DE-MCMC) algorithm to explore the posterior distribution and determine uncertainties of each fitted parameter.

In this section, we first describe our general fitting strategy, before describing
some deviations from the general strategy for specific targets.
We imposed Gaussian priors on the stellar spectroscopic properties $\Teff$, $\Rstar$, and $\feh$ based on the \texttt{SpecMatch-Emp} characterization described in \S\ref{ssec:spec_char}.
For the SED fit, we used broadband photometry from the \Gaia EDR3 \citep{GaiaEDR3_Brown2021,GaiaEDR3_Riello2021}, Tycho-2 \citep{Tycho2_Hog2000}, 2MASS \citep{TMASS_Cutri2003}, and WISE \citep{WISE_Cutri2012} catalogs.
We imposed a minimum uncertainty of 0.02~mag for the \Gaia and 2MASS photometry, and 0.03~mag for the WISE photometry.
We also imposed a Gaussian prior on the parallax from \Gaia EDR3, corrected for the parallax zero-point as described in \cite{GaiaEDR3_Lindegren2021}\footnote{\url{https://gitlab.com/icc-ub/public/gaiadr3_zeropoint}}, as well as an upper limit on line-of-sight extinction from \cite{Schlegel1998} and \cite{Schlafly2011}.

We fitted the available radial-velocities with an independent RV offset $\gamma$ and jitter $\jit$ terms for each instrument and target.
For those targets with only two observations from TRES (TOI-2567, -2570, -3540, -4137), we did not include the TRES RVs in the fit, as the introduction of two additional per-instrument free parameters did not justify their inclusion.
In all cases however, we find that the TRES measurements are consistent with the modelled RV semiamplitude from the global fit (Figures \ref{fig:toi2567_multiplot}, \ref{fig:toi2570_multiplot}, \ref{fig:toi3540_multiplot} \& \ref{fig:toi4137_multiplot}), giving us additional confidence in our results.
% Within the \Exofast code, the jitter term is fitted as $\jit^2$, which allows for negative jitter variances in case the radial-velocity uncertainties were overestimated.
We did not allow for any long-term radial-velocity trends, because our initial testing showed that trends were not required to achieve a good fit to the data for any of our systems, especially given the relatively short baseline of our RV observations.

We used both the \TESS and ground-based time-series photometry to constrain the transit model.
\Exofast fits an analytic transit model from \citet{Mandel02} and \citet{Agol2020} to the transit light-curve, with quadratic limb-darkening coefficients in each band constrained by the stellar properties and the tables from \citet{Claret2011} and \citet{Claret2017}.
\Exofast ensures that the constraint on the stellar mean density implicit in the transit model is consistent with the mean density implied by the MIST stellar evolution model.
We also fit for a separate flux baseline $F_0$ and added variance $\sigma^2$ for each transit light-curve.

While the fluxes from \TESS are already corrected for dilution from neighboring stars in the \TESS Input Catalog (TIC; \citealt{TIC_Stassun2018,TIC_Stassun2019}), we still fitted for a dilution factor $A_D$, to account for any inaccuracies and to propagate uncertainties, as recommended by \cite{ExoFASTv2_Eastman19}.
Here, $A_D = F_2/(F_1 + F_2)$ is the fractional contribution to the total flux ($F_1 + F_2$) from all neighboring stars ($F_2$).
We imposed a Gaussian prior on this dilution factor centered at zero and with a width equal to 10\% the contamination ratio found in the TIC.
For the ground-based photometry, we simultaneously detrended against the detrending vectors listed in Table \ref{tab:sg1_lcs}.
We normalized the detrend parameters to be between $[-1, +1]$, and used an additive detrending model for the light-curve.
The final light-curve model at time $i$ is then
\begin{equation}
    \mathrm{Model}_i = F_0 \left(T_i (1 - A_D) + A_D \right) + \sum C_j d_{i,j},
\end{equation}
where $T_i$ is the transit model with an out-of-transit baseline of 1, $d_{i,j}$ is the $j$-th detrending parameter at time $i$, and $C_j$ the additive coefficient for the $j$-th parameter.
% We used an additive detrending model, and all detrend parameters were first normalized to be between $[-1, +1]$.
For the \TESS long-cadence data, this model is integrated over the 30-minute or 10-minute exposure time to account for smearing of the light-curve over each exposure.

We performed our initial fits requiring circular orbits for each planet.
We also performed a second fit in which the eccentricity was allowed to be a free parameter.
In \Exofast, the eccentricity is parameterized in terms of $\sqrt{e}\cos{\omega}$ and $\sqrt{e}\sin{\omega}$.
In all but one case (TOI-2207\,b), the data are consistent with a circular orbit.
As such, we adopt the results from the circular fit for these objects, but also report 1-$\sigma$ upper limits on the eccentricity from the fit where we allowed eccentricity to float.
For TOI-2207\,b, our eccentric fit found that the median of the posterior distribution was more than $3$-$\sigma$ above zero, suggesting that this planet, with an orbital period of $P \approx 8.00$~days, has a detectably eccentric orbit.
We adopt the results from the eccentric fit for this target.

We ran the \Exofast DE-MCMC algorithm using the convergence criteria suggested by \cite{ExoFASTv2_Eastman19}, requiring the Gelman-Rubin statistic \citep{GelmanRubin} to be $< 1.01$, as well as $>1{,}000$ independent draws in each parameter. 
Tables \ref{tab:fitted_props} and \ref{tab:fitted_props_2} contain the median and 68\% confidence intervals of the marginalized one-dimensional posterior probability distributions for the fitted \added{stellar and planetary} parameters, from the adopted fit.
\added{Additional fitted parameters specific to the observations (e.g. RV offset, jitter, and flux dilution factors) are provided in Table \ref{tab:additional_fit_params} at the end of the paper.}
We also provide the full results from both circular and eccentric fits for all targets as a machine-readable companion to those tables.
The best-fit model for the transit, radial-velocities, and SED for each system are shown in Figures \ref{fig:toi2193_multiplot} and \ref{fig:toi2207_multiplot} through \ref{fig:toi4137_multiplot}.

\subsection{Target-Specific Notes}
In the case of TOI-2207 b, our eccentric fit found that the median of the posterior distribution was more than $3$-$\sigma$ above zero, suggesting that this planet, with an orbital period of $P \approx 8.00$~days, has a detectably eccentric orbit.
As such, in Table \ref{tab:fitted_props}, we report the median and 68\% confidence intervals for the eccentric fit for all parameters listed.

The objects TOI-2193A\,b and TOI-3540A\,b appear to be on orbits wherein the planetary transits graze the stellar limb.
In this regime, there is a strong degeneracy between the planet-to-star radius ratio $\Rp/\Rstar$ and the transit impact parameter $b$.
The resulting posterior distributions for these parameters have long tails allowing for extremely large and unphysical planet radii.
In these cases, we placed an upper limit of $\Rp/\Rstar < 0.5$ during the fit to ensure convergence of the MCMC fits within these limits.
As the medians of the posterior distributions would be heavily skewed by the long tails, we report the mode of the posterior distributions and the 95\% lower limit on the planet radius and 95\% upper limit on the orbital inclination (note that these limits also depend on the cutoff chosen during the fit).
% In these cases, instead of reporting the median of the distribution, which would be heavily skewed by the long tail, we report the mode of the posterior distribution and the 95\% lower limit on the planet radius, and upper limit on the orbital inclination.

In general, we did not fit for any dilution factors in the transit light-curves, except for the \TESS data, where we allowed $A_D$ to vary within a small range around zero.
However, for the targets described in Section \ref{ssec:companions} that have stellar companions, we also allowed for dilution of the ground-based light-curves, since the apertures used also contained the stellar companions.
In these cases, we used the best-fit multi-component SED model to derive dilution factors for each photometric filter to correct the light-curves, imposing a Gaussian prior with width equal to 10\% of the dilution factor around the mean value.

For targets observed in multiple \TESS sectors, we generally fit each sector of data separately, with a separate baseline flux $F_0$ and variance $\sigma^2$ per sector.
TOI-2567 was observed by \TESS in a total of 15 sectors (14--26, 40, 41), which would greatly increase the dimensionality of the fit were we to include these two additional free parameters for each.
In this case, we fit the Sectors 14--26 data, which were all at 30-min cadence, as a single light-curve.
The sectors 40 and 41 data were taken at 2-min cadence, and we fit these as a single light-curve too.
In general, our fitted values for $F_0$ and $\sigma^2$ for the \TESS data are close to 1.0 and 0.0 respectively, indicating minimal baseline offsets between sectors, so combining these consecutive sectors of data should not have had a significant impact on the fit.

\afterpage{
\begin{figure*}
\epsscale{0.9}
\plotone{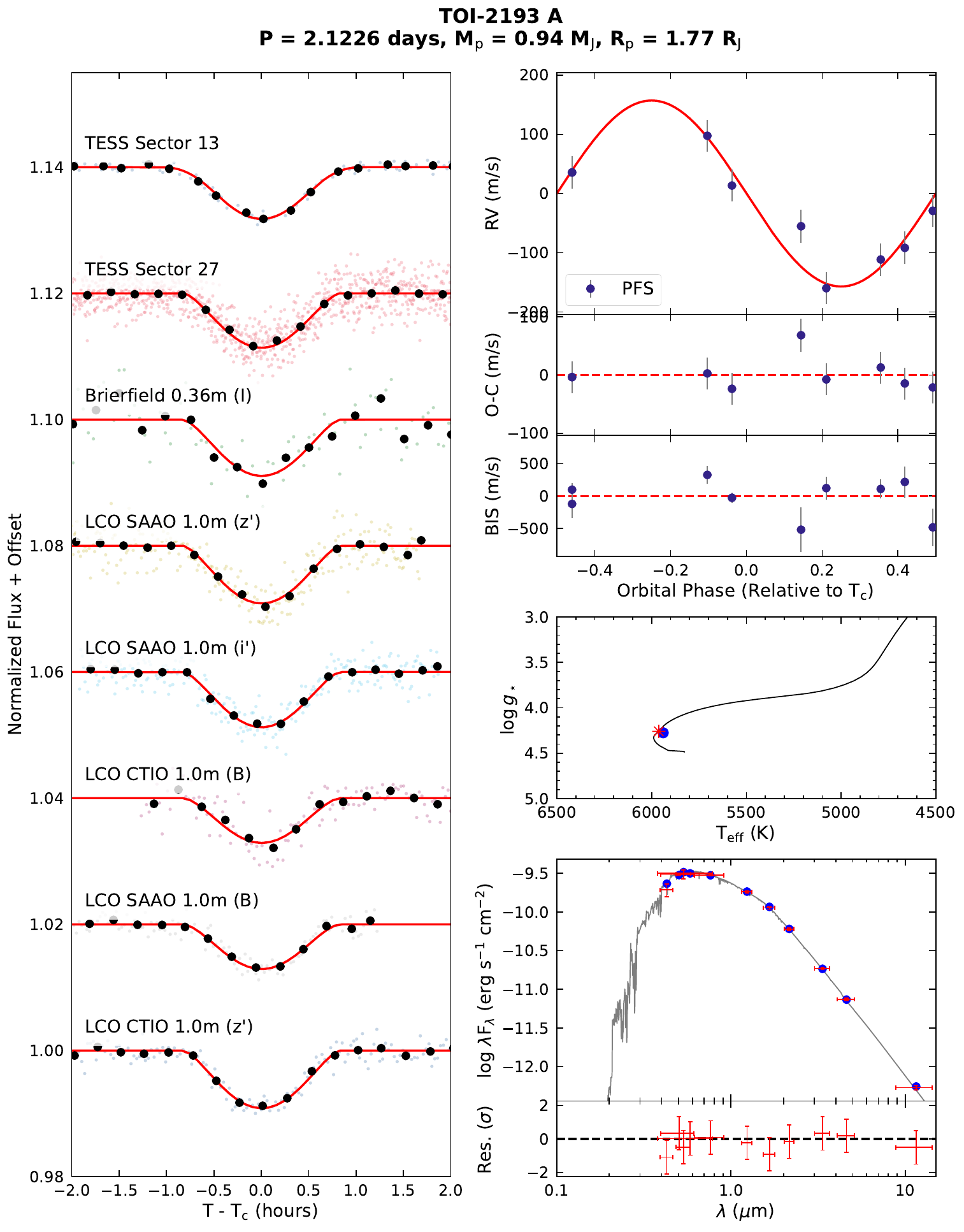}
\caption{Results of \Exofast fits for TOI-2193.
\textbf{Left:} \TESS and ground-based photometric observations, phased to the best-fit orbital period and time of conjunction. The black points are the photometric time-series data binned to 15-min cadence, while the faint colored points are the unbinned data.
The red line shows the best-fit transit model, corrected for flux dilution and additive detrending.
\textbf{Top right:} Radial-velocity observations of TOI-2193, phased to the best-fit orbital period.
The error bars reflect the internal measurement error added in quadrature to the fitted jitter \jit\,parameter for each instrument.
The two lower subpanels show the phased radial-velocity residuals and bisector span measurements.
\textbf{Middle right:} The best-fit MIST stellar evolution track (black line), which is fit simultaneously and self-consistently with the transit and RV model.
The blue point shows the best-fit stellar \Teff and \logg, while the red asterisk corresponds to the star's position along the track given its best-fit age.
The small discrepancies between the two are well within the fitted uncertainties in each parameter.
\textbf{Lower right:} Result of the SED fit, performed using the MIST bolometric correction grid.
The red points are the catalog broadband photometric measurements, corrected for the presence of any stellar companions (i.e. subtracting the yellow points in Figure \ref{fig:multi_sed_plots}), with vertical error bars showing the catalog uncertainty and horizontal error bars showing the bandpass width.
The model fluxes in each bandpass, derived from the MIST grid, are the blue points.
An atmospheric model from \cite{Kurucz1993} corresponding to the best-fit stellar parameters is plotted in gray for illustrative purposes only, and is not used directly in the fit.
The lower subpanel shows the residuals to the model fit, in units of the uncertainty on each measurement.
\label{fig:toi2193_multiplot}
}
\end{figure*}
\clearpage
}

\subsection{Potential False-Positive Scenarios} \label{ssec:false_pos}
Various astrophysical phenomena can lead to light-curves that appear similar to planetary transits, leading to false-positive planet detections (e.g. \citealt{Collins2018}).
The goal of our ground-based follow-up observations was to help rule out or reduce the likelihood of such scenarios.
For each of our ten hot Jupiter systems, the seeing-limited ground-based photometry confirm that the transits occur on the target stars, as opposed to being nearby eclipsing binaries that contaminate the \TESS photometric aperture.
The measured RV semi-amplitudes for all companions are also consistent with planetary mass objects, rather than brown dwarfs, which can have similar radii despite their significantly larger masses.
Furthermore, the spectra of each target showed no indications of secondary spectral lines.
To check for the possibility of unresolved blended eclipsing binaries causing line-profile variations that may appear as RV variations, we computed the Pearson-$r$ coefficient to check for correlations between the measured RV and the spectral line BIS.
For each of the ten target systems, we found no statistically significant correlations ($p>0.1$).

Given the extensive ground-based follow-up observations and lack of BIS variations, we are confident that all of our targets are confirmed as true planets.
However, we chose to pay closer attention to the two systems TOI-2193 and TOI-3540, which exhibit grazing transits and have close resolved stellar companions.
Although the measured RV reflex motions measured on the primary stars are indicative of planetary mass objects orbiting them, there may be concern that these may actually be blended eclipsing binary (EB) false positives, where the diluted light of the companion is the source of both the transit and apparent radial-velocity variations.
In the case of TOI-2193, the nearby companion TOI-2193B is $1\farcs885$ away.
A small aperture measurement ($1\farcs2$) of the LCO-CTIO light-curve taken on UT 2021-07-26 confirmed that the source of the transit signal is indeed TOI-2193A.
For TOI-3540, the companion star is just $0\farcs917$ away and could not be resolved by seeing-limited photometry.

In order to definitively rule out such scenarios for these two objects, we carried out a detailed blend modeling of each system following the procedures described in \citet{Hartman2019}, which is based on the work of \citet{Torres2004}.
In each case we jointly model the available \TESS and ground-based light curves, catalog broadband photometry, \Gaia parallaxes, and spectroscopically determined atmospheric parameters.
We include the resolved stellar companions in the modelling, jointly fitting for the masses, metallicities, distances, and ages of all stars that we assume contribute to the blended (or resolved) measurements, and we include the measured magnitude differences from high-resolution-imaging as observations to be fit in the modelling.
We use the MIST stellar models to constrain the properties of the stars.
For TOI-2193 we force the resolved stellar companion to have the same age, distance and metallicity as the primary star, while for TOI-3540 we allow the components to have independent values.

For each system we consider four scenarios:
(1) the primary object in the resolved pair is a single star with a transiting planet, while the secondary object is also a single star;
(2) the primary object in the resolved pair is a single star, while the secondary object is a two-component stellar eclipsing binary system;
(3) the primary object in the resolved pair is itself an unresolved hierarchical triple consisting of a bright non-eclipsing star and a fainter stellar eclipsing binary, and the secondary object in the resolved pair is a single star;
and (4) the primary object in the resolved pair is an unresolved blend between a bright non-eclipsing star and a line-of-sight background eclipsing binary, while the secondary object in the resolved pair is a single star.
For both TOI-2193 and TOI-3540 we find that scenario (1) provides the best (lowest $\chi^{2}$) fit to the observations, despite using the fewest model parameters.
In both cases the combination of the photometry and \Gaia parallax measurements favors scenarios where each of the resolved point sources is itself a single star, while the transit duration and depth, and the lack of secondary eclipses or ellipsoidal variations in the light curves favors a transiting planet around the primary star over scenarios involving eclipsing stellar binaries.
For both TOI-2193 and TOI-3540 we find that scenario (4) is the next-best-fitting scenario, and that this scenario has $\Delta \chi^{2} \approx 70$, and $\Delta \chi^{2} \approx 9$ compared to scenario (1) for TOI-2193 and TOI-3540, respectively.
The relative inability of the blend-models to fit the photometric data, together with the significant RV variations consistent with transiting giant planet companions, and the lack of any significant BIS variation in phase with the transit ephemerides, leads us to conclude that both TOI-2193A\,b and TOI-3540A\,b have been confirmed as transiting planets.

% Two of the systems merit further discussion -- TOI-2193 and TOI-3540, which show V-shaped light-curves.
% In the planetary interpretation, the V shapes are due to grazing transits; however, such a light-curve morphology is also characteristic of eclipsing binaries.
% Furthermore, the high resolution imaging of both systems revealed nearby stars, which may give rise to the concern that the observed radial-velocity variations is due to stray light from those objects.
% However, in both cases, the slit spectrographs used to obtain spectroscopic observations (PFS and HIRES for TOI-2193 and TOI-3540) respectively should have excluded the light of the nearby stars.

% In addition, in the case of TOI-2193, 
% We also measured the transit depth for each light-curve individually, and found that the depths in $B, i^\prime$, and $z^\prime$ bands did not differ at a level greater than the uncertainties.
% Significant differences in the transit depth would have indicated a blended EB scenario, where the different colors of the target star and EB lead to different levels of contamination and hence different eclipse depths.

% All things considered, we believe that the ten systems presented here are likely to be true planets. 

% \clearpage
% \global\pdfpageattr\expandafter{\the\pdfpageattr/Rotate 90}
\makeatletter\onecolumngrid@push\makeatother
\begin{rotatepage}
\movetableright=-1in
\movetabledown=2.3in
\begin{rotatetable}
\begin{deluxetable*}{l>{\centering}cccccc}
\tablecaption{Median Values and 68\% Confidence Intervals for Fitted Stellar and Planetary Parameters \label{tab:fitted_props}}
\tabletypesize{\footnotesize}
\tablecolumns{7}
\tablehead{
 & & \colhead{TOI-2193 b} & \colhead{TOI-2207 b} & \colhead{TOI-2236 b} & \colhead{TOI-2421 b} & \colhead{TOI-2567 b} 
}
\startdata
\multicolumn{7}{l}{\textbf{Planet Parameters}}\\
$P$ (days) & Period & $2.1225735 \pm 0.0000016$ & $8.001968^{+0.000024}_{-0.000025}$ & $3.5315902 \pm 0.0000026$ & $4.3474032^{+0.0000079}_{-0.0000078}$ & $5.983944 \pm 0.000013$\\
$T_c$ (BJD$_\mathrm{TDB}$) & Time of conjunction & $2459052.42122 \pm 0.00021$ & $2459283.82747 \pm 0.00046$ & $2459011.45704 \pm 0.00023$ & $2458957.02765^{+0.00064}_{-0.00063}$ & $2459007.78085 \pm 0.00052$\\
$T_{14}$ (days) & Transit duration & $0.06862^{+0.00087}_{-0.00084}$ & $0.1930^{+0.0015}_{-0.0013}$ & $0.127 \pm 0.001$ & $0.2026 \pm 0.0017$ & $0.2274 \pm 0.0016$\\
$\tau$ (days) & Ingress/egress duration & $0.03431^{+0.00043}_{-0.00042}$ & $0.01230^{+0.00140}_{-0.00047}$ & $0.0219 \pm 0.0012$ & $0.0125^{+0.0011}_{-0.0014}$ & $0.01365^{+0.00120}_{-0.00083}$\\
$a/R_\star$ & Planet-star separation & $5.71^{+0.17}_{-0.13}$ & $11.7 \pm 0.4$ & $6.84^{+0.16}_{-0.15}$ & $6.67^{+0.34}_{-0.24}$ & $8.64^{+0.24}_{-0.31}$\\
$\left(R_P / R_\star\right)^2$ & Transit depth & $0.0191\,(> 0.0169)$ & $0.004273^{+0.000098}_{-0.000090}$ & $0.007011^{+0.000092}_{-0.000090}$ & $0.002949^{+0.000071}_{-0.000075}$ & $0.003537^{+0.000073}_{-0.000068}$\\
$i$ (deg) & Inclination & $79.96\,(< 80.57)$ & $88.84^{+0.79}_{-0.85}$ & $83.58^{+0.26}_{-0.25}$ & $86.49^{+1.30}_{-0.76}$ & $88.30^{+1.10}_{-0.82}$\\
$K$ (m/s) & RV semi-amplitude & $136^{+25}_{-26}$ & $55.8^{+8.3}_{-10.0}$ & $172 \pm 43$ & $37.8^{+8.6}_{-8.9}$ & $20.6^{+3.2}_{-3.1}$\\
$a$ (AU) & Semimajor axis & $0.03319^{+0.00052}_{-0.00051}$ & $0.0854^{+0.0015}_{-0.0016}$ & $0.05009^{+0.00080}_{-0.00082}$ & $0.0543^{+0.0021}_{-0.0010}$ & $0.0672^{+0.0024}_{-0.0012}$\\
$R_P$ ($R_\mathrm{J}$) & Planet radius & $1.77\,(> 1.55)$ & $0.995^{+0.028}_{-0.027}$ & $1.282^{+0.032}_{-0.031}$ & $0.925^{+0.035}_{-0.034}$ & $0.975^{+0.031}_{-0.029}$\\
$M_P$ ($M_\mathrm{J}$) & Planet mass & $0.94 \pm 0.18$ & $0.64^{+0.10}_{-0.12}$ & $1.58^{+0.40}_{-0.39}$ & $0.333 \pm 0.079$ & $0.201^{+0.034}_{-0.031}$\\
$\rho_P$ (g cm$^{-3}$) & Planet density & $0.060\,(< 0.321)$ & $0.80 \pm 0.16$ & $0.93^{+0.25}_{-0.24}$ & $0.52^{+0.15}_{-0.13}$ & $0.269^{+0.052}_{-0.051}$\\
$\log{g_P}$ (cgs) & Planet surface gravity & $2.81\,(< 3.00)$ & $3.203^{+0.071}_{-0.092}$ & $3.38^{+0.10}_{-0.12}$ & $2.98^{+0.10}_{-0.12}$ & $2.720^{+0.071}_{-0.080}$\\
$b \equiv a\cos{i}/R_\star$ & Transit impact parameter & $0.990\,(> 0.967)$ & $0.20^{+0.16}_{-0.14}$ & $0.766^{+0.013}_{-0.014}$ & $0.409^{+0.071}_{-0.130}$ & $0.26^{+0.11}_{-0.16}$\\
$e$ & Eccentricity & 0.0 (fixed) & $0.174^{+0.048}_{-0.052}$ & 0.0 (fixed) & 0.0 (fixed) & 0.0 (fixed)\\
$e_\mathrm{lim}$\tablenotemark{a} & 1-$\sigma$ upper limit on eccentricity & $< 0.063$ & -- & $< 0.135$ & $< 0.136$ & $< 0.080$\\
$\tau_\mathrm{circ}$ (Gyr)\tablenotemark{b} & Tidal circularization timescale & $0.00036\,(< 0.00908)$ & $13.2^{+5.6}_{-4.6}$ & $0.40^{+0.12}_{-0.11}$ & $0.94^{+0.37}_{-0.28}$ & $1.78^{+0.44}_{-0.43}$\\
$\langle F \rangle$ (Gerg~s$^{-1}$~cm$^{-2}$) & Incident flux & $2.20 \pm 0.11$ & $0.553^{+0.026}_{-0.024}$ & $1.846^{+0.085}_{-0.088}$ & $1.257^{+0.065}_{-0.088}$ & $0.759^{+0.042}_{-0.041}$\\
$T_\mathrm{eq}$ (K) & Planet equilibirum temperature & $1763^{+21}_{-22}$ & $1259^{+16}_{-15}$ & $1688^{+19}_{-21}$ & $1534^{+20}_{-28}$ & $1352^{+18}_{-19}$\\
\multicolumn{7}{l}{\textbf{Stellar Parameters}}\\
$M_\star$ ($M_\odot$) & Stellar mass & $1.082^{+0.052}_{-0.049}$ & $1.296^{+0.069}_{-0.072}$ & $1.343^{+0.066}_{-0.065}$ & $1.131^{+0.130}_{-0.064}$ & $1.130^{+0.130}_{-0.059}$\\
$R_\star$ ($R_\odot$) & Stellar radius & $1.248^{+0.026}_{-0.028}$ & $1.564^{+0.041}_{-0.040}$ & $1.573 \pm 0.033$ & $1.750^{+0.051}_{-0.049}$ & $1.686^{+0.043}_{-0.044}$\\
$\log{g_\star}$ (cgs) & Stellar surface gravity & $4.278^{+0.030}_{-0.024}$ & $4.161^{+0.034}_{-0.036}$ & $4.172 \pm 0.024$ & $4.005^{+0.059}_{-0.037}$ & $4.044^{+0.036}_{-0.037}$\\
$\rho_\star$ (g cm$^{-3}$) & Stellar density & $0.781^{+0.072}_{-0.052}$ & $0.476^{+0.050}_{-0.047}$ & $0.486^{+0.034}_{-0.031}$ & $0.297^{+0.048}_{-0.031}$ & $0.340^{+0.030}_{-0.036}$\\
$L_\star$ ($L_\odot$) & Stellar luminosity & $1.779^{+0.075}_{-0.084}$ & $3.05 \pm 0.11$ & $3.41^{+0.14}_{-0.18}$ & $2.729^{+0.095}_{-0.091}$ & $2.54 \pm 0.11$\\
$T_\mathrm{eff}$ (K) & Stellar effective temperature & $5966^{+73}_{-72}$ & $6101^{+75}_{-73}$ & $6248^{+72}_{-77}$ & $5607 \pm 68$ & $5611^{+62}_{-65}$\\
$[\mathrm{Fe/H}]$ (dex) & Metallicity & $0.027^{+0.066}_{-0.052}$ & $0.181 \pm 0.086$ & $0.119^{+0.081}_{-0.078}$ & $0.190^{+0.083}_{-0.085}$ & $0.241^{+0.078}_{-0.079}$\\
$[\mathrm{Fe/H}]_0$ (dex)\tablenotemark{c} & Initial metallicity & $0.082^{+0.060}_{-0.052}$ & $0.245 \pm 0.081$ & $0.222 \pm 0.072$ & $0.210^{+0.076}_{-0.083}$ & $0.258^{+0.071}_{-0.073}$\\
Age (Gyr) & Stellar age & $5.5^{+1.9}_{-1.7}$ & $3.3^{+1.3}_{-1.1}$ & $2.52^{+0.89}_{-0.82}$ & $7.5^{+1.7}_{-2.8}$ & $7.6^{+1.8}_{-2.7}$\\
EEP\tablenotemark{d} & Equal evolutionary phase & $404^{+16}_{-28}$ & $383^{+30}_{-33}$ & $360^{+30}_{-15}$ & $455.6^{+4.5}_{-37.0}$ & $452.6^{+5.2}_{-33.0}$\\
$A_V$ (mag) & Visual extinction & $0.113^{+0.045}_{-0.060}$ & $0.049^{+0.031}_{-0.032}$ & $0.305^{+0.040}_{-0.064}$ & $0.024^{+0.013}_{-0.015}$ & $0.074^{+0.030}_{-0.042}$\\
d (pc) & Distance & $345.3 \pm 1.1$ & $380.6 \pm 2.9$ & $352.4 \pm 1.5$ & $328.1^{+4.2}_{-4.0}$ & $505.2^{+7.7}_{-7.4}$\\
\enddata
\tablecomments{
This table contains the fit results from the preferred fit for each target: circular fits ($e$ fixed at 0.0) for all targets apart from TOI-2207\,b, and an eccentric fit for TOI-2207\,b.
The results from both fits for each target are available as a machine-readable table.\\
For TOI-2193A\,b and TOI-3540A\,b, we provide the posterior mode and 95\% lower limits for the $(\Rp/\Rstar)^2$, \Rp, and $b$, and the posterior mode and 95\% upper  limits for $i$, $\rho_P$,$\log{g_P}$, and $\tau_\mathrm{circ}$.\\
$^{a-d}$ Additional notes are provided following Table 14.\\}
\end{deluxetable*}
\end{rotatetable}
\end{rotatepage}
\makeatletter\onecolumngrid@pop\makeatother

\global\pdfpageattr\expandafter{\the\pdfpageattr/Rotate 90}
\clearpage
\makeatletter\onecolumngrid@push\makeatother
\begin{rotatepage}
\movetabledown=2.3in
\begin{rotatetable}
\begin{deluxetable}{l>{\centering}cccccc}
\tablecaption{Median Values and 68\% Confidence Intervals for Fitted Stellar and Planetary Parameters (continued) \label{tab:fitted_props_2}}
\tabletypesize{\footnotesize}
\tablecolumns{7}
\tablehead{
 & & \colhead{TOI-2570 b} & \colhead{TOI-3331 b} & \colhead{TOI-3540 b} & \colhead{TOI-3693 b} & \colhead{TOI-4137 b} 
}
\startdata
\multicolumn{7}{l}{\textbf{Planet Parameters}}\\
$P$ (days) & Period & $2.9887615 \pm 0.0000022$ & $2.0180231^{+0.0000043}_{-0.0000044}$ & $3.1199990 \pm 0.0000079$ & $9.088516^{+0.000026}_{-0.000027}$ & $3.8016122 \pm 0.0000065$\\
$T_c$ (BJD$_\mathrm{TDB}$) & Time of conjunction & $2459393.14532 \pm 0.00021$ & $2459371.6530 \pm 0.0002$ & $2459109.14114 \pm 0.00089$ & $2458806.68164^{+0.00032}_{-0.00031}$ & $2458990.46651 \pm 0.00033$\\
$T_{14}$ (days) & Transit duration & $0.12527^{+0.00097}_{-0.00091}$ & $0.08846^{+0.00085}_{-0.00083}$ & $0.0792^{+0.0026}_{-0.0025}$ & $0.1482^{+0.0012}_{-0.0011}$ & $0.1429 \pm 0.0013$\\
$\tau$ (days) & Ingress/egress duration & $0.01423^{+0.00093}_{-0.00089}$ & $0.01397^{+0.00087}_{-0.00084}$ & $0.0396 \pm 0.0013$ & $0.01937^{+0.00100}_{-0.00052}$ & $0.0167^{+0.0012}_{-0.0011}$\\
$a/R_\star$ & Planet-star separation & $8.13^{+0.23}_{-0.22}$ & $7.09^{+0.18}_{-0.17}$ & $7.51^{+0.26}_{-0.25}$ & $22.13^{+0.30}_{-0.47}$ & $7.80 \pm 0.24$\\
$\left(R_P / R_\star\right)^2$ & Transit depth & $0.01304 \pm 0.00026$ & $0.01559 \pm 0.00033$ & $0.023\,(> 0.015)$ & $0.02134^{+0.00041}_{-0.00039}$ & $0.00749 \pm 0.00014$\\
$i$ (deg) & Inclination & $87.73^{+0.75}_{-0.57}$ & $85.40^{+0.38}_{-0.36}$ & $81.93\,(< 83.11)$ & $89.57^{+0.29}_{-0.28}$ & $85.7 \pm 0.4$\\
$K$ (m/s) & RV semi-amplitude & $111.2^{+6.9}_{-7.5}$ & $360 \pm 21$ & $155^{+17}_{-16}$ & $109^{+26}_{-24}$ & $156^{+17}_{-15}$\\
$a$ (AU) & Semimajor axis & $0.04145^{+0.00081}_{-0.00086}$ & $0.03144^{+0.00048}_{-0.00055}$ & $0.04289^{+0.00092}_{-0.00093}$ & $0.0813^{+0.0011}_{-0.0012}$ & $0.05222^{+0.00089}_{-0.00096}$\\
$R_P$ ($R_\mathrm{J}$) & Planet radius & $1.217^{+0.035}_{-0.034}$ & $1.158 \pm 0.043$ & $2.10\,(> 1.44)$ & $1.124^{+0.029}_{-0.023}$ & $1.211^{+0.040}_{-0.039}$\\
$M_P$ ($M_\mathrm{J}$) & Planet mass & $0.820^{+0.063}_{-0.065}$ & $2.27 \pm 0.16$ & $1.18 \pm 0.14$ & $1.02^{+0.24}_{-0.22}$ & $1.44^{+0.17}_{-0.15}$\\
$\rho_P$ (g cm$^{-3}$) & Planet density & $0.563^{+0.069}_{-0.063}$ & $1.82^{+0.24}_{-0.21}$ & $0.065\,(< 0.512)$ & $0.89^{+0.22}_{-0.20}$ & $1.01^{+0.17}_{-0.15}$\\
$\log{g_P}$ (cgs) & Planet surface gravity & $3.137^{+0.041}_{-0.043}$ & $3.624 \pm 0.041$ & $2.64\,(< 3.17)$ & $3.300^{+0.095}_{-0.110}$ & $3.387^{+0.057}_{-0.058}$\\
$b \equiv a\cos{i}/R_\star$ & Transit impact parameter & $0.32^{+0.07}_{-0.10}$ & $0.569^{+0.030}_{-0.034}$ & $1.073\,(> 0.936)$ & $0.17^{+0.10}_{-0.11}$ & $0.584^{+0.035}_{-0.038}$\\
$e$ & Eccentricity & 0.0 (fixed) & 0.0 (fixed) & 0.0 (fixed) & 0.0 (fixed) & 0.0 (fixed)\\
$e_\mathrm{lim}$\tablenotemark{a} & 1-$\sigma$ upper limit on eccentricity & $< 0.039$ & $< 0.188$ & $< 0.164$ & $< 0.054$ & $< 0.246$\\
$\tau_\mathrm{circ}$ (Gyr)\tablenotemark{b} & Tidal circularization timescale & $0.111^{+0.022}_{-0.018}$ & $0.071^{+0.015}_{-0.012}$ & $0.0027\,(< 0.0897)$ & $22.3^{+6.0}_{-5.4}$ & $0.66^{+0.16}_{-0.13}$\\
$\langle F \rangle$ (Gerg~s$^{-1}$~cm$^{-2}$) & Incident flux & $0.951^{+0.066}_{-0.060}$ & $1.114^{+0.096}_{-0.083}$ & $1.145^{+0.069}_{-0.066}$ & $0.0934^{+0.0064}_{-0.0056}$ & $1.379^{+0.092}_{-0.082}$\\
$T_\mathrm{eq}$ (K) & Planet equilibirum temperature & $1431^{+24}_{-23}$ & $1488^{+31}_{-29}$ & $1498 \pm 22$ & $801^{+13}_{-12}$ & $1570^{+25}_{-24}$\\
\multicolumn{7}{l}{\textbf{Stellar Parameters}}\\
$M_\star$ ($M_\odot$) & Stellar mass & $1.063^{+0.063}_{-0.065}$ & $1.016^{+0.047}_{-0.052}$ & $1.081^{+0.071}_{-0.069}$ & $0.867^{+0.036}_{-0.037}$ & $1.313^{+0.068}_{-0.071}$\\
$R_\star$ ($R_\odot$) & Stellar radius & $1.095^{+0.026}_{-0.025}$ & $0.952 \pm 0.029$ & $1.228 \pm 0.033$ & $0.791^{+0.017}_{-0.014}$ & $1.438^{+0.039}_{-0.038}$\\
$\log{g_\star}$ (cgs) & Stellar surface gravity & $4.385^{+0.030}_{-0.031}$ & $4.487^{+0.022}_{-0.023}$ & $4.293 \pm 0.036$ & $4.581^{+0.014}_{-0.021}$ & $4.240^{+0.031}_{-0.032}$\\
$\rho_\star$ (g cm$^{-3}$) & Stellar density & $1.138^{+0.100}_{-0.091}$ & $1.65^{+0.13}_{-0.12}$ & $0.823^{+0.090}_{-0.079}$ & $2.48^{+0.10}_{-0.16}$ & $0.621^{+0.059}_{-0.055}$\\
$L_\star$ ($L_\odot$) & Stellar luminosity & $1.199^{+0.086}_{-0.079}$ & $0.807^{+0.081}_{-0.068}$ & $1.549^{+0.079}_{-0.080}$ & $0.453^{+0.034}_{-0.030}$ & $2.75^{+0.20}_{-0.17}$\\
$T_\mathrm{eff}$ (K) & Stellar effective temperature & $5771^{+89}_{-87}$ & $5609^{+92}_{-87}$ & $5810 \pm 79$ & $5321^{+86}_{-82}$ & $6202^{+94}_{-90}$\\
$[\mathrm{Fe/H}]$ (dex) & Metallicity & $0.191 \pm 0.083$ & $0.148 \pm 0.084$ & $0.176 \pm 0.089$ & $0.071^{+0.079}_{-0.076}$ & $0.182^{+0.083}_{-0.085}$\\
$[\mathrm{Fe/H}]_0$ (dex)\tablenotemark{c} & Initial metallicity & $0.193 \pm 0.075$ & $0.13 \pm 0.08$ & $0.204^{+0.077}_{-0.078}$ & $0.050^{+0.078}_{-0.075}$ & $0.247 \pm 0.068$\\
Age (Gyr) & Stellar age & $4.4^{+3.3}_{-2.6}$ & $2.1^{+2.7}_{-1.5}$ & $6.0^{+3.0}_{-2.5}$ & $3.0^{+3.8}_{-2.1}$ & $2.1^{+1.3}_{-1.0}$\\
EEP\tablenotemark{d} & Equal evolutionary phase & $366^{+35}_{-32}$ & $328^{+21}_{-38}$ & $406^{+20}_{-37}$ & $323^{+22}_{-36}$ & $347^{+37}_{-18}$\\
$A_V$ (mag) & Visual extinction & $0.255^{+0.089}_{-0.092}$ & $0.21^{+0.14}_{-0.13}$ & $0.130^{+0.043}_{-0.065}$ & $0.2 \pm 0.1$ & $0.132^{+0.081}_{-0.072}$\\
d (pc) & Distance & $361.7 \pm 1.8$ & $224.6 \pm 1.3$ & $282.6^{+3.2}_{-3.1}$ & $176.5^{+1.5}_{-1.4}$ & $333.6^{+4.2}_{-4.1}$\\
\enddata

\tablecomments{Table 3 in \citet{ExoFASTv2_Eastman19} provides a detailed description of all derived and fitted parameters.\\ 
$^a$\,For those targets where we adopt a circular fit, we also provide the 68\% upper limit on eccentricity derived from the eccentric fits. \\
$^b$\,The tidal circularization timescale is computed with Equation (3) of \citet{Adams2006}, assuming a tidal quality factor $Q_S = 10^6$. \\
$^c$\,The stellar metallicity when the star was formed, that define the grid points for the MIST stellar evolutionary tracks.\\ 
$^d$\,The equal evolutionary phase (EEP) corresponds to specific points in the stellar evolutionary tracks, as described in \citet{MIST0_Dotter2016}.}
% \tablecomments{}
\end{deluxetable}
\end{rotatetable}
\end{rotatepage}
\makeatletter\onecolumngrid@pop\makeatother
% \pagebreak[4]
% \clearpage

\section{Discussion} \label{sec:discussion}
\global\pdfpageattr\expandafter{\the\pdfpageattr/Rotate 0}

\begin{figure}
\epsscale{1.2}
\plotone{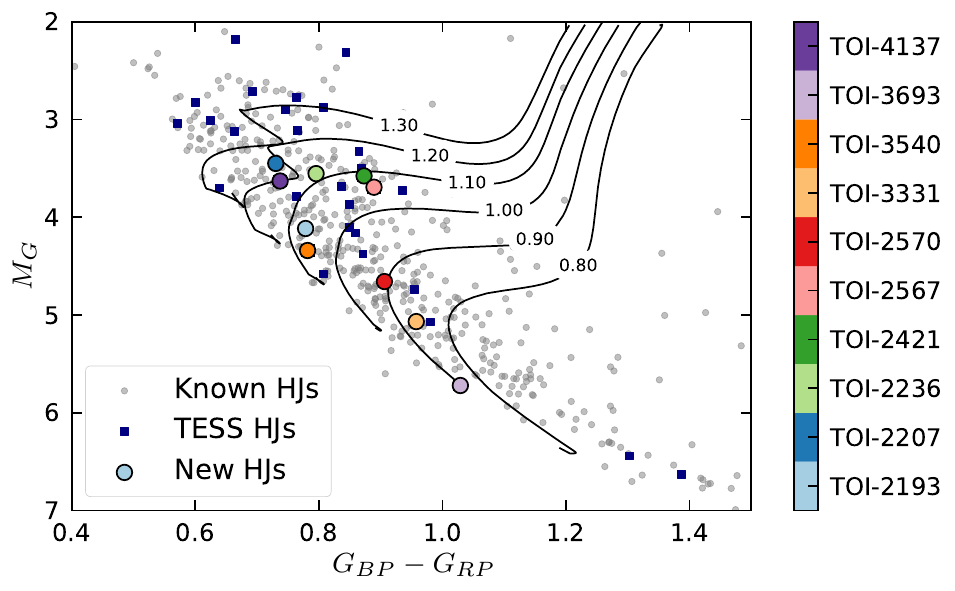}
\caption{\Gaia absolute $G$-magnitude and $G_\mathrm{BP} - G_\mathrm{RP}$ color-magnitude diagram for known hot Jupiter hosts \added{in the NASA Exoplanet Archive}.
The colored circles are the ten systems presented in this paper, identified by the color bar on the right (these colors are consistent with the following figures).
The navy blue squares show hot Jupiter systems discovered by \TESS, while the gray circles show the remaining hot Jupiter systems.
The black lines show the MIST evolutionary tracks at a metallicity of $\feh = +0.15$, close to the median metallicity of our sample, for stellar masses between 0.8 and 1.30~\Msun.
The points have not been corrected for interstellar extinction.
\label{fig:hj_pop_cmd}}
\end{figure}

\begin{figure}
\epsscale{1.2}
\plotone{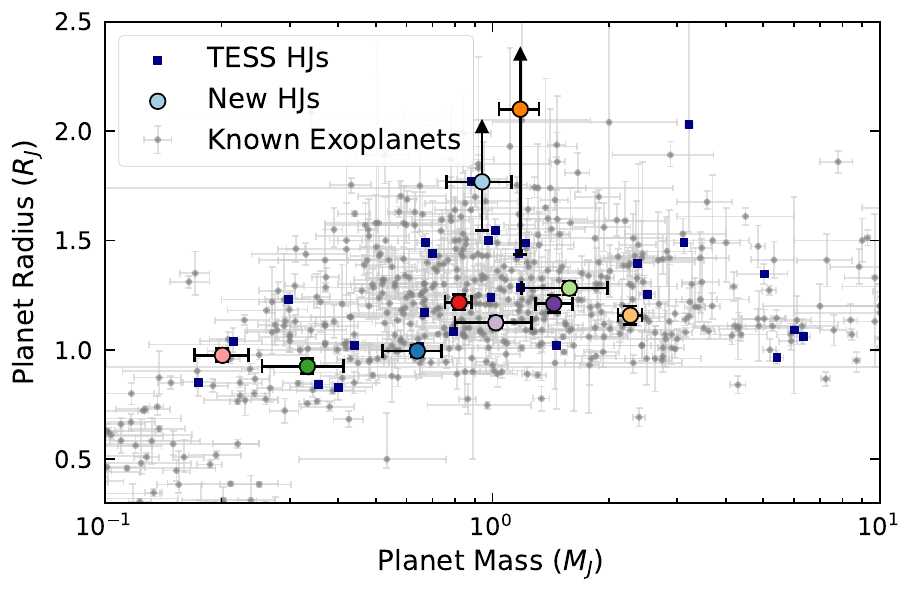}
\caption{
Mass-radius distribution for the systems presented in this paper (colored circles).
For TOI-2193A\,b and TOI-3540\,b, we plot the mode and lower limits on the planet radii.
The navy blue squares show hot Jupiter systems discovered by \TESS, while the gray circles show the masses and radii of all planets from the NASA Exoplanet Archive \added{(not just hot Jupiters)} with masses determined to better than 50\% and radii to better than 20\%.
\label{fig:hj_pop_mass_radius}}
\end{figure}

The ten planets presented in this paper have orbital periods between 2 and 10 days, and masses between 0.2 and 2.2 \Mjup. 
To put these newly-discovered planets into context, we downloaded data from the NASA Exoplanet Archive \citep{ExoplanetArchive_PSCompPars}\footnote{\url{https://exoplanetarchive.ipac.caltech.edu/}, accessed 14 Feb 2022}
\deleted{for all of the previously known planets with definite mass and radius measurements}.

\replaced{Figure \ref{fig:hj_pop_cmd} shows the distribution of the host stars in color-magnitude space.}
{In Figure \ref{fig:hj_pop_cmd}, we show the distribution of the stellar hosts of our ten planets in color-magnitude space,
in the context of other hot Jupiter hosts (orbital period $P < 10$~days, planet radius $8\,\Rearth<\Rp<24\,\Rearth$).}
Our ten planets orbit F and G stars, and all of the stars have metallicities similar to that of the Sun or higher, with a median $\feh$ of $+0.18$.
\replaced{This is in line with the well-known correlation between hot Jupiter occurrence and host star metallicity \citep{Santos2004,Valenti2005}}
{This is in line with the well-known preference for hot Jupiters to exist around stars with super-solar metallicities \citep{Santos2004,Valenti2005}.}

The planets in our sample have masses and radii generally consistent with the previously known population of hot Jupiters (Figure \ref{fig:hj_pop_mass_radius}).
The new planets are also consistent with the previously noted trend that
hot Jupiters with higher incident fluxes tend to have larger radii (Figure \ref{fig:hj_pop_teq};
see also \citealt{Demory2011}).
% The longest-period planet in our sample, TOI-3693\,b ($P = 9.09$~days), has a radius of $1.120^{+0.026}_{-0.020}\Rjup$ and is one of the largest known planets with equilibrium temperature $T_\mathrm{eq} < 800$~K.

\subsection{Two Inflated Saturns} \label{ssec:inflated_saturns}

\begin{figure}
\epsscale{1.2}
\plotone{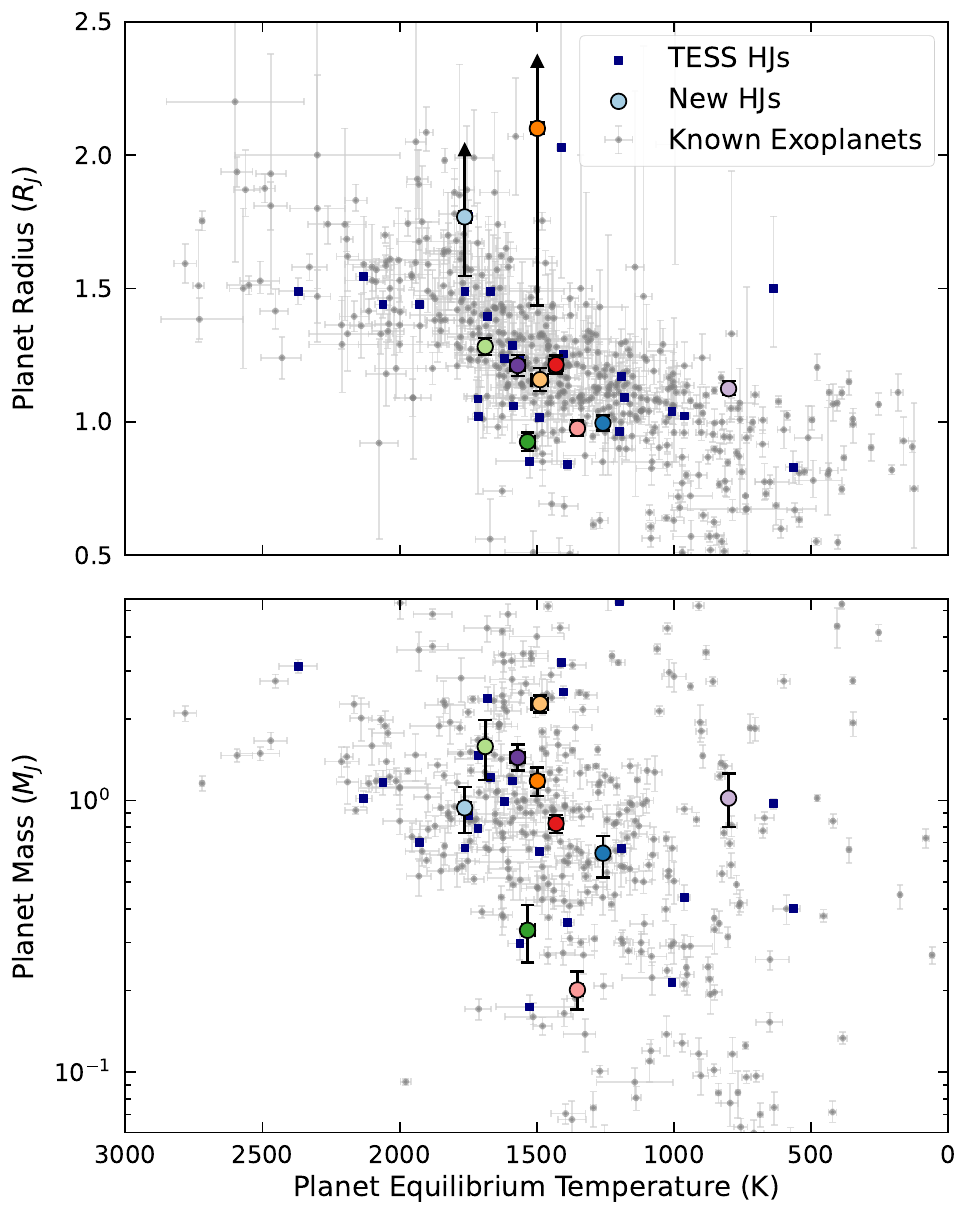}
% \plotone{hj_pop_planet_teq_mass}
\caption{
Distribution of planet radius (top panel) and mass (lower panel) as a function of planet insolation, expressed as the equilibrium temperature at the planet's orbital distance, assuming no albedo and perfect heat redistribution.
The two least massive planets in our sample, TOI-2421\,b and TOI-2567\,b, lie on the lower edge of the distribution of known hot Jupiters, just above the hot Neptune desert \citep{Mazeh2016}.
The navy blue squares and gray circles represent the same previously known systems as described in Figure \ref{fig:hj_pop_mass_radius}.
\label{fig:hj_pop_teq}}
\end{figure}

\begin{figure}
\epsscale{1.2}
\plotone{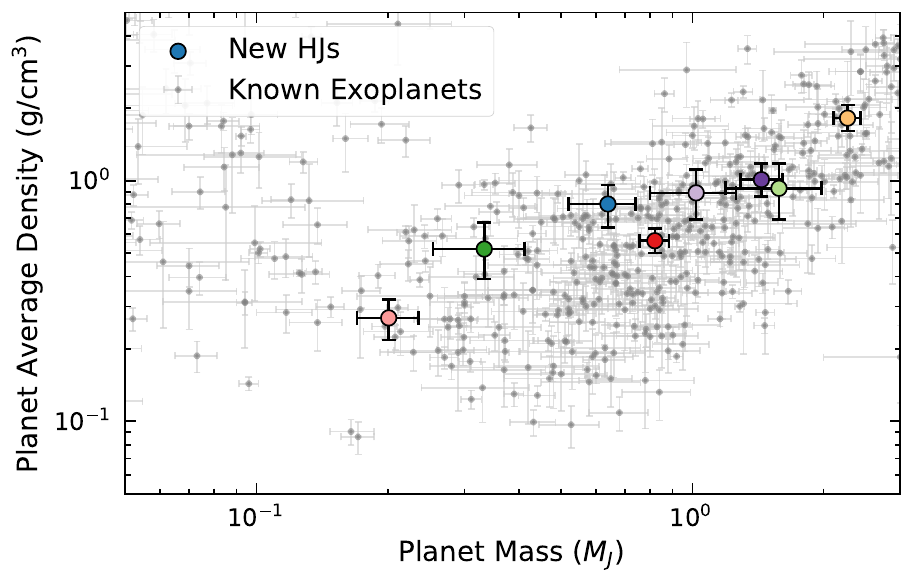}
\caption{
Bulk density of the planets in our sample, excluding the grazing planets TOI-2193A\,b and TOI-3540\,b.
\replaced{The gray points in the plot are previously known hot ($P < 10$~days) exoplanets from the NASA Exoplanet Archive, \replaced{with}{that have} masses and radii determined to better than 20\%.}
{The gray circles represent the same previously known systems as described in Figure \ref{fig:hj_pop_mass_radius}.}
\label{fig:hj_pop_density}
}
\end{figure}

Two of the host stars (TOI-2421, TOI-2567) appear to have recently evolved off the terminal age main sequence, with stellar radii $\Rstar = 1.75 \pm 0.05$\,\Rsun and $1.69 \pm 0.04$\,\Rsun respectively. Their masses are both about 1.13\,$\Msun$.
These two stars also host the lowest-mass planets in our sample -- TOI-2421\,b is a Saturn-mass planet, with a mass of $\Mp = 0.33 \pm 0.08 \Mjup$, and TOI-2567\,b is a sub-Saturn ($\Mp = 0.20\pm 0.03 \Mjup$).
The two planets lie on the upper boundary of the hot Neptune desert \citep{Mazeh2016}, as shown in the lower panel of Figure \ref{fig:hj_pop_teq}.

Both planets are inflated, with radii of $\Rp = 0.925^{+0.035}_{-0.034}\Rjup$ and $0.975^{+0.031}_{-0.029}\Rjup$ respectively.
Indeed, TOI-2567\,b has a bulk density of just $0.27\pm0.05$\gcc, making it one of the least dense planets known for its mass (Figure \ref{fig:hj_pop_density}).
The two planets join a small but growing collection of hot Saturns orbiting slightly evolved stars, including TOI-954\,b \citep{Sha2021}, TOI-1296\,b \citep{Moutou2021} and TOI-1842\,b \citep{Wittenmyer2022}.
Such planets can help test models of radius reinflation around evolved stars (e.g. \citealt{Lopez2016,Thorngren2021}) by helping to place constraints on the timescale of reinflation, and via comparison with such trends for their more massive counterparts (e.g. \citealt{Hartman2016}).

\subsection{Planet Eccentricities}

\begin{figure}
\epsscale{1.2}
\plotone{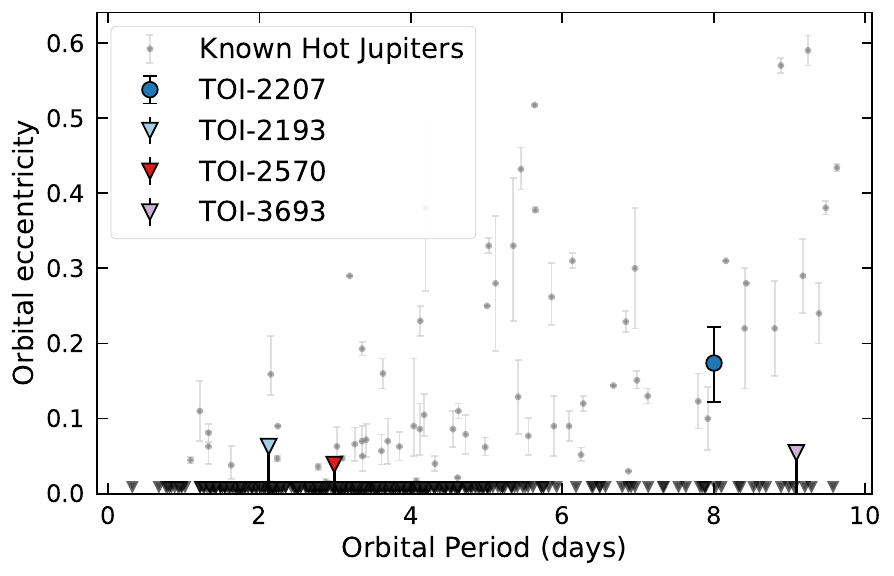}
\caption{
\label{fig:hj_pop_ecc}
Measured eccentricities and upper limits for the planets in our sample.
Here, we show the 68\% confidence interval measured for the eccentricity of TOI-2207\,b, along with 68\% upper limits on the eccentricity for TOI-2193\,b, TOI-2570\,b, and TOI-3693\,b, for which we can place a constraint better than $e < 0.1$ from the current data.
The gray circles show the measured eccentricities of \replaced{known exoplanets}{previously known giant exoplanets ($\Rp > 8 \Rearth$)} with eccentricities at least 2-$\sigma$ greater than zero, while gray triangles at the bottom of the plot show planets on orbits consistent with circular.
}
\end{figure}

While the shortest period hot Jupiters are expected to be on circular orbits due to tides raised by the star on the planet, the tidal circularization timescale increases rapidly with orbital distance.
Tidal circularization might be too slow to have affected planets with periods approaching 10 days or longer, with an extreme example being the recently discovered TOI-3362\,b, a potential proto-hot Jupiter on a $P=18.1$~day, $e=0.82$ orbit \citep{Dong2021a}.
Indeed, the two longest period planets in our sample, TOI-2207\,b ($P = 8.00$~days) and TOI-3693\,b ($P = 9.09$~days) have theoretical tidal circularization timescales of $12\pm5$ and $22\pm6$ Gyr, based on Equation 3 of \citet{Adams2006} which extended the work from \citet{Goldreich1966}, and assuming a tidal quality factor of $Q_P = 10^6$.

The measured orbital eccentricity of TOI-2207\,b is $e = 0.174^{+0.048}_{-0.052}$, which is greater than zero by more than 3-$\sigma$, although the significance of this result may be affected by the Lucy-Sweeney bias \citep{LucySweeney71}.
Given that the estimated stellar age is 4~Gyr, which is of the same order of magnitude as the theoretical circularization timescale, the current eccentricity might be a remnant of a high-eccentricity migration formation pathway for this planet.

In contrast, TOI-3693\,b, which has a longer orbital period, appears to have a more circular orbit.
The 68\% and 95\% upper limits for the eccentricity of this planet are $e < 0.054$ and $e < 0.13$ respectively.
More data, including the possible timing and detection of a secondary eclipse, would be required to obtain a more secure measurement of the planet eccentricity, but TOI-3693\,b is not likely to have an orbital eccentricity similar to that of TOI-2207\,b or some other warm Jupiters (e.g. TOI-640\,b \citep{Rodriguez2021}, TOI-559\,b \citep{Ikwut-Ukwa2021}).

These two longer period hot Jupiter systems are also potential targets for stellar obliquity measurements with the Rossiter-McLaughlin (RM) effect \citep{Rossiter1924,McLaughlin1924}.
The expected RM amplitudes for the two systems are $\approx30$\,\ms for TOI-2207\,b and $\approx90$\,\ms for TOI-3693\,b \citep{Gaudi2007}, which should be measurable on a large telescope given the relatively bright host stars ($V = 11.4, 12.0$ respectively).
The large planet-star separations of the two planets ($a/\Rstar \approx 12$ and $22$ respectively) result in long tidal realignment timescales.
Thus, misaligned orbits in these systems, particularly when correlated with orbital eccentricity, could be indicative of a high-eccentricity formation pathway, although they could also result from perturbations by an outer planets in the system.

\section{Conclusions} \label{sec:conclusion}

We presented the discovery and characterization of ten new hot Jupiters around F and G stars from NASA's \TESS mission.
These planets orbit relatively bright stars ($G < 12.5$) and are potential targets for atmospheric characterization, measurements of stellar obliquity, and other follow-up observations.
While we have drawn attention to some of the notable features of the new planets, including the low density of the sub-Saturn TOI-2567\,b and the detectable eccentricity of TOI-2207\,b,
the larger and longer-term purpose of the survey is to allow for more general conclusions to be drawn about the hot Jupiter population.
This will require more observations to detect and confirm new planets (the ``numerator'' of demographic calculations) as well as a detailed examination of the \TESS selection function and survey characteristics (the ``denominator'').
Based on the forecast of \citet{Yee2021b}, to assemble a sample of 400 hot Jupiters (an order-of-magnitude more planets than the \Kepler sample), a magnitude-limited survey would need to be complete down to $G = 12.5$.
The ten planets described here, along with the other new \TESS hot Jupiters that have
been described in the literature (\citealp[e.g.,][]{Rodriguez2019,Zhou2019a,Brahm2020,Davis2020,Nielsen2020,Ikwut-Ukwa2021,Rodriguez2021,Sha2021,Wong2021,Knudstrup2022,Rodriguez2022}) are steps toward realizing the promise of \TESS
for hot Jupiter demographics.

\newpage
\begin{acknowledgements}

We thank the anonymous reviewer whose comments helped improve the manuscript.
S.W.Y. thanks Gummi Stefansson for helpful conversations regarding the NEID observations.

This paper includes data collected by the \TESS mission that are publicly available from the Mikulski Archive for Space Telescopes (MAST). Funding for the \TESS mission is provided by NASA's Science Mission Directorate. We acknowledge the use of public \TESS data from pipelines at the \TESS Science Office and at the \TESS Science Processing Operations Center.
Resources supporting this work were provided by the NASA High-End Computing (HEC) Program through the NASA Advanced Supercomputing (NAS) Division at Ames Research Center for the production of the SPOC data products.
We also acknowledge the use of data from the Exoplanet Follow-up Observation Program website, which is operated by the California Institute of Technology, under contract with the National Aeronautics and Space Administration under the Exoplanet Exploration Program.
This research made use of Lightkurve, a Python package for \Kepler and \TESS data analysis \citep{Lightkurve18}.

Some of the data presented herein were obtained at the W. M. Keck Observatory, which is operated as a scientific partnership among the California Institute of Technology, the University of California and the National Aeronautics and Space Administration. The Observatory was made possible by the generous financial support of the W. M. Keck Foundation.
Keck telescope time was granted by NOIRLab (Prop. ID 2021B-0162, PI: Yee) through the Mid-Scale Innovations Program (MSIP). MSIP is funded by NSF.
The authors wish to recognize and acknowledge the very significant cultural role and reverence that the summit of Maunakea has always had within the indigenous Hawaiian community.  We are most fortunate to have the opportunity to conduct observations from this mountain.

This paper contains data taken with the NEID instrument, which was funded by the NASA-NSF Exoplanet Observational Research (NN-EXPLORE) partnership and built by Pennsylvania State University. NEID is installed on the WIYN telescope, which is operated by the National Optical Astronomy Observatory, and the NEID archive is operated by the NASA Exoplanet Science Institute at the California Institute of Technology. NN-EXPLORE is managed by the Jet Propulsion Laboratory, California Institute of Technology under contract with the National Aeronautics and Space Administration.
Data presented herein were obtained at the WIYN Observatory from telescope time allocated to NN-EXPLORE through the scientific partnership of the National Aeronautics and Space Administration, the National Science Foundation, and NOIRLab.
This work was supported by a NASA WIYN PI Data Award, administered by the NASA Exoplanet Science Institute.
The authors thank Sarah Logsdon and Heidi Schweiker for help with the NEID observations.
The authors are honored to be permitted to conduct astronomical research on Iolkam Du’ag (Kitt Peak), a mountain with particular significance to the Tohono O’odham.

This paper includes data gathered with the 6.5 meter Magellan Telescopes located at Las Campanas Observatory, Chile.

This research has used data from the CTIO/SMARTS 1.5m telescope, which is operated as part of the SMARTS Consortium by RECONS (\url{www.recons.org}) members Todd Henry, Hodari James, Wei-Chun Jao, and Leonardo Paredes. At the telescope, observations were carried out by Roberto Aviles and Rodrigo Hinojosa.
The CHIRON data were obtained from telescope time allocated under the NN-EXPLORE program with support from the National Aeronautics and Space Administration.

Some of the data presented herein were obtained at the \textsc{Minerva}-Australis facility from telescope time allocated under the NN-EXPLORE program with support from the National Aeronautics and Space Administration.
\textsc{Minerva}-Australis is supported by Australian Research Council LIEF Grant LE160100001, Discovery Grants DP180100972 and DP220100365, Mount Cuba Astronomical Foundation, and institutional partners University of Southern Queensland, UNSW Sydney, MIT, Nanjing University, George Mason University, University of Louisville, University of California Riverside, University of Florida, and The University of Texas at Austin.
We respectfully acknowledge the traditional custodians of all lands throughout Australia, and recognise their continued cultural and spiritual connection to the land, waterways, cosmos, and community. We pay our deepest respects to all Elders, ancestors and descendants of the Giabal, Jarowair, and Kambuwal nations, upon whose lands the \textsc{Minerva}-Australis facility at Mt Kent is situated.

This work makes use of observations from the LCOGT network. Part of the LCOGT telescope time was granted by NOIRLab through the Mid-Scale Innovations Program (MSIP). MSIP is funded by NSF.

This paper makes use of data from the MEarth Project, which is a collaboration between Harvard University and the Smithsonian Astrophysical Observatory. The MEarth Project acknowledges funding from the David and Lucile Packard Fellowship for Science and Engineering, the National Science Foundation under grants AST-0807690, AST-1109468, AST-1616624 and AST-1004488 (Alan T. Waterman Award), the National Aeronautics and Space Administration under Grant No. 80NSSC18K0476 issued through the XRP Program, and the John Templeton Foundation.

Adam Popowicz and Slawomir Lasota were responsible for data processing and automation of observations at SUTO observatories and were financed by grant BK-246/RAu-11/2022.
AJ acknowledges support from ANID -- Millennium  Science  Initiative -- ICN12\_009 and FONDECYT project 1210718.

J.H acknowledges support from NASA grants 80NSSC19K0386, 80NSSC19K1728, and 80NSSC21K0335.
\end{acknowledgements}

\facilities{TESS, MAST, Gaia, Keck:I (HIRES), WIYN (NEID), Magellan:Clay (PFS), CTIO:1.5m (CHIRON), Max Planck:2.2m (FEROS), FLWO:1.5m (TRES), LCOGT, Gemini}

\software{astropy \citep{Astropy13,Astropy18},
lightkurve \citep{Lightkurve18},
EXOFASTv2 \citep{ExoFASTv2_Eastman19},
SpecMatch-Emp \citep{SpecMatchEmp_Yee2017},
SpecMatch-Synth \citep{SpecMatchSynth_Petigura2015},
AstroImageJ \citep{AstroImageJ_Collins17},
TAPIR \citep{TAPIR_Jensen2013},
numpy \citep{Numpy}, scipy \citep{Scipy},
pandas \citep{Pandas20,Pandas_McKinney10}, matplotlib \citep{Matplotlib},
}

\clearpage
\makeatletter\onecolumngrid@push\makeatother
\def\ExToiMultiplot{2193}
\foreach \toi in \tois
{ 
\unless\ifnum\toi=\ExToiMultiplot{
\begin{figure}[p]
\includegraphics[width=500pt]{toi\toi_multiplot.pdf}
\caption{Same as Figure \ref{fig:toi2193_multiplot}, but for TOI-\toi\xspace b.}
\label{fig:toi\toi_multiplot}
\end{figure}
}
\fi
}
\clearpage
% \FloatBarrier

% \clearpage
% \appendix
% \section{Appendix}
% \section{TESS Light-Curves \label{appendix:tess_lcs}}

% \foreach \toi in \tois
% { 
% \begin{figure*}
% \includegraphics[width=0.45\linewidth]{tess_lcs/toi\toi_tess_lc}
% \includegraphics[width=500pt]{toi\toi_multiplot.png}
% \caption{Same as Figure \ref{fig:toi2193_multiplot}, but for TOI-\toi\xspace b.}
% \label{fig:toi\toi_multiplot}
% \end{figure*}
% }

% \begin{figure}
% \centering
% % \includegraphics{tess_lcs/toi2193_tess_lc}
% \includegraphics[width=0.425\textwidth]{tess_lcs/toi2193_tess_lc}
% \includegraphics[width=0.425\textwidth]{tess_lcs/toi2207_tess_lc}
% \includegraphics[width=0.45\linewidth]{tess_lcs/toi2236_tess_lc}
% \includegraphics[width=0.45\linewidth]{tess_lcs/toi2374_tess_lc}
% % \caption{TESS SPOC light-curves for TOI-2193, TOI-2207, TOI-2236, TOI-2374.}
% \end{figure}
% \begin{figure}
% \centering
% % \includegraphics[width=0.45\textwidth]{tess_lcs/toi2421_tess_lc}
% \includegraphics[width=0.45\textwidth]{tess_lcs/toi2567_tess_lc}
% \includegraphics[width=0.45\textwidth]{tess_lcs/toi3693_tess_lc}
% \includegraphics[width=0.45\textwidth]{tess_lcs/toi4137_tess_lc}
% \caption{TESS SPOC light-curves for TOI-2421, TOI-2567, TOI-3693, TOI-4137.}
% \end{figure}
% \begin{figure}
% \centering
% \includegraphics[width=0.45\textwidth]{tess_lcs/toi3331_tess_lc}
% \includegraphics[width=0.45\textwidth]{tess_lcs/toi3540_tess_lc}
% \caption{QLP light-curves for TOI-3331, TOI-3540.}
% \end{figure}

% \clearpage

% \begin{figure}
% \centering
% \includegraphics[width=0.45\linewidth]{toi3331_speckle}
% \includegraphics[width=0.45\linewidth]{toi3540_speckle} \\
% \includegraphics[width=0.45\linewidth]{toi3540_ao_Brgamma}
% \includegraphics[width=0.45\linewidth]{toi3540_ao_Hcont} \\
% \includegraphics[width=0.45\linewidth]{toi3693_speckle} 
% \includegraphics[width=0.45\linewidth]{toi4137_speckle}
% \caption{High-Resolution imaging of hot Jupiter hosts described in this paper.
% \textbf{Top row:}
% SOAR HRCam Observation of TOI-3331. A companion was detected at an angular separation of $2"$, visible in the inset ACF image to the NE and SW of the primary star;
% SOAR HRCam Observation of TOI-3540;
% \textbf{Middle row:}
% Palomar PHARO Observation (Br$\gamma$) of TOI-3540;
% Palomar PHARO Observation ($H$cont) of TOI-3540. The observations of TOI-3540 reveal a close companion at $0\farcs92$ from the primary star, visible in the inset images;
% \textbf{Bottom row:}
% SAI-2.5m Speckle Polarimeter Observation of TOI-3693;
% SAI-2.5m Speckle Polarimeter Observation of TOI-4137.
% \label{fig:high_res_imaging_1}}
% \end{figure}
\makeatletter\onecolumngrid@pop\makeatother

\appendix
\section{Additional Fit Parameters}
We present in Table \ref{tab:additional_fit_params} the median and 68\% confidence intervals for additional fit parameters not listed in Tables \ref{tab:fitted_props} and \ref{tab:fitted_props_2} for the adopted fits.
These are the linear and quadratic limb-darkening parameters $(u_1, u_2)$ in each band; additional flux dilution from neighboring stars in each band $(D)$; the relative RV offset for each instrument $\gamma_\mathrm{rel}$ (\ms); and the RV jitter for each instrument $\sigma_J$ (\ms).
% ; the added variance for each light-curve $\sigma^2$; the baseline flux for each light-curve $F_0$; and additive detrend coefficients for each light-curve ($C_i$).
% \movetabledown=0.7in
% \begin{longrotatetable}
\startlongtable
\begin{deluxetable*}{lccccc} \label{tab:additional_fit_params}
\tablecaption{Additional Fit Parameters (Median and 68\% Confidence Intervals)}
\tabletypesize{\footnotesize}
\tablehead{\colhead{Parameter}}
\startdata
\\[-\normalbaselineskip]\multicolumn{2}{l}{\textbf{TOI-2193}}\\
 &B&I&i'&z'&TESS\\
~~~~$u_{1}$ &$0.605\pm0.035$&$0.252^{+0.049}_{-0.050}$&$0.256^{+0.043}_{-0.044}$&$0.211\pm0.033$&$0.299\pm0.031$\\
~~~~$u_{2}$ &$0.192\pm0.035$&$0.287\pm0.049$&$0.275\pm0.045$&$0.279\pm0.033$&$0.310\pm0.031$\\
~~~~$A_D$ &$0.00634^{+0.00064}_{-0.00063}$&$0.0312\pm0.0031$&$0.0273\pm0.0028$&$0.0366\pm0.0037$&$0.0004\pm0.0034$\\
 &PFS\\
~~~~$\gamma_{\rm rel}$ &$25\pm15$\\
~~~~$\sigma_J$ &$37^{+20}_{-11}$\\
\hline\\[-\normalbaselineskip]\multicolumn{2}{l}{\textbf{TOI-2207}}\\
 &I&z'&TESS\\
~~~~$u_{1}$ &$0.247\pm0.044$&$0.164\pm0.044$&$0.261\pm0.045$\\
~~~~$u_{2}$ &$0.304\pm0.047$&$0.265^{+0.047}_{-0.048}$&$0.305\pm0.048$\\
~~~~$A_D$ &--&--&$-0.0000\pm0.0029$\\
 &CHIRON&FEROS&PFS\\
~~~~$\gamma_{\rm rel}$ &$-35056\pm11$&$-35676\pm40$&$-10.8^{+8.0}_{-8.1}$\\
~~~~$\sigma_J$ &$10^{+21}_{-11}$&$92^{+65}_{-31}$&$17.4^{+15}_{-7.7}$\\
\hline\\[-\normalbaselineskip]\multicolumn{2}{l}{\textbf{TOI-2236}}\\
 &B&I&z'&TESS\\
~~~~$u_{1}$ &$0.549\pm0.052$&$0.227\pm0.050$&$0.192\pm0.050$&$0.232\pm0.025$\\
~~~~$u_{2}$ &$0.219\pm0.051$&$0.306^{+0.050}_{-0.049}$&$0.302^{+0.050}_{-0.049}$&$0.302\pm0.025$\\
~~~~$A_D$ &--&--&--&$-0.0008\pm0.0046$\\
 &CHIRON\\
~~~~$\gamma_{\rm rel}$ &$45127^{+40}_{-35}$\\
~~~~$\sigma_J$ &$58^{+70}_{-58}$\\
\hline\\[-\normalbaselineskip]\multicolumn{2}{l}{\textbf{TOI-2421}}\\
 &g'&i'&TESS\\
~~~~$u_{1}$ &$0.615^{+0.051}_{-0.052}$&$0.349\pm0.031$&$0.327\pm0.033$\\
~~~~$u_{2}$ &$0.146\pm0.051$&$0.277\pm0.029$&$0.260\pm0.035$\\
~~~~$A_D$ &--&--&$-0.000000^{+0.00010}_{-0.000100}$\\
 &CHIRON&MINERVA-A-3&MINERVA-A-6\\
~~~~$\gamma_{\rm rel}$ &$6432.2^{+7.8}_{-7.6}$&$7020\pm19$&$7110^{+160}_{-170}$\\
~~~~$\sigma_J$ &$8.3^{+17}_{-8.3}$&$41^{+27}_{-17}$&$260^{+400}_{-180}$\\
\hline\\[-\normalbaselineskip]\multicolumn{2}{l}{\textbf{TOI-2567}}\\
 &i'&TESS\\
~~~~$u_{1}$ &$0.346^{+0.049}_{-0.050}$&$0.329^{+0.031}_{-0.032}$\\
~~~~$u_{2}$ &$0.271\pm0.050$&$0.259\pm0.035$\\
~~~~$A_D$ &--&$0.0009\pm0.0088$\\
 &HIRES\\
~~~~$\gamma_{\rm rel}$ &$-3.2^{+2.2}_{-2.0}$\\
~~~~$\sigma_J$ &$3.8^{+5.1}_{-2.9}$\\
\hline\\[-\normalbaselineskip]\multicolumn{2}{l}{\textbf{TOI-2570}}\\
 &B&g'&i'&TESS\\
~~~~$u_{1}$ &$0.677^{+0.055}_{-0.056}$&$0.589\pm0.053$&$0.309\pm0.044$&$0.299\pm0.028$\\
~~~~$u_{2}$ &$0.141\pm0.053$&$0.186\pm0.052$&$0.264\pm0.047$&$0.273\pm0.029$\\
~~~~$A_D$ &--&--&--&$-0.062\pm0.016$\\
 &NEID\\
~~~~$\gamma_{\rm rel}$ &$-45135.4\pm5.0$\\
~~~~$\sigma_J$ &$10.0^{+9.0}_{-4.9}$\\
\hline\\[-\normalbaselineskip]\multicolumn{2}{l}{\textbf{TOI-3331}}\\
 &I&g'&i'&TESS\\
~~~~$u_{1}$ &$0.303\pm0.050$&$0.639^{+0.040}_{-0.041}$&$0.343\pm0.028$&$0.324\pm0.051$\\
~~~~$u_{2}$ &$0.249\pm0.049$&$0.178\pm0.040$&$0.263\pm0.026$&$0.251\pm0.050$\\
~~~~$A_D$ &$0.123\pm0.011$&$0.0506\pm0.0053$&$0.1159^{+0.0097}_{-0.0098}$&$0.107\pm0.039$\\
 &PFS\\
~~~~$\gamma_{\rm rel}$ &$66\pm13$\\
~~~~$\sigma_J$ &$27^{+28}_{-11}$\\
\hline\\[-\normalbaselineskip]\multicolumn{2}{l}{\textbf{TOI-3540}}\\
 &I&R&TESS\\
~~~~$u_{1}$ &$0.329^{+0.050}_{-0.051}$&$0.370^{+0.049}_{-0.050}$&$0.268^{+0.048}_{-0.049}$\\
~~~~$u_{2}$ &$0.317\pm0.050$&$0.281\pm0.049$&$0.248^{+0.048}_{-0.049}$\\
~~~~$A_D$ &$0.167\pm0.016$&$0.143\pm0.014$&$0.151\pm0.015$\\
 &HIRES\\
~~~~$\gamma_{\rm rel}$ &$-19\pm12$\\
~~~~$\sigma_J$ &$25^{+25}_{-10.}$\\
\hline\\[-\normalbaselineskip]\multicolumn{2}{l}{\textbf{TOI-3693}}\\
 &R&TESS\\
~~~~$u_{1}$ &$0.406\pm0.051$&$0.367\pm0.035$\\
~~~~$u_{2}$ &$0.197^{+0.051}_{-0.052}$&$0.238\pm0.036$\\
~~~~$A_D$ &--&$-0.0045\pm0.0098$\\
 &HIRES&TRES\\
~~~~$\gamma_{\rm rel}$ &$-13\pm13$&$197^{+93}_{-91}$\\
~~~~$\sigma_J$ &$36.4^{+16}_{-9.8}$&$164^{+96}_{-79}$\\
\hline\\[-\normalbaselineskip]\multicolumn{2}{l}{\textbf{TOI-4137}}\\
 &i'&TESS\\
~~~~$u_{1}$ &$0.251\pm0.048$&$0.237\pm0.033$\\
~~~~$u_{2}$ &$0.315\pm0.049$&$0.293\pm0.034$\\
~~~~$A_D$ &--&$0.007\pm0.011$\\
 &NEID\\
~~~~$\gamma_{\rm rel}$ &$-41121^{+13}_{-11}$\\
~~~~$\sigma_J$ &$19^{+27}_{-19}$\\
\hline
\enddata
\end{deluxetable*}
% \end{longrotatetable}

% \clearpage
\bibliography{manuscript,instruments,software,catalogs}

\begin{thebibliography}{}
\expandafter\ifx\csname natexlab\endcsname\relax\def\natexlab#1{#1}\fi
\providecommand{\url}[1]{\href{#1}{#1}}
\providecommand{\dodoi}[1]{doi:~\href{http://doi.org/#1}{\nolinkurl{#1}}}
\providecommand{\doeprint}[1]{\href{http://ascl.net/#1}{\nolinkurl{http://ascl.net/#1}}}
\providecommand{\doarXiv}[1]{\href{https://arxiv.org/abs/#1}{\nolinkurl{https://arxiv.org/abs/#1}}}

\bibitem[{Adams \& Laughlin(2006)}]{Adams2006}
Adams, F.~C., \& Laughlin, G. 2006, The Astrophysical Journal, 649, 1004,
  \dodoi{10.1086/506145}

\bibitem[{Addison {et~al.}(2019)Addison, Wright, Wittenmyer, Horner, Mengel,
  Johns, Marti, Nicholson, Soutter, Bowler, Crossfield, Kane, Kielkopf,
  Plavchan, Tinney, Zhang, Clark, Clerte, Eastman, Swift, Bottom, Muirhead,
  McCrady, Herzig, Hogstrom, Wilson, Sliski, Johnson, Wright, Johnson, Blake,
  Riddle, Lin, Cornachione, Bedding, Stello, Huber, Marsden, \&
  Carter}]{MinervaAustralis_Addison2019}
Addison, B., Wright, D.~J., Wittenmyer, R.~A., {et~al.} 2019, Publications of
  the Astronomical Society of the Pacific, 131, 115003,
  \dodoi{10.1088/1538-3873/ab03aa}

\bibitem[{Addison {et~al.}(2021)Addison, Wright, Nicholson, Cale, Mocnik,
  Huber, Plavchan, Wittenmyer, Vanderburg, Chaplin, Chontos, Clark, Eastman,
  Ziegler, Brahm, Carter, Clerte, Espinoza, Horner, Bentley, Jord{\'a}n, Kane,
  Kielkopf, Laychock, Mengel, Okumura, Stassun, Bedding, Bowler, Burnelis,
  {Blanco-Cuaresma}, Collins, Crossfield, Davis, Evensberget, Heitzmann,
  Howell, Law, Mann, Marsden, Matson, O'Connor, Shporer, Stevens, Tinney,
  Tylor, Wang, Zhang, Henning, Kossakowski, Ricker, Sarkis, Schlecker, Torres,
  Vanderspek, Latham, Seager, Winn, Jenkins, Mireles, Rowden, Pepper, Daylan,
  Schlieder, Collins, Collins, Tan, Ball, Basu, Buzasi, Campante, Corsaro,
  {Gonz{\'a}lez-Cuesta}, Davies, {de Almeida}, Nascimento, {a}, Guo, Handberg,
  Hekker, Hey, Kallinger, Kawaler, Kayhan, Kuszlewicz, Lund, Lyttle, Mathur,
  Miglio, Mosser, Nielsen, Serenelli, Aguirre, \& Theme{\ss}l}]{Addison2021}
Addison, B.~C., Wright, D.~J., Nicholson, B.~A., {et~al.} 2021,
  arXiv:2001.07345 [astro-ph], \dodoi{10.1093/mnras/staa3960}

\bibitem[{Agol {et~al.}(2020)Agol, Luger, \& {Foreman-Mackey}}]{Agol2020}
Agol, E., Luger, R., \& {Foreman-Mackey}, D. 2020, \aj, 159, 123,
  \dodoi{10.3847/1538-3881/ab4fee}

\bibitem[{Alonso {et~al.}(2004)Alonso, Brown, Torres, Latham, Sozzetti,
  Mandushev, Belmonte, Charbonneau, Deeg, Dunham, O'Donovan, \&
  Stefanik}]{TrES_Alonso2004}
Alonso, R., Brown, T.~M., Torres, G., {et~al.} 2004, The Astrophysical Journal,
  613, L153, \dodoi{10.1086/425256}

\bibitem[{{Astropy Collaboration} {et~al.}(2013){Astropy Collaboration},
  {Robitaille}, {Tollerud}, {Greenfield}, {Droettboom}, {Bray}, {Aldcroft},
  {Davis}, {Ginsburg}, {Price-Whelan}, {Kerzendorf}, {Conley}, {Crighton},
  {Barbary}, {Muna}, {Ferguson}, {Grollier}, {Parikh}, {Nair}, {Unther},
  {Deil}, {Woillez}, {Conseil}, {Kramer}, {Turner}, {Singer}, {Fox}, {Weaver},
  {Zabalza}, {Edwards}, {Azalee Bostroem}, {Burke}, {Casey}, {Crawford},
  {Dencheva}, {Ely}, {Jenness}, {Labrie}, {Lim}, {Pierfederici}, {Pontzen},
  {Ptak}, {Refsdal}, {Servillat}, \& {Streicher}}]{Astropy13}
{Astropy Collaboration}, {Robitaille}, T.~P., {Tollerud}, E.~J., {et~al.} 2013,
  \aap, 558, A33, \dodoi{10.1051/0004-6361/201322068}

\bibitem[{{Astropy Collaboration} {et~al.}(2018){Astropy Collaboration},
  {Price-Whelan}, {Sip{\H{o}}cz}, {G{\"u}nther}, {Lim}, {Crawford}, {Conseil},
  {Shupe}, {Craig}, {Dencheva}, {Ginsburg}, {Vand erPlas}, {Bradley},
  {P{\'e}rez-Su{\'a}rez}, {de Val-Borro}, {Aldcroft}, {Cruz}, {Robitaille},
  {Tollerud}, {Ardelean}, {Babej}, {Bach}, {Bachetti}, {Bakanov}, {Bamford},
  {Barentsen}, {Barmby}, {Baumbach}, {Berry}, {Biscani}, {Boquien}, {Bostroem},
  {Bouma}, {Brammer}, {Bray}, {Breytenbach}, {Buddelmeijer}, {Burke},
  {Calderone}, {Cano Rodr{\'\i}guez}, {Cara}, {Cardoso}, {Cheedella}, {Copin},
  {Corrales}, {Crichton}, {D'Avella}, {Deil}, {Depagne}, {Dietrich}, {Donath},
  {Droettboom}, {Earl}, {Erben}, {Fabbro}, {Ferreira}, {Finethy}, {Fox},
  {Garrison}, {Gibbons}, {Goldstein}, {Gommers}, {Greco}, {Greenfield},
  {Groener}, {Grollier}, {Hagen}, {Hirst}, {Homeier}, {Horton}, {Hosseinzadeh},
  {Hu}, {Hunkeler}, {Ivezi{\'c}}, {Jain}, {Jenness}, {Kanarek}, {Kendrew},
  {Kern}, {Kerzendorf}, {Khvalko}, {King}, {Kirkby}, {Kulkarni}, {Kumar},
  {Lee}, {Lenz}, {Littlefair}, {Ma}, {Macleod}, {Mastropietro}, {McCully},
  {Montagnac}, {Morris}, {Mueller}, {Mumford}, {Muna}, {Murphy}, {Nelson},
  {Nguyen}, {Ninan}, {N{\"o}the}, {Ogaz}, {Oh}, {Parejko}, {Parley}, {Pascual},
  {Patil}, {Patil}, {Plunkett}, {Prochaska}, {Rastogi}, {Reddy Janga},
  {Sabater}, {Sakurikar}, {Seifert}, {Sherbert}, {Sherwood-Taylor}, {Shih},
  {Sick}, {Silbiger}, {Singanamalla}, {Singer}, {Sladen}, {Sooley},
  {Sornarajah}, {Streicher}, {Teuben}, {Thomas}, {Tremblay}, {Turner},
  {Terr{\'o}n}, {van Kerkwijk}, {de la Vega}, {Watkins}, {Weaver}, {Whitmore},
  {Woillez}, {Zabalza}, \& {Astropy Contributors}}]{Astropy18}
{Astropy Collaboration}, {Price-Whelan}, A.~M., {Sip{\H{o}}cz}, B.~M., {et~al.}
  2018, \aj, 156, 123, \dodoi{10.3847/1538-3881/aabc4f}

\bibitem[{Bakos {et~al.}(2004)Bakos, Noyes, Kov{\'a}cs, Stanek, Sasselov, \&
  Domsa}]{HAT_Bakos2004}
Bakos, G., Noyes, R.~W., Kov{\'a}cs, G., {et~al.} 2004, Publications of the
  Astronomical Society of the Pacific, 116, 266, \dodoi{10.1086/382735}

\bibitem[{Bakos {et~al.}(2013)Bakos, Csubry, Penev, Bayliss, Jord{\'a}n,
  Afonso, Hartman, Henning, Kov{\'a}cs, Noyes, B{\'e}ky, Suc, Cs{\'a}k, Rabus,
  L{\'a}z{\'a}r, Papp, S{\'a}ri, Conroy, Zhou, Sackett, Schmidt, Mancini,
  Sasselov, \& Ueltzhoeffer}]{HATS_Bakos2013}
Bakos, G.~{\'A}., Csubry, Z., Penev, K., {et~al.} 2013, Publications of the
  Astronomical Society of the Pacific, 125, 154, \dodoi{10.1086/669529}

\bibitem[{Baranne {et~al.}(1996)Baranne, Queloz, Mayor, Adrianzyk, Knispel,
  Kohler, Lacroix, Meunier, Rimbaud, \& Vin}]{Baranne1996}
Baranne, A., Queloz, D., Mayor, M., {et~al.} 1996, Astronomy and Astrophysics
  Supplement Series, 119, 373

\bibitem[{{Berta} {et~al.}(2012){Berta}, {Irwin}, {Charbonneau}, {Burke}, \&
  {Falco}}]{MEarth_Berta2012}
{Berta}, Z.~K., {Irwin}, J., {Charbonneau}, D., {Burke}, C.~J., \& {Falco},
  E.~E. 2012, \aj, 144, 145, \dodoi{10.1088/0004-6256/144/5/145}

\bibitem[{Brahm {et~al.}(2017)Brahm, Jord{\'a}n, \& Espinoza}]{CERES_Brahm2017}
Brahm, R., Jord{\'a}n, A., \& Espinoza, N. 2017, \pasp, 129, 034002,
  \dodoi{10.1088/1538-3873/aa5455}

\bibitem[{{Brahm} {et~al.}(2019){Brahm}, {Espinoza}, {Jord{\'a}n}, {Henning},
  {Sarkis}, {Jones}, {D{\'\i}az}, {Jenkins}, {Vanzi}, {Zapata}, {Petrovich},
  {Kossakowski}, {Rabus}, {Rojas}, \& {Torres}}]{WINE_Brahm2019}
{Brahm}, R., {Espinoza}, N., {Jord{\'a}n}, A., {et~al.} 2019, \aj, 158, 45,
  \dodoi{10.3847/1538-3881/ab279a}

\bibitem[{Brahm {et~al.}(2020)Brahm, Nielsen, Wittenmyer, Wang, Rodriguez,
  Espinoza, Jones, Jord{\'a}n, Henning, Hobson, Kossakowski, Rojas, Sarkis,
  Schlecker, Trifonov, Shahaf, Ricker, Vanderspek, Latham, Seager, Winn,
  Jenkins, Addison, Bakos, Bhatti, Bayliss, Berlind, Bieryla, Bouchy, Bowler,
  Brice{\~n}o, Brown, Bryant, Caldwell, Charbonneau, Collins, Davis, Esquerdo,
  Fulton, Guerrero, Henze, Hogan, Horner, Huang, Irwin, Kane, Kielkopf, Mann,
  Mazeh, McCormac, McCully, Mengel, Mireles, Okumura, Plavchan, Quinn, Rabus,
  Saesen, Schlieder, Segransan, Shiao, Shporer, Siverd, Stassun, Suc, Tan,
  Torres, Tinney, Udry, Vanzi, Vezie, Vines, Vuckovic, Wright, Yahalomi,
  Zapata, Zhang, \& Ziegler}]{Brahm2020}
Brahm, R., Nielsen, L.~D., Wittenmyer, R.~A., {et~al.} 2020, The Astronomical
  Journal, 160, 235, \dodoi{10.3847/1538-3881/abba3b}

\bibitem[{Brown {et~al.}(2021)Brown, Vallenari, Prusti, de~Bruijne, Babusiaux,
  Biermann, Creevey, Evans, Eyer, Hutton, Jansen, Jordi, Klioner, Lammers,
  Lindegren, Luri, Mignard, Panem, Pourbaix, Randich, Sartoretti, Soubiran,
  Walton, Arenou, {Bailer-Jones}, Bastian, Cropper, Drimmel, Katz, Lattanzi,
  van Leeuwen, Bakker, Cacciari, Casta{\~n}eda, Angeli, Ducourant, Fabricius,
  Fouesneau, Fr{\'e}mat, Guerra, Guerrier, Guiraud, Piccolo, Masana, Messineo,
  Mowlavi, Nicolas, Nienartowicz, Pailler, Panuzzo, Riclet, Roux, Seabroke,
  Sordo, Tanga, Th{\'e}venin, {Gracia-Abril}, Portell, Teyssier, Altmann,
  Andrae, {Bellas-Velidis}, Benson, Berthier, Blomme, Brugaletta, Burgess,
  Busso, Carry, Cellino, Cheek, Clementini, Damerdji, Davidson, Delchambre,
  Dell'Oro, {Fern{\'a}ndez-Hern{\'a}ndez}, Galluccio, {Garc{\'i}a-Lario},
  {Garcia-Reinaldos}, {Gonz{\'a}lez-N{\'u}{\~n}ez}, Gosset, Haigron, Halbwachs,
  Hambly, Harrison, Hatzidimitriou, Heiter, Hern{\'a}ndez, Hestroffer, Hodgkin,
  Holl, Jan{\ss}en, de~Fombelle, Jordan, {Krone-Martins}, Lanzafame,
  L{\"o}ffler, Lorca, Manteiga, Marchal, Marrese, Moitinho, Mora, Muinonen,
  Osborne, Pancino, Pauwels, Petit, {Recio-Blanco}, Richards, Riello,
  Rimoldini, Robin, Roegiers, Rybizki, Sarro, Siopis, Smith, Sozzetti, Ulla,
  Utrilla, van Leeuwen, van Reeven, Abbas, Aramburu, Accart, Aerts, Aguado,
  Ajaj, Altavilla, {\'A}lvarez, {Cid-Fuentes}, Alves, Anderson, Varela, Antoja,
  Audard, Baines, Baker, {Balaguer-N{\'u}{\~n}ez}, Balbinot, Balog, Barache,
  Barbato, Barros, Barstow, Bartolom{\'e}, Bassilana, Bauchet,
  {Baudesson-Stella}, Becciani, Bellazzini, Bernet, Bertone, Bianchi,
  {Blanco-Cuaresma}, Boch, Bombrun, Bossini, Bouquillon, Bragaglia, Bramante,
  Breedt, Bressan, Brouillet, Bucciarelli, Burlacu, Busonero, Butkevich, Buzzi,
  Caffau, Cancelliere, C{\'a}novas, {Cantat-Gaudin}, Carballo, Carlucci,
  Carnerero, Carrasco, Casamiquela, Castellani, {Castro-Ginard}, Sampol,
  Chaoul, Charlot, Chemin, Chiavassa, Cioni, Comoretto, Cooper, Cornez, Cowell,
  Crifo, Crosta, Crowley, Dafonte, Dapergolas, David, David, de~Laverny, Luise,
  March, Ridder, de~Souza, de~Teodoro, de~Torres, del Peloso, del Pozo, Delbo,
  Delgado, Delgado, Delisle, Matteo, Diakite, Diener, Distefano, Dolding,
  Eappachen, Edvardsson, Enke, Esquej, Fabre, Fabrizio, Faigler, Fedorets,
  Fernique, Fienga, Figueras, Fouron, Fragkoudi, Fraile, Franke, Gai, Garabato,
  {Garcia-Gutierrez}, {Garc{\'i}a-Torres}, Garofalo, Gavras, Gerlach, Geyer,
  Giacobbe, Gilmore, Girona, Giuffrida, Gomel, Gomez, {Gonzalez-Santamaria},
  {Gonz{\'a}lez-Vidal}, Granvik, {Guti{\'e}rrez-S{\'a}nchez}, Guy, Hauser,
  Haywood, Helmi, Hidalgo, Hilger, H{\l}adczuk, Hobbs, Holland, Huckle,
  Jasniewicz, Jonker, Campillo, Julbe, Karbevska, Kervella, Khanna, Kochoska,
  Kontizas, Kordopatis, Korn, {Kostrzewa-Rutkowska}, Kruszy{\'n}ska, Lambert,
  Lanza, Lasne, Campion, Fustec, Lebreton, Lebzelter, Leccia, Leclerc,
  {Lecoeur-Taibi}, Liao, Licata, Lindstr{\o}m, Lister, Livanou, Lobel, Pardo,
  Managau, Mann, Marchant, Marconi, Santos, Marinoni, Marocco, Marshall, Polo,
  {Mart{\'i}n-Fleitas}, Masip, Massari, {Mastrobuono-Battisti}, Mazeh,
  McMillan, Messina, Michalik, Millar, Mints, Molina, Molinaro, Moln{\'a}r,
  Montegriffo, Mor, Morbidelli, Morel, Morris, Mulone, Munoz, Muraveva, Murphy,
  Musella, Noval, Ord{\'e}novic, Orr{\`u}, Osinde, Pagani, Pagano, Palaversa,
  Palicio, Panahi, Pawlak, Esteller, Penttil{\"a}, Piersimoni, Pineau, Plachy,
  Plum, Poggio, Poretti, Poujoulet, Pr{\v s}a, Pulone, Racero, Ragaini, Rainer,
  Raiteri, Rambaux, Ramos, {Ramos-Lerate}, Fiorentin, Regibo, Reyl{\'e},
  Ripepi, Riva, Rixon, Robichon, Robin, Roelens, Rohrbasser,
  {Romero-G{\'o}mez}, Rowell, Royer, Rybicki, Sadowski, Sell{\'e}s, Sahlmann,
  Salgado, Salguero, Samaras, Gimenez, Sanna, Santove{\~n}a, Sarasso,
  Schultheis, Sciacca, Segol, Segovia, S{\'e}gransan, Semeux, Shahaf, Siddiqui,
  Siebert, Siltala, Slezak, Smart, Solano, Solitro, Souami, Souchay, Spagna,
  Spoto, Steele, Steidelm{\"u}ller, Stephenson, S{\"u}veges, Szabados,
  {Szegedi-Elek}, Taris, Tauran, Taylor, Teixeira, Thuillot, Tonello, Torra,
  Torra, Turon, Unger, Vaillant, van Dillen, Vanel, Vecchiato, Viala, Vicente,
  Voutsinas, Weiler, Wevers, Wyrzykowski, Yoldas, Yvard, Zhao, Zorec, Zucker,
  Zurbach, \& Zwitter}]{GaiaEDR3_Brown2021}
Brown, A. G.~A., Vallenari, A., Prusti, T., {et~al.} 2021, Astronomy \&
  Astrophysics, 649, A1, \dodoi{10.1051/0004-6361/202039657}

\bibitem[{Brown {et~al.}(2013)Brown, Baliber, Bianco, Bowman, Burleson, Conway,
  Crellin, Depagne, Vera, Dilday, Dragomir, Dubberley, Eastman, Elphick,
  Falarski, Foale, Ford, Fulton, Garza, Gomez, Graham, Greene, Haldeman,
  Hawkins, Haworth, Haynes, Hidas, Hjelstrom, Howell, Hygelund, Lister,
  Lobdill, Martinez, Mullins, Norbury, Parrent, Paulson, Petry, Pickles,
  Posner, Rosing, Ross, Sand, Saunders, Shobbrook, Shporer, Street, Thomas,
  Tsapras, Tufts, Valenti, Horst, Walker, White, \& Willis}]{LCOGT_Brown2013}
Brown, T.~M., Baliber, N., Bianco, F.~B., {et~al.} 2013, Publications of the
  Astronomical Society of the Pacific, 125, 1031, \dodoi{10.1086/673168}

\bibitem[{Buchhave {et~al.}(2010)Buchhave, Bakos, Hartman, Torres, Kov{\'a}cs,
  Latham, Noyes, Esquerdo, Everett, Howard, Marcy, Fischer, Johnson, Andersen,
  F{\textbackslash}Hur{\'e}sz, Perumpilly, Sasselov, Stefanik, B{\'e}ky,
  L{\'a}z{\'a}r, Papp, \& S{\'a}ri}]{Buchhave2010}
Buchhave, L.~A., Bakos, G.~{\'A}., Hartman, J.~D., {et~al.} 2010, The
  Astrophysical Journal, 720, 1118, \dodoi{10.1088/0004-637X/720/2/1118}

\bibitem[{Buchhave {et~al.}(2012)Buchhave, Latham, Johansen, Bizzarro, Torres,
  Rowe, Batalha, Borucki, Brugamyer, Caldwell, Bryson, Ciardi, Cochran, Endl,
  Esquerdo, Ford, Geary, Gilliland, Hansen, Isaacson, Laird, Lucas, Marcy,
  Morse, Robertson, Shporer, Stefanik, Still, \& Quinn}]{SPC_Buchhave2012}
Buchhave, L.~A., Latham, D.~W., Johansen, A., {et~al.} 2012, Nature, 486, 375,
  \dodoi{10.1038/nature11121}

\bibitem[{{Butler} {et~al.}(1996){Butler}, {Marcy}, {Williams}, {McCarthy},
  {Dosanjh}, \& {Vogt}}]{PFS_Butler1996}
{Butler}, R.~P., {Marcy}, G.~W., {Williams}, E., {et~al.} 1996, \pasp, 108,
  500, \dodoi{10.1086/133755}

\bibitem[{Caldwell {et~al.}(2020)Caldwell, Tenenbaum, Twicken, Jenkins, Ting,
  Smith, Hedges, Fausnaugh, Rose, \& Burke}]{TESS_SPOC_Caldwell2020}
Caldwell, D.~A., Tenenbaum, P., Twicken, J.~D., {et~al.} 2020, Research Notes
  of the AAS, 4, 201, \dodoi{10.3847/2515-5172/abc9b3}

\bibitem[{Choi {et~al.}(2016)Choi, Dotter, Conroy, Cantiello, Paxton, \&
  Johnson}]{MISTI_Choi2016}
Choi, J., Dotter, A., Conroy, C., {et~al.} 2016, The Astrophysical Journal,
  823, 102, \dodoi{10.3847/0004-637X/823/2/102}

\bibitem[{Claret(2017)}]{Claret2017}
Claret, A. 2017, Astronomy \& Astrophysics, 600, A30,
  \dodoi{10.1051/0004-6361/201629705}

\bibitem[{Claret \& Bloemen(2011)}]{Claret2011}
Claret, A., \& Bloemen, S. 2011, Astronomy \& Astrophysics, 529, A75,
  \dodoi{10.1051/0004-6361/201116451}

\bibitem[{{Coelho} {et~al.}(2005){Coelho}, {Barbuy}, {Mel{\'e}ndez},
  {Schiavon}, \& {Castilho}}]{Coelho2005}
{Coelho}, P., {Barbuy}, B., {Mel{\'e}ndez}, J., {Schiavon}, R.~P., \&
  {Castilho}, B.~V. 2005, \aap, 443, 735, \dodoi{10.1051/0004-6361:20053511}

\bibitem[{{Collins} {et~al.}(2018){Collins}, {Quinn}, {Latham}, {Christiansen},
  {Ciardi}, {Dragomir}, {Crossfield}, \& {Seager}}]{TFOP_Collins2018}
{Collins}, K., {Quinn}, S.~N., {Latham}, D.~W., {et~al.} 2018, in American
  Astronomical Society Meeting Abstracts, Vol. 231, American Astronomical
  Society Meeting Abstracts \#231, 439.08

\bibitem[{Collins {et~al.}(2017)Collins, Kielkopf, Stassun, \&
  Hessman}]{AstroImageJ_Collins17}
Collins, K.~A., Kielkopf, J.~F., Stassun, K.~G., \& Hessman, F.~V. 2017, \aj,
  153, 77, \dodoi{10.3847/1538-3881/153/2/77}

\bibitem[{Collins {et~al.}(2018)Collins, Collins, Pepper, {Labadie-Bartz},
  Stassun, Gaudi, Bayliss, Bento, COL{\'O}N, Feliz, James, Johnson, Kuhn, Lund,
  Penny, Rodriguez, Siverd, Stevens, Yao, Zhou, Akshay, Aldi, Ashcraft,
  Awiphan, Ba{\c s}t{\"u}rk, Baker, Beatty, Benni, Berlind, Berriman,
  {Berta-Thompson}, Bieryla, Bozza, Novati, Calkins, Cann, Ciardi, Clark,
  Cochran, Cohen, Conti, Crepp, Curtis, D'Ago, Diazeguigure, Dressing, Dubois,
  Ellingson, Ellis, Esquerdo, Evans, Friedli, Fukui, Fulton, Gonzales, Good,
  Gregorio, Gumusayak, Hancock, Harada, Hart, Hintz, {Jang-Condell}, Jeffery,
  Jensen, Jofr{\'e}, Joner, Kar, Kasper, Keten, Kielkopf, Komonjinda, Kotnik,
  Latham, Leuquire, Lewis, Logie, Lowther, Macqueen, Martin, Mawet, Mcleod,
  Murawski, Narita, Nordhausen, Oberst, Odden, Panka, Petrucci, Plavchan,
  Quinn, Rau, Reed, Relles, Renaud, Scarpetta, Sorber, Spencer, Spencer,
  Stephens, Stockdale, Tan, Trueblood, Trueblood, Vanaverbeke, Villanueva,
  Warner, West, Yal{\c c}{\i}nkaya, Yeigh, \& Zambelli}]{Collins2018}
Collins, K.~A., Collins, K.~I., Pepper, J., {et~al.} 2018, The Astronomical
  Journal, 156, 234, \dodoi{10.3847/1538-3881/aae582}

\bibitem[{{Crane} {et~al.}(2006){Crane}, {Shectman}, \&
  {Butler}}]{PFS_Crane2006}
{Crane}, J.~D., {Shectman}, S.~A., \& {Butler}, R.~P. 2006, in Society of
  Photo-Optical Instrumentation Engineers (SPIE) Conference Series, Vol. 6269,
  Society of Photo-Optical Instrumentation Engineers (SPIE) Conference Series,
  ed. I.~S. {McLean} \& M.~{Iye}, 626931, \dodoi{10.1117/12.672339}

\bibitem[{{Crane} {et~al.}(2010){Crane}, {Shectman}, {Butler}, {Thompson},
  {Birk}, {Jones}, \& {Burley}}]{PFS_Crane2010}
{Crane}, J.~D., {Shectman}, S.~A., {Butler}, R.~P., {et~al.} 2010, in Society
  of Photo-Optical Instrumentation Engineers (SPIE) Conference Series, Vol.
  7735, Ground-based and Airborne Instrumentation for Astronomy III, ed. I.~S.
  {McLean}, S.~K. {Ramsay}, \& H.~{Takami}, 773553, \dodoi{10.1117/12.857792}

\bibitem[{{Crane} {et~al.}(2008){Crane}, {Shectman}, {Butler}, {Thompson}, \&
  {Burley}}]{PFS_Crane2008}
{Crane}, J.~D., {Shectman}, S.~A., {Butler}, R.~P., {Thompson}, I.~B., \&
  {Burley}, G.~S. 2008, in Society of Photo-Optical Instrumentation Engineers
  (SPIE) Conference Series, Vol. 7014, Ground-based and Airborne
  Instrumentation for Astronomy II, ed. I.~S. {McLean} \& M.~M. {Casali},
  701479, \dodoi{10.1117/12.789637}

\bibitem[{Cutri {et~al.}(2003)Cutri, Skrutskie, {van Dyk}, Beichman, Carpenter,
  Chester, Cambresy, Evans, Fowler, Gizis, Howard, Huchra, Jarrett, Kopan,
  Kirkpatrick, Light, Marsh, McCallon, Schneider, Stiening, Sykes, Weinberg,
  Wheaton, Wheelock, \& Zacarias}]{TMASS_Cutri2003}
Cutri, R.~M., Skrutskie, M.~F., {van Dyk}, S., {et~al.} 2003, VizieR Online
  Data Catalog, II/246

\bibitem[{Cutri(2012)}]{WISE_Cutri2012}
Cutri, R. M.~e. 2012, VizieR Online Data Catalog, II/311

\bibitem[{Dalba {et~al.}(2020)Dalba, Fulton, Isaacson, Kane, \&
  Howard}]{CPS_Dalba2020a}
Dalba, P.~A., Fulton, B., Isaacson, H., Kane, S.~R., \& Howard, A.~W. 2020, The
  Astronomical Journal, 160, 149, \dodoi{10.3847/1538-3881/abad27}

\bibitem[{Davis {et~al.}(2020)Davis, Wang, Jones, Eastman, G{\"u}nther,
  Stassun, Addison, Collins, Quinn, Latham, Trifonov, Shahaf, Mazeh, Kane,
  Narita, Wang, Tan, Ciardi, Tokovinin, Ziegler, Tronsgaard, Millholland, Cruz,
  Berlind, Calkins, Esquerdo, Collins, Conti, Murgas, Evans, Lewin, Radford,
  Paredes, Henry, {Hodari-Sadiki}, Lund, Christiansen, Law, Mann, Brice{\~n}o,
  Parviainen, Palle, Watanabe, Ricker, Vanderspek, Seager, Winn, Jenkins,
  Krishnamurthy, Batalha, Burt, Col{\'o}n, Dynes, Caldwell, Morris, Henze, \&
  Fischer}]{Davis2020}
Davis, A.~B., Wang, S., Jones, M., {et~al.} 2020, The Astronomical Journal,
  160, 229, \dodoi{10.3847/1538-3881/aba49d}

\bibitem[{Dawson \& Johnson(2018)}]{Dawson2018}
Dawson, R.~I., \& Johnson, J.~A. 2018, Annual Review of Astronomy and
  Astrophysics, 56, 175, \dodoi{10.1146/annurev-astro-081817-051853}

\bibitem[{Demory \& Seager(2011)}]{Demory2011}
Demory, B.-O., \& Seager, S. 2011, The Astrophysical Journal Supplement Series,
  197, 12, \dodoi{10.1088/0067-0049/197/1/12}

\bibitem[{Donati {et~al.}(1997)Donati, Semel, Carter, Rees, \&
  Cameron}]{CHIRON_Donati1997}
Donati, J.-F., Semel, M., Carter, B.~D., Rees, D.~E., \& Cameron, A.~C. 1997,
  Monthly Notices of the Royal Astronomical Society, 291, 658,
  \dodoi{10.1093/mnras/291.4.658}

\bibitem[{Dong {et~al.}(2021)Dong, Huang, Zhou, Dawson, Rodriguez, Eastman,
  Collins, Quinn, Shporer, Triaud, Wang, Beatty, Jackson, Collins, Abe, Suarez,
  Crouzet, MeKarnia, Dransfield, Jensen, Stockdale, Barkaoui, Heitzmann,
  Wright, Addison, Wittenmyer, Okumura, Bowler, Horner, Kane, Kielkopf, Liu,
  Plavchan, Mengel, Ricker, Vanderspek, Latham, Seager, Winn, Jenkins,
  Christiansen, \& Paegert}]{Dong2021a}
Dong, J., Huang, C.~X., Zhou, G., {et~al.} 2021, arXiv:2109.03771 [astro-ph].
\newblock \doarXiv{2109.03771}

\bibitem[{Dotter(2016)}]{MIST0_Dotter2016}
Dotter, A. 2016, The Astrophysical Journal Supplement Series, 222, 8,
  \dodoi{10.3847/0067-0049/222/1/8}

\bibitem[{Eastman {et~al.}(2013)Eastman, Gaudi, \& Agol}]{ExoFAST_Eastman2013}
Eastman, J., Gaudi, B.~S., \& Agol, E. 2013, Publications of the Astronomical
  Society of the Pacific, 125, 83, \dodoi{10.1086/669497}

\bibitem[{{Eastman} {et~al.}(2019){Eastman}, {Rodriguez}, {Agol}, {Stassun},
  {Beatty}, {Vanderburg}, {Gaudi}, {Collins}, \& {Luger}}]{ExoFASTv2_Eastman19}
{Eastman}, J.~D., {Rodriguez}, J.~E., {Agol}, E., {et~al.} 2019, arXiv
  e-prints, arXiv:1907.09480.
\newblock \doarXiv{1907.09480}

\bibitem[{{ExoFOP}(2019)}]{ExoFoPTESS}
{ExoFOP}. 2019, Exoplanet Follow-up Observing Program - TESS,  IPAC,
  \dodoi{10.26134/EXOFOP3}

\bibitem[{Fabrycky \& Tremaine(2007)}]{Fabrycky2007}
Fabrycky, D., \& Tremaine, S. 2007, The Astrophysical Journal, 669, 1298,
  \dodoi{10.1086/521702}

\bibitem[{Fausnaugh {et~al.}(2020)Fausnaugh, Burke, Ricker, \&
  Vanderspek}]{TESS_QLP_Fausnaugh2020}
Fausnaugh, M.~M., Burke, C.~J., Ricker, G.~R., \& Vanderspek, R. 2020, Research
  Notes of the AAS, 4, 251, \dodoi{10.3847/2515-5172/abd63a}

\bibitem[{Feinstein {et~al.}(2019)Feinstein, Montet, {Foreman-Mackey}, Bedell,
  Saunders, Bean, Christiansen, Hedges, Luger, Scolnic, \&
  Cardoso}]{Feinstein2019}
Feinstein, A.~D., Montet, B.~T., {Foreman-Mackey}, D., {et~al.} 2019,
  Publications of the Astronomical Society of the Pacific, 131, 094502,
  \dodoi{10.1088/1538-3873/ab291c}

\bibitem[{{F\H{u}r{\'e}sz}(2008)}]{TRES_Furesz2008}
{F\H{u}r{\'e}sz}, G. 2008, PhD thesis, University of Szeged, Hungary

\bibitem[{Fortney {et~al.}(2021)Fortney, Dawson, \& Komacek}]{Fortney2021}
Fortney, J.~J., Dawson, R.~I., \& Komacek, T.~D. 2021, Journal of Geophysical
  Research: Planets, 126, e2020JE006629, \dodoi{10.1029/2020JE006629}

\bibitem[{Gaudi \& Winn(2007)}]{Gaudi2007}
Gaudi, B.~S., \& Winn, J.~N. 2007, The Astrophysical Journal, 655, 550,
  \dodoi{10.1086/509910}

\bibitem[{Gavel {et~al.}(2014)Gavel, Kupke, Dillon, Norton, Ratliff, Cabak,
  Phillips, Rockosi, McGurk, Srinath, Peck, Deich, Lanclos, Gates, Saylor,
  Ward, \& Pfister}]{ShaneAO_Gavel2014}
Gavel, D., Kupke, R., Dillon, D., {et~al.} 2014, in Adaptive {{Optics Systems
  IV}}, Vol. 9148 ({SPIE}), 38--48, \dodoi{10.1117/12.2055256}

\bibitem[{{Gelman} \& {Rubin}(1992)}]{GelmanRubin}
{Gelman}, A., \& {Rubin}, D.~B. 1992, Statistical Science, 7, 457,
  \dodoi{10.1214/ss/1177011136}

\bibitem[{Goldreich \& Soter(1966)}]{Goldreich1966}
Goldreich, P., \& Soter, S. 1966, Icarus, 5, 375,
  \dodoi{10.1016/0019-1035(66)90051-0}

\bibitem[{Guerrero {et~al.}(2021)Guerrero, Seager, Huang, Vanderburg, Soto,
  Mireles, Hesse, Fong, Glidden, Shporer, Latham, Collins, Quinn, Burt,
  Dragomir, Crossfield, Vanderspek, Fausnaugh, Burke, Ricker, Daylan, Essack,
  G{\"u}nther, Osborn, Pepper, Rowden, Sha, Jr, Yahalomi, Yu, Ballard, Batalha,
  Berardo, Chontos, Dittmann, Esquerdo, {Mikal-Evans}, Jayaraman,
  Krishnamurthy, Louie, Mehrle, Niraula, Rackham, Rodriguez, Rowden,
  {Sousa-Silva}, Watanabe, Wong, Zhan, Zivanovic, Christiansen, Ciardi, Swain,
  Lund, Mullally, Fleming, Rodriguez, Boyd, Quintana, Barclay, Col{\'o}n,
  Rinehart, Schlieder, Clampin, Jenkins, Twicken, Caldwell, Coughlin, Henze,
  Lissauer, Morris, Rose, Smith, Tenenbaum, Ting, Wohler, Bakos, Bean,
  {Berta-Thompson}, Bieryla, Bouma, Buchhave, Butler, Charbonneau, Doty, Ge,
  Holman, Howard, Kaltenegger, Kane, Kjeldsen, Kreidberg, Lin, Minsky, Narita,
  Paegert, P{\'a}l, Palle, Sasselov, Spencer, Sozzetti, Stassun, Torres, Udry,
  \& Winn}]{TESS_TOIs_Guerrero2021}
Guerrero, N.~M., Seager, S., Huang, C.~X., {et~al.} 2021, The Astrophysical
  Journal Supplement Series, 254, 39, \dodoi{10.3847/1538-4365/abefe1}

\bibitem[{Halverson {et~al.}(2016)Halverson, Terrien, Mahadevan, Roy, Bender,
  Stef{\'a}nsson, Monson, Levi, Hearty, Blake, McElwain, Schwab, Ramsey,
  Wright, Wang, Gong, \& Roberston}]{NEID_Halverson2016}
Halverson, S., Terrien, R., Mahadevan, S., {et~al.} 2016, in Ground-Based and
  {{Airborne Instrumentation}} for {{Astronomy VI}}, Vol. 9908 ({SPIE}),
  2022--2041, \dodoi{10.1117/12.2232761}

\bibitem[{Harris {et~al.}(2020)Harris, Millman, van~der Walt, Gommers,
  Virtanen, Cournapeau, Wieser, Taylor, Berg, Smith, Kern, Picus, Hoyer, van
  Kerkwijk, Brett, Haldane, del R{'{\i}}o, Wiebe, Peterson,
  G{'{e}}rard-Marchant, Sheppard, Reddy, Weckesser, Abbasi, Gohlke, \&
  Oliphant}]{Numpy}
Harris, C.~R., Millman, K.~J., van~der Walt, S.~J., {et~al.} 2020, Nature, 585,
  357, \dodoi{10.1038/s41586-020-2649-2}

\bibitem[{Hartman {et~al.}(2016)Hartman, Bakos, Bhatti, Penev, Bieryla, Latham,
  Kov{\'a}cs, Torres, Csubry, de~{Val-Borro}, Buchhave, Kov{\'a}cs, Quinn,
  Howard, Isaacson, Fulton, Everett, Esquerdo, B{\'e}ky, Szklenar, Falco,
  Santerne, Boisse, H{\'e}brard, Burrows, L{\'a}z{\'a}r, Papp, \&
  S{\'a}ri}]{Hartman2016}
Hartman, J.~D., Bakos, G.~{\'A}., Bhatti, W., {et~al.} 2016, The Astronomical
  Journal, 152, 182, \dodoi{10.3847/0004-6256/152/6/182}

\bibitem[{Hartman {et~al.}(2019)Hartman, Bakos, Bayliss, Bento, Bhatti, Brahm,
  Csubry, Espinoza, Henning, Jord{\'a}n, Mancini, Penev, Rabus, Sarkis, Suc,
  de~{Val-Borro}, Zhou, Addison, Arriagada, Butler, Crane, Durkan, Shectman,
  Tan, Thompson, Tinney, Wright, L{\'a}z{\'a}r, Papp, \&
  S{\'a}ri}]{Hartman2019}
Hartman, J.~D., Bakos, G.~{\'A}., Bayliss, D., {et~al.} 2019, The Astronomical
  Journal, 157, 55, \dodoi{10.3847/1538-3881/aaf8b6}

\bibitem[{Hayward {et~al.}(2001)Hayward, Brandl, Pirger, Blacken, Gull,
  Schoenwald, \& Houck}]{PHARO_Hayward2001}
Hayward, T.~L., Brandl, B., Pirger, B., {et~al.} 2001, Publications of the
  Astronomical Society of the Pacific, 113, 105, \dodoi{10.1086/317969}

\bibitem[{H{\o}g {et~al.}(2000)H{\o}g, Fabricius, Makarov, Urban, Corbin,
  Wycoff, Bastian, Schwekendiek, \& Wicenec}]{Tycho2_Hog2000}
H{\o}g, E., Fabricius, C., Makarov, V.~V., {et~al.} 2000, Astronomy and
  Astrophysics, 355, L27

\bibitem[{Horch {et~al.}(2014)Horch, Howell, Everett, \& Ciardi}]{Horch2014}
Horch, E.~P., Howell, S.~B., Everett, M.~E., \& Ciardi, D.~R. 2014, The
  Astrophysical Journal, 795, 60, \dodoi{10.1088/0004-637X/795/1/60}

\bibitem[{Howard \& Fulton(2016)}]{CPS_Howard2016}
Howard, A.~W., \& Fulton, B.~J. 2016, Publications of the Astronomical Society
  of the Pacific, 128, 114401, \dodoi{10.1088/1538-3873/128/969/114401}

\bibitem[{Howard {et~al.}(2010)Howard, Johnson, Marcy, Fischer, Wright, Bernat,
  Henry, Peek, Isaacson, Apps, Endl, Cochran, Valenti, Anderson, \&
  Piskunov}]{CPS_Howard2010}
Howard, A.~W., Johnson, J.~A., Marcy, G.~W., {et~al.} 2010, The Astrophysical
  Journal, 721, 1467, \dodoi{10.1088/0004-637X/721/2/1467}

\bibitem[{Huang {et~al.}(2020{\natexlab{a}})Huang, Vanderburg, P{\'a}l, Sha,
  Yu, Fong, Fausnaugh, Shporer, Guerrero, Vanderspek, \&
  Ricker}]{TESS_QLP_Huang2020a}
Huang, C.~X., Vanderburg, A., P{\'a}l, A., {et~al.} 2020{\natexlab{a}},
  Research Notes of the AAS, 4, 204, \dodoi{10.3847/2515-5172/abca2e}

\bibitem[{Huang {et~al.}(2020{\natexlab{b}})Huang, Vanderburg, P{\'a}l, Sha,
  Yu, Fong, Fausnaugh, Shporer, Guerrero, Vanderspek, \&
  Ricker}]{TESS_QLP_Huang2020b}
---. 2020{\natexlab{b}}, Research Notes of the AAS, 4, 206,
  \dodoi{10.3847/2515-5172/abca2d}

\bibitem[{{Hunter}(2007)}]{Matplotlib}
{Hunter}, J.~D. 2007, Computing in Science Engineering, 9, 90,
  \dodoi{10.1109/MCSE.2007.55}

\bibitem[{{Ikwut-Ukwa} {et~al.}(2021){Ikwut-Ukwa}, Rodriguez, Quinn, Zhou,
  Vanderburg, Ali, Bunten, Gaudi, Latham, Howell, Huang, Bieryla, Collins,
  Carmichael, Rabus, Eastman, Collins, Tan, Schwarz, Myers, Stockdale,
  Kielkopf, Radford, Oelkers, Jenkins, Ricker, Seager, Vanderspek, Winn, Burt,
  Butler, Calkins, Crane, Gnilka, Esquerdo, Fong, Kreidberg, Mink, Rodriguez,
  Schlieder, Shectman, Shporer, Teske, Ting, Villase{\~n}or, \&
  Yahalomi}]{Ikwut-Ukwa2021}
{Ikwut-Ukwa}, M., Rodriguez, J.~E., Quinn, S.~N., {et~al.} 2021, The
  Astronomical Journal, 163, 9, \dodoi{10.3847/1538-3881/ac2ee1}

\bibitem[{Irwin {et~al.}(2007)Irwin, Irwin, Aigrain, Hodgkin, Hebb, \&
  Moraux}]{MEarth_Irwin2007}
Irwin, J., Irwin, M., Aigrain, S., {et~al.} 2007, Monthly Notices of the Royal
  Astronomical Society, 375, 1449, \dodoi{10.1111/j.1365-2966.2006.11408.x}

\bibitem[{Jenkins {et~al.}(2016)Jenkins, Twicken, McCauliff, Campbell,
  Sanderfer, Lung, {Mansouri-Samani}, Girouard, Tenenbaum, Klaus, Smith,
  Caldwell, Chacon, Henze, Heiges, Latham, Morgan, Swade, Rinehart, \&
  Vanderspek}]{TESS_SPOC_Jenkins2016}
Jenkins, J.~M., Twicken, J.~D., McCauliff, S., {et~al.} 2016, in Software and
  {{Cyberinfrastructure}} for {{Astronomy IV}}, Vol. 9913 ({SPIE}), 1232--1251,
  \dodoi{10.1117/12.2233418}

\bibitem[{{Jensen}(2013)}]{TAPIR_Jensen2013}
{Jensen}, E. 2013, {Tapir: A web interface for transit/eclipse observability},
  Astrophysics Source Code Library.
\newblock \doeprint{1306.007}

\bibitem[{{Kaufer} {et~al.}(1999){Kaufer}, {Stahl}, {Tubbesing},
  {N{\o}rregaard}, {Avila}, {Francois}, {Pasquini}, \&
  {Pizzella}}]{FEROS_Kaufer1999}
{Kaufer}, A., {Stahl}, O., {Tubbesing}, S., {et~al.} 1999, The Messenger, 95, 8

\bibitem[{Knudstrup {et~al.}(2022)Knudstrup, Serrano, Gandolfi, Albrecht,
  Cochran, Endl, Macqueen, Tronsgaard, Bieryla, Buchhave, Stassun, Collins,
  Nowak, Deeg, Barkaoui, Safonov, Strakhov, Belinski, Twicken, Jenkins, Howard,
  Isaacson, Winn, Collins, Conti, Furesz, Gan, Kielkopf, Massey, Murgas,
  Murphy, Palle, Quinn, Reed, Ricker, Seager, Shiao, Schwartz, Srdoc, \&
  Watanabe}]{Knudstrup2022}
Knudstrup, E., Serrano, L.~M., Gandolfi, D., {et~al.} 2022, arXiv:2204.13956
  [astro-ph], \dodoi{10.48550/arXiv.2204.13956}

\bibitem[{Kostov {et~al.}(2019)Kostov, Mullally, Quintana, Coughlin, Mullally,
  Barclay, Col{\'o}n, Schlieder, Barentsen, \& Burke}]{DAVE_Kostov2019}
Kostov, V.~B., Mullally, S.~E., Quintana, E.~V., {et~al.} 2019, The
  Astronomical Journal, 157, 124, \dodoi{10.3847/1538-3881/ab0110}

\bibitem[{Kov{\'a}cs {et~al.}(2002)Kov{\'a}cs, Zucker, \&
  Mazeh}]{BLS_Kovacs2002}
Kov{\'a}cs, G., Zucker, S., \& Mazeh, T. 2002, Astronomy \& Astrophysics, 391,
  369, \dodoi{10.1051/0004-6361:20020802}

\bibitem[{Kruse {et~al.}(2019)Kruse, Agol, Luger, \&
  {Foreman-Mackey}}]{QATS_Kruse2019}
Kruse, E., Agol, E., Luger, R., \& {Foreman-Mackey}, D. 2019, The Astrophysical
  Journal Supplement Series, 244, 11, \dodoi{10.3847/1538-4365/ab346b}

\bibitem[{Kunimoto {et~al.}(2021)Kunimoto, Huang, Tey, Fong, Hesse, Shporer,
  Guerrero, Fausnaugh, Vanderspek, \& Ricker}]{TESS_QLP_Kunimoto2021}
Kunimoto, M., Huang, C., Tey, E., {et~al.} 2021, \rnaas, 5, 234,
  \dodoi{10.3847/2515-5172/ac2ef0}

\bibitem[{Kunimoto {et~al.}(2022)Kunimoto, Daylan, Guerrero, Fong, Bryson,
  Ricker, Fausnaugh, Huang, Sha, Shporer, Vanderburg, Vanderspek, \&
  Yu}]{TESS_Faint_Kunimoto2021b}
Kunimoto, M., Daylan, T., Guerrero, N., {et~al.} 2022, The Astrophysical
  Journal Supplement Series, 259, 33, \dodoi{10.3847/1538-4365/ac5688}

\bibitem[{Kupke {et~al.}(2012)Kupke, Gavel, Roskosi, Cabak, Cowley, Dillon,
  Gates, McGurk, Norton, Peck, Ratliff, \& Reinig}]{ShaneAO_Kupke2012}
Kupke, R., Gavel, D., Roskosi, C., {et~al.} 2012, in Adaptive {{Optics Systems
  III}}, Vol. 8447 ({SPIE}), 1190--1196, \dodoi{10.1117/12.926470}

\bibitem[{{Kurucz}(1993)}]{Kurucz1993}
{Kurucz}, R.~L. 1993, {SYNTHE spectrum synthesis programs and line data}

\bibitem[{{Lightkurve Collaboration} {et~al.}(2018){Lightkurve Collaboration},
  {Cardoso}, {Hedges}, {Gully-Santiago}, {Saunders}, {Cody}, {Barclay}, {Hall},
  {Sagear}, {Turtelboom}, {Zhang}, {Tzanidakis}, {Mighell}, {Coughlin}, {Bell},
  {Berta-Thompson}, {Williams}, {Dotson}, \& {Barentsen}}]{Lightkurve18}
{Lightkurve Collaboration}, {Cardoso}, J.~V.~d.~M., {Hedges}, C., {et~al.}
  2018, {Lightkurve: Kepler and TESS time series analysis in Python},
  Astrophysics Source Code Library.
\newblock \doeprint{1812.013}

\bibitem[{Lin \& Papaloizou(1986)}]{Lin1986}
Lin, D. N.~C., \& Papaloizou, J. 1986, The Astrophysical Journal, 309, 846,
  \dodoi{10.1086/164653}

\bibitem[{Lindegren {et~al.}(2021)Lindegren, Klioner, Hern{\'a}ndez, Bombrun,
  {Ramos-Lerate}, Steidelm{\"u}ller, Bastian, Biermann, de~Torres, Gerlach,
  Geyer, Hilger, Hobbs, Lammers, McMillan, Stephenson, Casta{\~n}eda, Davidson,
  Fabricius, {Gracia-Abril}, Portell, Rowell, Teyssier, Torra, Bartolom{\'e},
  Clotet, Garralda, {Gonz{\'a}lez-Vidal}, Torra, Abbas, Altmann, Varela,
  {Balaguer-N{\'u}{\~n}ez}, Balog, Barache, Becciani, Bernet, Bertone, Bianchi,
  Bouquillon, Brown, Bucciarelli, Busonero, Butkevich, Buzzi, Cancelliere,
  Carlucci, Charlot, Cioni, Crosta, Crowley, del Peloso, del Pozo, Drimmel,
  Esquej, Fienga, Fraile, Gai, {Garcia-Reinaldos}, Guerra, Hambly, Hauser,
  Jan{\ss}en, Jordan, {Kostrzewa-Rutkowska}, Lattanzi, Liao, Licata, Lister,
  L{\"o}ffler, Marchant, Masip, Mignard, Mints, Molina, Mora, Morbidelli,
  Murphy, Pagani, Panuzzo, Esteller, Poggio, Fiorentin, Riva, Sell{\'e}s,
  Gimenez, Sarasso, Sciacca, Siddiqui, Smart, Souami, Spagna, Steele, Taris,
  Utrilla, van Reeven, \& Vecchiato}]{GaiaEDR3_Lindegren2021}
Lindegren, L., Klioner, S.~A., Hern{\'a}ndez, J., {et~al.} 2021, Astronomy \&
  Astrophysics, 649, A2, \dodoi{10.1051/0004-6361/202039709}

\bibitem[{Lopez \& Fortney(2016)}]{Lopez2016}
Lopez, E.~D., \& Fortney, J.~J. 2016, The Astrophysical Journal, 818, 4,
  \dodoi{10.3847/0004-637X/818/1/4}

\bibitem[{{Lucy} \& {Sweeney}(1971)}]{LucySweeney71}
{Lucy}, L.~B., \& {Sweeney}, M.~A. 1971, \aj, 76, 544, \dodoi{10.1086/111159}

\bibitem[{{Mandel} \& {Agol}(2002)}]{Mandel02}
{Mandel}, K., \& {Agol}, E. 2002, \apjl, 580, L171, \dodoi{10.1086/345520}

\bibitem[{Matson {et~al.}(2018)Matson, Howell, Horch, \& Everett}]{Matson2018}
Matson, R.~A., Howell, S.~B., Horch, E.~P., \& Everett, M.~E. 2018, The
  Astronomical Journal, 156, 31, \dodoi{10.3847/1538-3881/aac778}

\bibitem[{Mayor \& Queloz(1995)}]{Mayor1995}
Mayor, M., \& Queloz, D. 1995, Nature, 378, 355, \dodoi{10.1038/378355a0}

\bibitem[{Mazeh {et~al.}(2016)Mazeh, Holczer, \& Faigler}]{Mazeh2016}
Mazeh, T., Holczer, T., \& Faigler, S. 2016, Astronomy \& Astrophysics, 589,
  A75, \dodoi{10.1051/0004-6361/201528065}

\bibitem[{McCullough {et~al.}(2005)McCullough, Stys, Valenti, Fleming, Janes,
  \& Heasley}]{XO_McCullough2005}
McCullough, P.~R., Stys, J.~E., Valenti, J.~A., {et~al.} 2005, Publications of
  the Astronomical Society of the Pacific, 117, 783, \dodoi{10.1086/432024}

\bibitem[{{McGurk} {et~al.}(2014){McGurk}, {Rockosi}, {Gavel}, {Kupke}, {Peck},
  {Pfister}, {Ward}, {Deich}, {Gates}, {Gates}, {Alcott}, {Cowley}, {Dillon},
  {Lanclos}, {Sandford}, {Saylor}, {Srinath}, {Weiss}, \&
  {Norton}}]{ShaneAO_McGurk2014}
{McGurk}, R., {Rockosi}, C., {Gavel}, D., {et~al.} 2014, in Society of
  Photo-Optical Instrumentation Engineers (SPIE) Conference Series, Vol. 9148,
  Adaptive Optics Systems IV, ed. E.~{Marchetti}, L.~M. {Close}, \& J.-P.
  {Vran}, 91483A, \dodoi{10.1117/12.2057027}

\bibitem[{McLaughlin(1924)}]{McLaughlin1924}
McLaughlin, D.~B. 1924, The Astrophysical Journal, 60, 22,
  \dodoi{10.1086/142826}

\bibitem[{Mireles {et~al.}(2021)Mireles, Hesse, Guerrero, Kunimoto, Huang, \&
  Fong}]{TESS_cTOIs_Mireles2021}
Mireles, I., Hesse, K., Guerrero, N., {et~al.} 2021, Posters from the TESS
  Science Conference II (TSC2), 128, \dodoi{10.5281/zenodo.5130661}

\bibitem[{Montalto {et~al.}(2020)Montalto, Borsato, Granata, Lacedelli,
  Malavolta, Manthopoulou, Nardiello, Nascimbeni, \& Piotto}]{Montalto2020}
Montalto, M., Borsato, L., Granata, V., {et~al.} 2020, Monthly Notices of the
  Royal Astronomical Society, 498, 1726, \dodoi{10.1093/mnras/staa2438}

\bibitem[{{Morton}(2015)}]{Isochrones_Morton2015}
{Morton}, T.~D. 2015, {isochrones: Stellar model grid package}.
\newblock \doeprint{1503.010}

\bibitem[{Moutou {et~al.}(2021)Moutou, Almenara, H{\'e}brard, Santos, Stassun,
  Deheuvels, Barros, Benni, Bieryla, Boisse, Bonfils, Boyd, Collins, Baker,
  {Cort{\'e}s-Zuleta}, Dalal, Debras, Deleuil, Delfosse, Demangeon, Essack,
  Forveille, Girardin, Guerra, Heidari, Hesse, Hoyer, Jenkins, Kiefer,
  K{\"o}nig, Laloum, Latham, Lopez, Martioli, Osborn, Ricker, Seager,
  Vanderspek, Vezie, Villase{\~n}or, Winn, Wohler, \& Ziegler}]{Moutou2021}
Moutou, C., Almenara, J.~M., H{\'e}brard, G., {et~al.} 2021, Astronomy \&
  Astrophysics, 653, A147, \dodoi{10.1051/0004-6361/202141151}

\bibitem[{{NASA Exoplanet Archive}(2022)}]{ExoplanetArchive_PSCompPars}
{NASA Exoplanet Archive}. 2022, Planetary Systems Composite Parameters,
  Version: 2022-02-14,  NExScI-Caltech/IPAC, \dodoi{10.26133/NEA13}

\bibitem[{Nielsen {et~al.}(2019)Nielsen, Bouchy, Turner, Giles, Mascare{\~n}o,
  Lovis, Marmier, Pepe, S{\'e}gransan, Udry, Otegi, Ottoni, Stalport, Ricker,
  Vanderspek, Latham, Seager, Winn, Jenkins, Kane, Wittenmyer, Bowler,
  Crossfield, Horner, Kielkopf, Morton, Plavchan, Tinney, Zhang, Wright,
  Mengel, Clark, Okumura, Addison, Caldwell, Cartwright, Collins, Francis,
  Guerrero, Huang, Matthews, Pepper, Rose, Villase{\~n}or, Wohler, Stassun,
  Howell, Ciardi, Gonzales, Matson, Beichman, \& Schlieder}]{Nielsen2019}
Nielsen, L.~D., Bouchy, F., Turner, O., {et~al.} 2019, Astronomy \&
  Astrophysics, 623, A100, \dodoi{10.1051/0004-6361/201834577}

\bibitem[{Nielsen {et~al.}(2020)Nielsen, Brahm, Bouchy, Espinoza, Turner,
  Rappaport, Pearce, Ricker, Vanderspek, Latham, Seager, Winn, Jenkins, Acton,
  Bakos, Barclay, Barkaoui, Bhatti, Brice{\~n}o, Bryant, Burleigh, Ciardi,
  Collins, Collins, Cooke, Csubry, dos Santos, Eigm{\"u}ller, Fausnaugh, Gan,
  Gillon, Goad, Guerrero, Hagelberg, Hart, Henning, Huang, Jehin, Jenkins,
  Jord{\'a}n, Kielkopf, Kossakowski, Lavie, Law, Lendl, de~Leon, Lovis, Mann,
  Marmier, McCormac, Mori, Moyano, Narita, Osip, Otegi, Pepe, Pozuelos,
  Raynard, Relles, Sarkis, S{\'e}gransan, Seidel, Shporer, Stalport, Stockdale,
  Suc, Tamura, Tan, Tilbrook, Ting, Trifonov, Udry, Vanderburg, Wheatley,
  Wingham, Zhan, \& Ziegler}]{Nielsen2020}
Nielsen, L.~D., Brahm, R., Bouchy, F., {et~al.} 2020, Astronomy \&
  Astrophysics, 639, A76, \dodoi{10.1051/0004-6361/202037941}

\bibitem[{Nutzman \& Charbonneau(2008)}]{MEarth_Nutzman2008}
Nutzman, P., \& Charbonneau, D. 2008, Publications of the Astronomical Society
  of the Pacific, 120, 317, \dodoi{10.1086/533420}

\bibitem[{Olmschenk {et~al.}(2021)Olmschenk, Silva, Rau, Barry, Kruse,
  Cacciapuoti, Kostov, Powell, Wyrwas, Schnittman, \& Barclay}]{Olmschenk2021}
Olmschenk, G., Silva, S.~I., Rau, G., {et~al.} 2021, The Astronomical Journal,
  161, 273, \dodoi{10.3847/1538-3881/abf4c6}

\bibitem[{pandas~development team(2020)}]{Pandas20}
pandas~development team, T. 2020, pandas-dev/pandas: Pandas, latest,  Zenodo,
  \dodoi{10.5281/zenodo.3509134}

\bibitem[{Paredes {et~al.}(2021)Paredes, Henry, Quinn, Gies,
  {Hinojosa-Go{\~n}i}, James, Jao, \& White}]{CHIRON_Paredes2021}
Paredes, L.~A., Henry, T.~J., Quinn, S.~N., {et~al.} 2021, The Astronomical
  Journal, 162, 176, \dodoi{10.3847/1538-3881/ac082a}

\bibitem[{Pepe {et~al.}(2002)Pepe, Mayor, Galland, Naef, Queloz, Santos, Udry,
  \& Burnet}]{Pepe2002}
Pepe, F., Mayor, M., Galland, F., {et~al.} 2002, Astronomy \& Astrophysics,
  388, 632, \dodoi{10.1051/0004-6361:20020433}

\bibitem[{Pepper {et~al.}(2012)Pepper, Kuhn, Siverd, James, \&
  Stassun}]{KELT_Pepper2012}
Pepper, J., Kuhn, R.~B., Siverd, R., James, D., \& Stassun, K. 2012,
  Publications of the Astronomical Society of the Pacific, 124, 230,
  \dodoi{10.1086/665044}

\bibitem[{Pepper {et~al.}(2007)Pepper, Pogge, DePoy, Marshall, Stanek, Stutz,
  Poindexter, Siverd, O'Brien, Trueblood, \& Trueblood}]{KELT_Pepper2007}
Pepper, J., Pogge, R.~W., DePoy, D.~L., {et~al.} 2007, Publications of the
  Astronomical Society of the Pacific, 119, 923, \dodoi{10.1086/521836}

\bibitem[{{Petigura}(2015)}]{SpecMatchSynth_Petigura2015}
{Petigura}, E.~A. 2015, PhD thesis, University of California, Berkeley, United
  States

\bibitem[{Pollacco {et~al.}(2006)Pollacco, Skillen, Cameron, Christian,
  Hellier, Irwin, Lister, Street, West, Anderson, Clarkson, Deeg, Enoch, Evans,
  Fitzsimmons, Haswell, Hodgkin, Horne, Kane, Keenan, Maxted, Norton, Osborne,
  Parley, Ryans, Smalley, Wheatley, \& Wilson}]{WASP_Pollacco2006}
Pollacco, D.~L., Skillen, I., Cameron, A.~C., {et~al.} 2006, Publications of
  the Astronomical Society of the Pacific, 118, 1407, \dodoi{10.1086/508556}

\bibitem[{Quinn {et~al.}(2012)Quinn, White, Latham, Buchhave, Cantrell, Dahm,
  F{\textbackslash}Hur{\'e}sz, Szentgyorgyi, Geary, Torres, Bieryla, Berlind,
  Calkins, Esquerdo, \& Stefanik}]{Quinn2012}
Quinn, S.~N., White, R.~J., Latham, D.~W., {et~al.} 2012, The Astrophysical
  Journal, 756, L33, \dodoi{10.1088/2041-8205/756/2/L33}

\bibitem[{Rasio \& Ford(1996)}]{Rasio1996}
Rasio, F.~A., \& Ford, E.~B. 1996, Science, 274, 954,
  \dodoi{10.1126/science.274.5289.954}

\bibitem[{{Ricker} {et~al.}(2015){Ricker}, {Winn}, {Vanderspek}, {Latham},
  {Bakos}, {Bean}, {Berta-Thompson}, {Brown}, {Buchhave}, {Butler}, {Butler},
  {Chaplin}, {Charbonneau}, {Christensen-Dalsgaard}, {Clampin}, {Deming},
  {Doty}, {De Lee}, {Dressing}, {Dunham}, {Endl}, {Fressin}, {Ge}, {Henning},
  {Holman}, {Howard}, {Ida}, {Jenkins}, {Jernigan}, {Johnson}, {Kaltenegger},
  {Kawai}, {Kjeldsen}, {Laughlin}, {Levine}, {Lin}, {Lissauer}, {MacQueen},
  {Marcy}, {McCullough}, {Morton}, {Narita}, {Paegert}, {Palle}, {Pepe},
  {Pepper}, {Quirrenbach}, {Rinehart}, {Sasselov}, {Sato}, {Seager},
  {Sozzetti}, {Stassun}, {Sullivan}, {Szentgyorgyi}, {Torres}, {Udry}, \&
  {Villasenor}}]{TESS_Ricker15}
{Ricker}, G.~R., {Winn}, J.~N., {Vanderspek}, R., {et~al.} 2015, Journal of
  Astronomical Telescopes, Instruments, and Systems, 1, 014003,
  \dodoi{10.1117/1.JATIS.1.1.014003}

\bibitem[{Riello {et~al.}(2021)Riello, Angeli, Evans, Montegriffo, Carrasco,
  Busso, Palaversa, Burgess, Diener, Davidson, Rowell, Fabricius, Jordi,
  Bellazzini, Pancino, Harrison, Cacciari, van Leeuwen, Hambly, Hodgkin,
  Osborne, Altavilla, Barstow, Brown, Castellani, Cowell, Luise, Gilmore,
  Giuffrida, Hidalgo, Holland, Marinoni, Pagani, Piersimoni, Pulone, Ragaini,
  Rainer, Richards, Sanna, Walton, Weiler, \& Yoldas}]{GaiaEDR3_Riello2021}
Riello, M., Angeli, F.~D., Evans, D.~W., {et~al.} 2021, Astronomy \&
  Astrophysics, 649, A3, \dodoi{10.1051/0004-6361/202039587}

\bibitem[{Rodriguez {et~al.}(2019)Rodriguez, Quinn, Huang, Vanderburg, Penev,
  Brahm, Jord{\'a}n, {Ikwut-Ukwa}, Tsirulik, Latham, Stassun, Shporer, Ziegler,
  Matthews, Eastman, Gaudi, Collins, Guerrero, Relles, Barclay, Batalha,
  Berlind, Bieryla, Bouma, Boyd, Burt, Calkins, Christiansen, Ciardi,
  Col{\'o}n, Conti, Crossfield, Daylan, Dittmann, Dragomir, Dynes, Espinoza,
  Esquerdo, Essack, Soto, Glidden, G{\"u}nther, Henning, Jenkins, Kielkopf,
  Krishnamurthy, Law, Levine, Lewin, Mann, Morgan, Morris, Oelkers, Paegert,
  Pepper, Quintana, Ricker, Rowden, Seager, Sarkis, Schlieder, Sha, Tokovinin,
  Torres, Vanderspek, Villanueva, Villase{\~n}or, Winn, Wohler, Wong, Yahalomi,
  Yu, Zhan, \& Zhou}]{Rodriguez2019}
Rodriguez, J.~E., Quinn, S.~N., Huang, C.~X., {et~al.} 2019, The Astronomical
  Journal, 157, 191, \dodoi{10.3847/1538-3881/ab11d9}

\bibitem[{Rodriguez {et~al.}(2021)Rodriguez, Quinn, Zhou, Vanderburg, Nielsen,
  Wittenmyer, Brahm, Reed, Huang, Vach, Ciardi, Oelkers, Stassun, Hellier,
  Gaudi, Eastman, Collins, Bieryla, Christian, Latham, Carleo, Wright,
  Matthews, Gonzales, Ziegler, Dressing, Howell, Tan, Wittrock, Plavchan,
  McLeod, Baker, Wang, Radford, Schwarz, Esposito, Ricker, Vanderspek, Seager,
  Winn, Jenkins, Addison, Anderson, Barclay, Beatty, Berlind, Bouchy, Bowen,
  Bowler, Brasseur, Brice{\~n}o, Caldwell, Calkins, Cartwright, Chaturvedi,
  Chaverot, Chimaladinne, Christiansen, Collins, Crossfield, Eastridge,
  Espinoza, Esquerdo, Feliz, Fenske, Fong, Gan, Giacalone, Gill, Gordon,
  Granados, Grieves, Guenther, Guerrero, Henning, Henze, Hesse, Hobson, Horner,
  James, Jensen, Jimenez, Jord{\'a}n, Kane, Kielkopf, Kim, Kuhn, Latouf, Law,
  Levine, Lund, Mann, Mao, Matson, Mengel, Mink, Newman, O'Dwyer, Okumura,
  Palle, Pepper, Quintana, Sarkis, Savel, Schlieder, Schnaible, Shporer,
  Sefako, Seidel, Siverd, Skinner, Stalport, Stevens, Stibbards, Tinney, West,
  Yahalomi, \& Zhang}]{Rodriguez2021}
Rodriguez, J.~E., Quinn, S.~N., Zhou, G., {et~al.} 2021, The Astronomical
  Journal, 161, 194, \dodoi{10.3847/1538-3881/abe38a}

\bibitem[{Rodriguez {et~al.}(2022)Rodriguez, Quinn, Vanderburg, Zhou, Eastman,
  Thygesen, Cale, Ciardi, Reed, Oelkers, Collins, Bieryla, Latham, Gaudi,
  Hellier, Sokolovsky, Schulte, Srdoc, Kielkopf, Horta, Massey, Evans,
  Stephens, McLeod, Chazov, Krushinsky, Ghachoui, Safonov, Dedrick, Conti,
  Laloum, Giacalone, Ziegler, Serra, Nogues, Murgas, Michaels, Ricker,
  Vanderspek, Winn, Jenkins, Addison, Alfaro, Anderson, Ayad, Bedding,
  Belinsky, Benkhaldoun, Berlind, Blake, Bowen, Bowler, Boyle, Branson,
  Briceno, Calkins, Campbell, Chomiuk, Collins, Cornachione, Daassou, Dressing,
  Esquerdo, Feliz, Fong, Gan, Gill, Goliguzova, Hansen, Hintz, Horner, Huang,
  James, Jensen, Johnson, Kane, Barkaoui, Kim, Kim, Kuhn, Law, Lewin, Liu,
  Lund, Mann, McCrady, Mengel, Mink, Murphy, Narita, Newman, Okumura, Osborn,
  Paegert, Palle, Pepper, Plavchan, Popov, Rabus, Ranshaw, Rodriguez, Roh,
  Reefe, Savel, Schwarz, Shporer, Siverd, Sliski, Stassun, Stevens, Soubkiou,
  Ting, Tinney, Vowell, West, Wilson, Wittenmyer, Wittrock, Wright, Zhang, \&
  Zobel}]{Rodriguez2022}
Rodriguez, J.~E., Quinn, S.~N., Vanderburg, A., {et~al.} 2022, arXiv:2205.05709
  [astro-ph].
\newblock \doarXiv{2205.05709}

\bibitem[{Rossiter(1924)}]{Rossiter1924}
Rossiter, R.~A. 1924, The Astrophysical Journal, 60, 15, \dodoi{10.1086/142825}

\bibitem[{Safonov {et~al.}(2017)Safonov, Lysenko, \& Dodin}]{SAI_Safonov2017}
Safonov, B.~S., Lysenko, P.~A., \& Dodin, A.~V. 2017, Astronomy Letters, 43,
  344, \dodoi{10.1134/S1063773717050036}

\bibitem[{Santos {et~al.}(2004)Santos, Israelian, \& Mayor}]{Santos2004}
Santos, N.~C., Israelian, G., \& Mayor, M. 2004, Astronomy \& Astrophysics,
  415, 1153, \dodoi{10.1051/0004-6361:20034469}

\bibitem[{{Savel} {et~al.}(2020){Savel}, {Dressing}, {Hirsch}, {Ciardi},
  {Fleming}, {Giacalone}, {Mayo}, \& {Christiansen}}]{SIMmER_Savel2020}
{Savel}, A.~B., {Dressing}, C.~D., {Hirsch}, L.~A., {et~al.} 2020, \aj, 160,
  287, \dodoi{10.3847/1538-3881/abc47d}

\bibitem[{Schlafly \& Finkbeiner(2011)}]{Schlafly2011}
Schlafly, E.~F., \& Finkbeiner, D.~P. 2011, The Astrophysical Journal, 737,
  103, \dodoi{10.1088/0004-637X/737/2/103}

\bibitem[{Schlegel {et~al.}(1998)Schlegel, Finkbeiner, \& Davis}]{Schlegel1998}
Schlegel, D.~J., Finkbeiner, D.~P., \& Davis, M. 1998, The Astrophysical
  Journal, 500, 525, \dodoi{10.1086/305772}

\bibitem[{Schwab {et~al.}(2016)Schwab, Rakich, Gong, Mahadevan, Halverson, Roy,
  Terrien, Robertson, Hearty, Levi, Monson, Wright, McElwain, Bender, Blake,
  St{\"u}rmer, Gurevich, Chakraborty, \& Ramsey}]{NEID_Schwab2016}
Schwab, C., Rakich, A., Gong, Q., {et~al.} 2016, in Ground-Based and {{Airborne
  Instrumentation}} for {{Astronomy VI}}, Vol. 9908 ({SPIE}), 2220--2225,
  \dodoi{10.1117/12.2234411}

\bibitem[{Scott \& Howell(2018)}]{Gemini_NESSI_Alopeke_Scott2018}
Scott, N.~J., \& Howell, S.~B. 2018, in Optical and {{Infrared Interferometry}}
  and {{Imaging VI}}, Vol. 10701 ({SPIE}), 99--103, \dodoi{10.1117/12.2311539}

\bibitem[{Scott {et~al.}(2021)Scott, Howell, Gnilka, Stephens, Salinas, Matson,
  Furlan, Horch, Everett, Ciardi, Mills, \&
  Quigley}]{Gemini_Zorro_Alopeke_Scott2021}
Scott, N.~J., Howell, S.~B., Gnilka, C.~L., {et~al.} 2021, Frontiers in
  Astronomy and Space Sciences, 8, 138, \dodoi{10.3389/fspas.2021.716560}

\bibitem[{Sha {et~al.}(2021)Sha, Huang, Shporer, Rodriguez, Vanderburg, Brahm,
  Hagelberg, Matthews, Ziegler, Livingston, Stassun, Wright, Crane, Espinoza,
  Bouchy, Bakos, Collins, Zhou, Bieryla, Hartman, Wittenmyer, Nielsen,
  Plavchan, Bayliss, Sarkis, Tan, Cloutier, Mancini, Jord{\'a}n, Wang, Henning,
  Narita, Penev, Teske, Kane, Mann, Addison, Tamura, Horner, Barbieri, Burt,
  D{\'i}az, Crossfield, Dragomir, Drass, Feinstein, Zhang, Hart, Kielkopf,
  Jensen, Montet, Ottoni, Schwarz, Rojas, Nespral, Torres, Mengel, Udry,
  Zapata, Snoddy, Okumura, Ricker, Vanderspek, Latham, Winn, Seager, Jenkins,
  Col{\'o}n, Henze, Krishnamurthy, Ting, Vezie, \& Villanueva}]{Sha2021}
Sha, L., Huang, C.~X., Shporer, A., {et~al.} 2021, The Astronomical Journal,
  161, 82, \dodoi{10.3847/1538-3881/abd187}

\bibitem[{Shallue \& Vanderburg(2018)}]{Keplerspline_Shallue2018}
Shallue, C.~J., \& Vanderburg, A. 2018, The Astronomical Journal, 155, 94,
  \dodoi{10.3847/1538-3881/aa9e09}

\bibitem[{Smith {et~al.}(2012)Smith, Stumpe, Cleve, Jenkins, Barclay, Fanelli,
  Girouard, Kolodziejczak, McCauliff, Morris, \& Twicken}]{TESS_PDC_Smith2012}
Smith, J.~C., Stumpe, M.~C., Cleve, J. E.~V., {et~al.} 2012, Publications of
  the Astronomical Society of the Pacific, 124, 1000, \dodoi{10.1086/667697}

\bibitem[{Stassun {et~al.}(2018)Stassun, Oelkers, Pepper, Paegert, DeLee,
  Torres, Latham, Charpinet, Dressing, Huber, Kane, Lepine, Mann, Muirhead,
  {Rojas-Ayala}, Silvotti, Fleming, Levine, Plavchan, \&
  Group}]{TIC_Stassun2018}
Stassun, K.~G., Oelkers, R.~J., Pepper, J., {et~al.} 2018, The Astronomical
  Journal, 156, 102, \dodoi{10.3847/1538-3881/aad050}

\bibitem[{Stassun {et~al.}(2019)Stassun, Oelkers, Paegert, Torres, Pepper, Lee,
  Collins, Latham, Muirhead, Chittidi, {Rojas-Ayala}, Fleming, Rose, Tenenbaum,
  Ting, Kane, Barclay, Bean, Brassuer, Charbonneau, Ge, Lissauer, Mann, McLean,
  Mullally, Narita, Plavchan, Ricker, Sasselov, Seager, Sharma, Shiao,
  Sozzetti, Stello, Vanderspek, Wallace, \& Winn}]{TIC_Stassun2019}
Stassun, K.~G., Oelkers, R.~J., Paegert, M., {et~al.} 2019, The Astronomical
  Journal, 158, 138, \dodoi{10.3847/1538-3881/ab3467}

\bibitem[{Stumpe {et~al.}(2014)Stumpe, Smith, Catanzarite, Cleve, Jenkins,
  Twicken, \& Girouard}]{TESS_PDC_Stumpe2014}
Stumpe, M.~C., Smith, J.~C., Catanzarite, J.~H., {et~al.} 2014, Publications of
  the Astronomical Society of the Pacific, 126, 100, \dodoi{10.1086/674989}

\bibitem[{Stumpe {et~al.}(2012)Stumpe, Smith, Cleve, Twicken, Barclay, Fanelli,
  Girouard, Jenkins, Kolodziejczak, McCauliff, \& Morris}]{TESS_PDC_Stumpe2012}
Stumpe, M.~C., Smith, J.~C., Cleve, J. E.~V., {et~al.} 2012, Publications of
  the Astronomical Society of the Pacific, 124, 985, \dodoi{10.1086/667698}

\bibitem[{Teske {et~al.}(2018)Teske, Wang, Wolfgang, Dai, Shectman, Butler,
  Crane, \& Thompson}]{Teske2018}
Teske, J.~K., Wang, S., Wolfgang, A., {et~al.} 2018, The Astronomical Journal,
  155, 148, \dodoi{10.3847/1538-3881/aaab56}

\bibitem[{Thorngren {et~al.}(2021)Thorngren, Fortney, Lopez, Berger, \&
  Huber}]{Thorngren2021}
Thorngren, D.~P., Fortney, J.~J., Lopez, E.~D., Berger, T.~A., \& Huber, D.
  2021, The Astrophysical Journal Letters, 909, L16,
  \dodoi{10.3847/2041-8213/abe86d}

\bibitem[{Tokovinin(2018)}]{SOAR_Tokovinin2018}
Tokovinin, A. 2018, Publications of the Astronomical Society of the Pacific,
  130, 035002, \dodoi{10.1088/1538-3873/aaa7d9}

\bibitem[{Tokovinin \& Cantarutti(2008)}]{SOAR_Tokovinin2008}
Tokovinin, A., \& Cantarutti, R. 2008, Publications of the Astronomical Society
  of the Pacific, 120, 170, \dodoi{10.1086/528809}

\bibitem[{Tokovinin {et~al.}(2013)Tokovinin, Fischer, Bonati, Giguere, Moore,
  Schwab, Spronck, \& Szymkowiak}]{CHIRON_Tokovinin2013}
Tokovinin, A., Fischer, D.~A., Bonati, M., {et~al.} 2013, Publications of the
  Astronomical Society of the Pacific, 125, 1336, \dodoi{10.1086/674012}

\bibitem[{Torres {et~al.}(2004)Torres, Konacki, Sasselov, \& Jha}]{Torres2004}
Torres, G., Konacki, M., Sasselov, D.~D., \& Jha, S. 2004, The Astrophysical
  Journal, 614, 979, \dodoi{10.1086/423734}

\bibitem[{{Trifonov} {et~al.}(2021){Trifonov}, {Brahm}, {Espinoza}, {Henning},
  {Jord{\'a}n}, {Nesvorny}, {Dawson}, {Lissauer}, {Lee}, {Kossakowski},
  {Rojas}, {Hobson}, {Sarkis}, {Schlecker}, {Bitsch}, {Bakos}, {Barbieri},
  {Bhatti}, {Butler}, {Crane}, {Nandakumar}, {D{\'\i}az}, {Shectman}, {Teske},
  {Torres}, {Suc}, {Vines}, {Wang}, {Ricker}, {Shporer}, {Vanderburg},
  {Dragomir}, {Vanderspek}, {Burke}, {Daylan}, {Shiao}, {Jenkins}, {Wohler},
  {Seager}, \& {Winn}}]{WINE_Trifonov2021}
{Trifonov}, T., {Brahm}, R., {Espinoza}, N., {et~al.} 2021, \aj, 162, 283,
  \dodoi{10.3847/1538-3881/ac1bbe}

\bibitem[{Valenti \& Fischer(2005)}]{Valenti2005}
Valenti, J.~A., \& Fischer, D.~A. 2005, The Astrophysical Journal Supplement
  Series, 159, 141, \dodoi{10.1086/430500}

\bibitem[{Vanderburg \& Johnson(2014)}]{Keplerspline_Vanderburg2014}
Vanderburg, A., \& Johnson, J.~A. 2014, Publications of the Astronomical
  Society of the Pacific, 126, 948, \dodoi{10.1086/678764}

\bibitem[{Virtanen {et~al.}(2020)Virtanen, Gommers, Oliphant, Haberland, Reddy,
  Cournapeau, Burovski, Peterson, Weckesser, Bright, {van der Walt}, Brett,
  Wilson, Millman, Mayorov, Nelson, Jones, Kern, Larson, Carey, Polat, Feng,
  Moore, {VanderPlas}, Laxalde, Perktold, Cimrman, Henriksen, Quintero, Harris,
  Archibald, Ribeiro, Pedregosa, {van Mulbregt}, \& {SciPy 1.0
  Contributors}}]{Scipy}
Virtanen, P., Gommers, R., Oliphant, T.~E., {et~al.} 2020, Nature Methods, 17,
  261, \dodoi{10.1038/s41592-019-0686-2}

\bibitem[{{Vogt} {et~al.}(1994){Vogt}, {Allen}, {Bigelow}, {Bresee}, {Brown},
  {Cantrall}, {Conrad}, {Couture}, {Delaney}, {Epps}, {Hilyard}, {Hilyard},
  {Horn}, {Jern}, {Kanto}, {Keane}, {Kibrick}, {Lewis}, {Osborne},
  {Pardeilhan}, {Pfister}, {Ricketts}, {Robinson}, {Stover}, {Tucker}, {Ward},
  \& {Wei}}]{HIRES_Vogt94}
{Vogt}, S.~S., {Allen}, S.~L., {Bigelow}, B.~C., {et~al.} 1994, in Society of
  Photo-Optical Instrumentation Engineers (SPIE) Conference Series, Vol. 2198,
  Instrumentation in Astronomy VIII, ed. D.~L. {Crawford} \& E.~R. {Craine},
  362, \dodoi{10.1117/12.176725}

\bibitem[{{W}es {M}c{K}inney(2010)}]{Pandas_McKinney10}
{W}es {M}c{K}inney. 2010, in {P}roceedings of the 9th {P}ython in {S}cience
  {C}onference, ed. {S}t\'efan van~der {W}alt \& {J}arrod {M}illman, 56 -- 61,
  \dodoi{10.25080/Majora-92bf1922-00a}

\bibitem[{Wheatley {et~al.}(2018)Wheatley, West, Goad, Jenkins, Pollacco,
  Queloz, Rauer, Udry, Watson, Chazelas, Eigm{\"u}ller, Lambert, Genolet,
  McCormac, Walker, Armstrong, Bayliss, Bento, Bouchy, Burleigh, Cabrera,
  Casewell, Chaushev, Chote, Csizmadia, Erikson, Faedi, Foxell, G{\"a}nsicke,
  Gillen, Grange, G{\"u}nther, Hodgkin, Jackman, Jord{\'a}n, Louden,
  Metrailler, Moyano, Nielsen, Osborn, Poppenhaeger, Raddi, Raynard, Smith,
  Soto, \& {Titz-Weider}}]{NGTS_Wheatley2018}
Wheatley, P.~J., West, R.~G., Goad, M.~R., {et~al.} 2018, Monthly Notices of
  the Royal Astronomical Society, 475, 4476, \dodoi{10.1093/mnras/stx2836}

\bibitem[{Wittenmyer {et~al.}(2018)Wittenmyer, Horner, Carter, Kane, Plavchan,
  Ciardi, \& {consortium}}]{MinervaAustralis_Wittenmyer2018}
Wittenmyer, R.~A., Horner, J., Carter, B.~D., {et~al.} 2018, arXiv:1806.09282
  [astro-ph].
\newblock \doarXiv{1806.09282}

\bibitem[{Wittenmyer {et~al.}(2022)Wittenmyer, Clark, Trifonov, Addison,
  Wright, Stassun, Horner, Lowson, Kielkopf, Kane, Plavchan, Shporer, Zhang,
  Bowler, Mengel, Okumura, Rabus, Johnson, Harbeck, Tronsgaard, Buchhave,
  Collins, Collins, Gan, Jensen, Howell, Furlan, Gnilka, Lester, Matson, Scott,
  Ricker, Vanderspek, Latham, Seager, Winn, Jenkins, Rudat, Quintana,
  Rodriguez, Caldwell, Quinn, Essack, \& Bouma}]{Wittenmyer2022}
Wittenmyer, R.~A., Clark, J.~T., Trifonov, T., {et~al.} 2022, The Astronomical
  Journal, 163, 82, \dodoi{10.3847/1538-3881/ac3f39}

\bibitem[{Wong {et~al.}(2021)Wong, Shporer, Zhou, Kitzmann, Komacek, Tan,
  Tronsgaard, Buchhave, Vissapragada, {Greklek-McKeon}, Rodriguez, Ahlers,
  Quinn, Furlan, Howell, Bieryla, Heng, Knutson, Collins, McLeod, Berlind,
  Brown, Calkins, de~Leon, {Esparza-Borges}, Esquerdo, Fukui, Gan, Girardin,
  Gnilka, Ikoma, Jensen, Kielkopf, Kodama, Kurita, Lester, Lewin, Marino,
  Murgas, Narita, Pall{\'e}, Schwarz, Stassun, Tamura, Watanabe, Benneke,
  Ricker, Latham, Vanderspek, Seager, Winn, Jenkins, Caldwell, Fong, Huang,
  Mireles, Schlieder, Shiao, \& Villase{\~n}or}]{Wong2021}
Wong, I., Shporer, A., Zhou, G., {et~al.} 2021, The Astronomical Journal, 162,
  256, \dodoi{10.3847/1538-3881/ac26bd}

\bibitem[{Yee {et~al.}(2017)Yee, Petigura, \& {von
  Braun}}]{SpecMatchEmp_Yee2017}
Yee, S.~W., Petigura, E.~A., \& {von Braun}, K. 2017, The Astrophysical
  Journal, 836, 77, \dodoi{10.3847/1538-4357/836/1/77}

\bibitem[{Yee {et~al.}(2021)Yee, Winn, \& Hartman}]{Yee2021b}
Yee, S.~W., Winn, J.~N., \& Hartman, J.~D. 2021, The Astronomical Journal, 162,
  240, \dodoi{10.3847/1538-3881/ac2958}

\bibitem[{Zhou {et~al.}(2019)Zhou, Huang, Bakos, Hartman, Latham, Quinn,
  Collins, Winn, Wong, Kov{\'a}cs, Csubry, Bhatti, Penev, Bieryla, Esquerdo,
  Berlind, Calkins, de~{Val-Borro}, Noyes, L{\'a}z{\'a}r, Papp, S{\'a}ri,
  Kov{\'a}cs, Buchhave, Szklenar, B{\'e}ky, Johnson, Cochran, Kniazev, Stassun,
  Fulton, Shporer, Espinoza, Bayliss, Everett, Howell, Hellier, Anderson,
  Cameron, West, Brown, Schanche, Barkaoui, Pozuelos, Gillon, Jehin,
  Benkhaldoun, Daassou, Ricker, Vanderspek, Seager, Jenkins, Lissauer,
  Armstrong, Collins, Gan, Hart, Horne, Kielkopf, Nielsen, Nishiumi, Narita,
  Palle, Relles, Sefako, Tan, Davies, Goeke, Guerrero, Haworth, \&
  Villanueva}]{Zhou2019a}
Zhou, G., Huang, C.~X., Bakos, G.~{\'A}., {et~al.} 2019, The Astronomical
  Journal, 158, 141, \dodoi{10.3847/1538-3881/ab36b5}

\bibitem[{Zhou {et~al.}(2020)Zhou, Quinn, Irwin, Huang, Collins, Bouma, Khan,
  Landrigan, Vanderburg, Rodriguez, Latham, Torres, Douglas, Bieryla, Esquerdo,
  Berlind, Calkins, Buchhave, Charbonneau, Collins, Kielkopf, Jensen, Tan,
  Hart, Carter, Stockdale, Ziegler, Law, Mann, Howell, Matson, Scott, Furlan,
  White, Hellier, Anderson, West, Ricker, Vanderspek, Seager, Jenkins, Winn,
  Mireles, Rowden, Yahalomi, Wohler, Brasseur, Daylan, \&
  Col{\'o}n}]{CHIRON_Zhou2020}
Zhou, G., Quinn, S.~N., Irwin, J., {et~al.} 2020, The Astronomical Journal,
  161, 2, \dodoi{10.3847/1538-3881/abba22}

\bibitem[{Ziegler {et~al.}(2019)Ziegler, Tokovinin, Brice{\~n}o, Mang, Law, \&
  Mann}]{SOAR_TESS_Ziegler2019}
Ziegler, C., Tokovinin, A., Brice{\~n}o, C., {et~al.} 2019, \aj, 159, 19,
  \dodoi{10.3847/1538-3881/ab55e9}

\bibitem[{Ziegler {et~al.}(2021)Ziegler, Tokovinin, Latiolais, Brice{\~n}o,
  Law, \& Mann}]{SOAR_TESS_Ziegler2021}
Ziegler, C., Tokovinin, A., Latiolais, M., {et~al.} 2021, The Astronomical
  Journal, 162, 192, \dodoi{10.3847/1538-3881/ac17f6}

\end{thebibliography}
\bibliographystyle{aasjournal}

\end{document}